\documentclass[12pt]{article}

\usepackage{amsmath}
\usepackage{amssymb}
\usepackage{epsfig}
\usepackage{graphicx}
\newcommand{\capdef}{}
\newcommand{\mycaption}[2][\capdef]{\renewcommand{\capdef}{#2}%
        \caption[#1]{{\itshape #2}}}
\makeatletter
\renewcommand{\fnum@table}{\textbf{\tablename~\thetable}}
\renewcommand{\fnum@figure}{\textbf{\figurename~\thefigure}}
\makeatother

\newcounter{myenumi}

\renewcommand{\themyenumi}{\roman{myenumi}}
{\end{list}}

\newlength{\myem}
\newcounter{mysubequation}[equation]

\makeatletter
\renewcommand{\section}{\@startsection{section}{1}{0em}{-\baselineskip}%
{\baselineskip}{\normalfont\large\bfseries}}
\renewcommand{\subsection}%
{\@startsection{subsection}{2}{0em}{-0.7\baselineskip}%
{0.7\baselineskip}{\normalfont\bfseries}}
\makeatother

\newcommand{\bea}{\begin{eqnarray*}}
\newcommand{\eea}{\end{eqnarray*}}

\newcommand{\deltacp}{\delta_\mathrm{CP}}

\newcommand{\GeV}{\,\mathrm{GeV}}

\newcommand{\eV}{\,\mathrm{eV}}

\newcommand{\reu}{{\nu_e\rightarrow\nu_\mu}}
\newcommand{\ree}{{\nu_e\rightarrow\nu_e}}
\newcommand{\rue}{{\nu_\mu\rightarrow\nu_e}}
\newcommand{\reub}{{\bar{\nu}_e\rightarrow\bar{\nu}_\mu}}
\newcommand{\ruu}{{\nu_\mu\rightarrow\nu_\mu}}
\newcommand{\rux}{{\nu_\mu\rightarrow\nu_x}}
\newcommand{\ruxb}{{\bar{\nu}_\mu\rightarrow\bar{\nu}_x}}
\newcommand{\ruub}{{\bar{\nu}_\mu\rightarrow\bar{\nu}_\mu}}
\newcommand{\reeb}{{\bar{\nu}_e\rightarrow\bar{\nu}_e}}

\newcommand{\dm}[1]{{\Delta m^2_{#1}}}

\DeclareMathOperator{\im}{Im}
\DeclareMathOperator{\re}{Re}

\newcommand{\ie}{{\it i.e.}}

\newcommand{\eg}{{\it e.g.}}

\newcommand{\cf}{{\it cf.}}

\newcommand{\eq}{Equation}
\newcommand{\eqs}{Equations}

\newcommand{\fig}{Figure}
\newcommand{\Fig}{Figure}
\newcommand{\figs}{Figures}
\newcommand{\Figs}{Figures}
\newcommand{\Ref}{Ref.}
\newcommand{\Refs}{Refs.}
\newcommand{\Sec}{Section}
\newcommand{\Secs}{Sections}
\newcommand{\App}{Appendix}
\newcommand{\Apps}{Appendices}
\newcommand{\Tab}{Table}

\newcommand{\JHFHK}{JHF-HK}
\newcommand{\JHFSK}{JHF-SK}
\newcommand{\NuFactI}{NuFact-I}
\newcommand{\NuFactII}{NuFact-II}

\begin{document}
%%%%%%%%%%%%%%%%%%%%%%%%%%%%%%%%%%%%%%%%%%%%%%%%%%%%%%%%%%%%%%%%%%%%%
%%%%                     Title-page                              %%%%
%%%%%%%%%%%%%%%%%%%%%%%%%%%%%%%%%%%%%%%%%%%%%%%%%%%%%%%%%%%%%%%%%%%%%

\begin{titlepage}

% the footnote symbols are only redefined for the title page !
\renewcommand{\thefootnote}{\alph{footnote}}

\vspace*{-3.cm}
\begin{flushright}
TUM-HEP-462/02\\
MPI-PhT/02-15\\
%hep-ph/0204352
\end{flushright}

\vspace*{0.5cm}

\renewcommand{\thefootnote}{\fnsymbol{footnote}}
\setcounter{footnote}{-1}

{\begin{center}
{\large\bf Superbeams versus Neutrino
Factories$^*$\footnote{\hspace*{-1.6mm}$^*$Work supported by
``Sonderforschungsbereich 375 f\"ur Astro-Teilchenphysik'' der Deutschen
Forschungsgemeinschaft and the ``Studienstiftung des deutschen Volkes'' (German
National Merit Foundation) [W.W.].}} \end{center}}
\renewcommand{\thefootnote}{\alph{footnote}}

\vspace*{.8cm}
\vspace*{.3cm}
{\begin{center} {\large{\sc
                P.~Huber\footnote[1]{\makebox[1.cm]{Email:}
                phuber@ph.tum.de},~
                M.~Lindner\footnote[2]{\makebox[1.cm]{Email:}
                lindner@ph.tum.de},~and~
                W.~Winter\footnote[3]{\makebox[1.cm]{Email:}
                wwinter@ph.tum.de}~
                }}
\end{center}}
\vspace*{0cm}
{\it
\begin{center}

\footnotemark[1]${}^,$\footnotemark[2]${}^,$\footnotemark[3]%
       Institut f\"ur theoretische Physik, Physik--Department,
       Technische Universit\"at M\"unchen,\\
       James--Franck--Strasse, D--85748 Garching, Germany

\footnotemark[1]%
       Max-Planck-Institut f\"ur Physik, Postfach 401212,
       D--80805 M\"unchen, Germany

\end{center}}

\vspace*{1.5cm}

{\Large \bf
\begin{center} Abstract \end{center}  }

We compare the physics potential of planned superbeams with the
one of neutrino factories. Therefore, the experimental
setups as well as the most relevant uncertainties and errors are
considered on the same footing as much as possible. We
use an improved analysis including the full parameter correlations, as well as
statistical, systematical, and degeneracy errors. Especially, degeneracies have
so far not been taken into account in a numerical analysis. We furthermore
include external input, such as improved knowledge of the solar oscillation
parameters from the KamLAND experiment. This allows us to determine the
limiting uncertainties in all cases. For a specific comparison, we choose two
representatives of each class:
For the superbeam, we take the first conceivable setup, namely the
JHF to SuperKamiokande experiment, as well as, on a longer time scale, the
JHF to HyperKamiokande experiment. For the neutrino factory,
we choose an initially conceivable setup and an advanced machine.
We determine the potential to measure the small mixing angle
$\sin^2 2 \theta_{13}$, the sign of $\Delta m^2_{31}$, and the
leptonic CP phase $\deltacp$, which also implies that we compare the
limitations of the different setups. We find interesting results, such as the
complete loss of the sensitivity to the sign of $\Delta m^2_{31}$ due to
degeneracies in many cases.

\vspace*{.5cm}

\end{titlepage}

\newpage

\renewcommand{\thefootnote}{\arabic{footnote}}
\setcounter{footnote}{0}

%%%%%%%%%%%%%%%%%%%%%%%%%%%%%%%%%%%%%%%%%%%%%%%%%%%%%%%%%%%%%%%%%%%%%
%                     Introduction                                  %
%%%%%%%%%%%%%%%%%%%%%%%%%%%%%%%%%%%%%%%%%%%%%%%%%%%%%%%%%%%%%%%%%%%%%

\section{Introduction \label{sec:SEC-intro}}

Atmospheric neutrino oscillations are now well established
and imply for energies of some $\mathrm{GeV}$ neutrino oscillations on scales of
about $100$ to $10,000 \, \mathrm{km}$. The combination of all available
data strongly favors the LMA solution for solar oscillations, which allows a
sizable leptonic CP violation in long baseline oscillations.
So--called neutrino factories are therefore very promising for
precise measurements of the neutrino mass squared differences, the mixing
angles, and the leptonic CP violation. The physics potential of neutrino
factories has been studied for different setups, such as in
\Refs~\cite{DeRujula:1998hd,Barger:1999jj,FLPR,CERVERA,Barger:2000yn,FHL,Albright:2000xi,Burguet-Castell:2001ez,Freund:2001ui,Yasuda:2001ip,Bueno:2001jd,Yasuda:2002jk}. However, there are still
technological challenges to be solved, which means that the first neutrino
factory can not be expected to be completed within this decade. Until then,
upgraded conventional beams, so--called superbeams, can be used as alternative
neutrino sources, especially since much of their technology is anyway needed for
neutrino factories. The physics potential of superbeams has, for example,
been investigated in
\Refs~\cite{Barger:2000nf,Gomez-Cadenas:2001eu,Itow:2001ee,Minakata:2001qm,Barenboim:2002zx}. Thus,
superbeams are the next logical step in staging scenarios towards neutrino
factories. However, one may wonder where the physics potential of superbeams
ends and if they could make neutrino factories obsolete. It is very hard to
compare superbeams and neutrino factories on the basis of the specific studies
above, because numerous setups and simplifications have been used and different
levels of sophistication were employed, which means that it is
not clear how these options can be compared when all effects are included and
the same methods are applied. The aim of this paper is to study these two
options in a comparative way and to demonstrate that, at least for the case of
the LMA solution, neutrino factories are required for high precision
measurements. Since the leading oscillation parameters can in all cases be
determined with a very good precision, we focus on the most difficult parameters
to measure in our comparison, which are $\sin^2 2 \theta_{13}$, $\mathrm{sgn}
(\Delta m_{31}^2 )$, and $\deltacp$.

As mentioned above, an important issue in the comparison of superbeams and
neutrino factories is the use of reasonable and comparable parameters.
It does, for example, not make much sense to compare 20 years
of running of a very sophisticated JHF to HyperKamiokande experiment
with an initial stage neutrino factory running only for one year.
We therefore define scenarios with comparable running times
and similar sophistication within the respective framework.
Specifically, we choose the following reference scenarios:
First of all, the JHF to SuperKamiokande experiment (to be called \JHFSK)
with a $2^\circ$ off--axis beam represents the initial stage
superbeam. Its neutrino factory counterpart (labeled \NuFactI)
is chosen to have an equal running time with a moderate number of
stored muons. In addition, we take, as an advanced stage superbeam
experiment, the JHF to HyperKamiokande experiment (named \JHFHK),
again with a $2^\circ$ off--axis beam, compared to an advanced
neutrino factory (\NuFactII) with equal running time and a
large number of stored muons.
The ranges defined by \JHFSK\ to \JHFHK\ and \NuFactI\ to \NuFactII\ cover the 
conceivable experimental setups, which means that any superbeam or neutrino
factory experiment similar to ours can be expected to be found within these
ranges. Whenever applicable, we will therefore show the results of
the different scenarios in one plot to allow visual interpolation.
Our results are based on a combined event rate analysis of all
accessible oscillation channels with a statistical method which
treats all parameters equally instead of arbitrarily selecting
specific parameter subsets. This method was already applied
in neutrino factory studies in \Ref~\cite{Freund:2001ui} and it
was very successful in finding any two--parameter correlation
in the multi--dimensional parameter space. We use in this paper an
improved version, which allows us to include even multi--parameter
correlations. In addition, we include systematical errors, external input, and
degeneracy problems.

The paper is organized as follows: First we motivate and define the used
scenarios in \Sec~\ref{sec:experiments}. In \Sec~\ref{sec:classerrors}, we
introduce the different measurement errors which are included in our analysis.
We then present in \Sec~\ref{sec:framework} the oscillation framework, which
will allow us to understand most of the results analytically, followed by a
discussion of oscillation parameter degeneracies in \Sec~\ref{sec:degerrors},
which plays a major role in the analysis. In \Secs~\ref{sec:results1} and
\ref{sec:results2} we then present our results, where we especially focus on the
measurements of $\sin^2 2 \theta_{13}$ and $\deltacp$. In
\Sec~\ref{sec:conclusion}, we finally conclude with an overall comparison of the
main results. Details of the event rate analysis and the used statistical
methods are given in \Apps~\ref{sec:appendix}, \ref{sec:det}, and
\ref{sec:statistics}.

%%%%%%%%%%%%%%%%%%%%%%%%%%%%%%%%%%%%%%%%%%%%%%%%%%%%%%%%%%%%%
\section{Long baseline oscillation experiments}
\label{sec:experiments}

Superbeams and neutrino factory beams have many things in
common, but also have many differences in their setups. We therefore highlight
the most important features and differences in this section.

%%%%%%%%%%%%%%%%%%%%%%%%%%%%%%%%%%%%%%%%%%%%%%%%%%%%%%%%%%%%%
\subsection{Superbeams}
\label{sec:JHF}

Superbeams are based on conventional beam dump techniques for
producing neutrino beams \cite{CNGS1,CNGS2,MINOS,Nakamura:2000uu}.
An intense proton beam is typically directed onto a massive
target producing mostly pions, which are captured by an optical
system of magnets in order to obtain a pion beam. These pions
decay in a decay pipe, yielding a muon neutrino beam.
The neutrino beam, however, is contaminated by approximately
$0.5\%$ of electron neutrinos. Reversing the electrical current
in the lens system results in an antineutrino beam. The energy
spectrum of the muon beam can be controlled over a wide range:
it depends on the incident proton energy, the optical system,
and the precise direction of the beam axis compared to the
direction of the detector. It is possible to produce broad band
high energy beams, such as the CNGS beam~\cite{CNGS1,CNGS2}, or
 narrow band lower energy beams, such as in some configurations
of the NuMI beam~\cite{MINOS}.

Superbeams employ the same technique as conventional beams, but use a proton
beam intensity closer to the mechanical stability limit of the
target at a typical thermal power of $0.7\,\mathrm{MW}$ to $4\,\mathrm{MW}$. The
much higher neutrino luminosity allows the use of the decay kinematics of
pions to produce a so--called ``off--axis beam'', where the detector
is located some degrees off the main beam axis. This reduces the
overall neutrino flux, but leads to an important relative suppression
of the electron neutrino contamination. Several off--axis
superbeams with energies of about $1\,\mathrm{GeV}$ to $2\,
\mathrm{GeV}$ have been proposed in  Japan~\cite{Itow:2001ee,Aoki:2002ks},
America~\cite{Para:2001cu}, and Europe~\cite{Gomez-Cadenas:2001eu,Dydak}.

The most sensitive neutrino oscillation channel for the parameters we are
interested in, is the $\nu_\mu \rightarrow \nu_e$ appearance channel. Therefore,
the detector should have excellent capabilities to separate electron and muon
charged current events. In addition, an efficient rejection of neutral current
events is required, because the neutral current interaction mode is flavor
blind. With low statistics, the magnitude of the contamination itself limits the
sensitivity to the $\nu_\mu \rightarrow \nu_e $ transition severely, while with
high statistics the insufficient knowledge of its magnitude constrains the
sensitivity. A near detector allows, for example, a substantial reduction of the
background uncertainty~\cite{Itow:2001ee,Szleper:2001nj} and plays a crucial
role in controlling other systematical errors, such as the flux normalization,
the spectral shape of the beam, and the neutrino cross section at low energies.
At energies of about $1\,\mathrm{GeV}$, the dominant charge current interaction
mode is quasi--elastic scattering, which suggests that the water Cherenkov
detector is the optimal type of detector. At these energies, a baseline length
of about $300\,\mathrm{km}$ would be optimal to measure at the first
maximum of the oscillation. At about $2\,\mathrm{GeV}$, there is already a
considerable contribution of inelastic scattering to the charged current
interactions, which means that it would be useful to measure the energy of the
hadronic part of the cross section. This favors low--Z hadron calorimeters,
which also have a factor of ten better neutral current rejection capability
compared to water Cherenkov detectors~\cite{Para:2001cu}. In this case, the
optimum baseline is around $600\,\mathrm{km}$. The matter effects are expected
to be small for these experiments for two reasons. First of all, an
energy of about $1\,\mathrm{GeV}$ to $2\,\mathrm{GeV}$ is small compared to the
MSW resonance energy of approximately $13\,\mathrm{GeV}$ in the upper mantle.
The second reason is that the baseline is too short to build up significant
matter effects.

%%%%%%%%%%%%%%%%%%%%%%%%%%%%%%%%%%%%%%%%%%%%%%%%%%%%%%%%%%%%%
\subsection{Neutrino factories}
%%%%%%%%%%%%%%%%%%%%%%%%%%%%%%%%%%%%%%%%%%%%%%%%%%%%%%%%%%%%%
\label{sec:nufact}

At neutrino factories, decaying muons are stored in the long straight sections
of a storage ring and are producing muon and electron anti neutrinos of equal
numbers~\cite{Geer:1998iz}. The muons are produced by pion decays, where the
pions are produced by the same technique as for superbeams. After being
collected, they have to be cooled and reaccelerated very quickly. This has not
yet been demonstrated and is the major technological challenge for
neutrino factories~\cite{FNAL}. The spectrum and flavor content of
the beam are completely characterized by the muon decay and are
therefore very precisely known~\cite{PDG}. The only adjustable
parameter is the muon energy $E_\mu$, which is usually considered to be at
about $20$ to $50\,\mathrm{GeV}$. It is also possible
to produce and store anti muons in order to obtain a CP--conjugated beam. The
symmetric operation of both beams leads to the cancellation or
drastical reduction of many errors and systematical effects. Unless otherwise
stated, we will further on refer to the neutrino beam and leave it to the
reader to CP--conjugate all quantities for the anti neutrino beam.

Amongst all flavors and interaction types, muon charged current events are
the easiest to detect. The appearance channel with the best sensitivity
therefore is the $\reub$ transition, which produces so called ``wrong sign
muons''. Therefore, a detector must be able to identify the charge of a muon in
order to distinguish appearance and disappearance channels. The dominant charge
current interaction in the multi--GeV range is deep--inelastic scattering,
making a good energy resolution for the hadronic energy deposition necessary.
Magnetized iron calorimeters are thus the favored choice for neutrino factory
detectors. In order to achieve the required muon charge separation, it is
necessary to impose a minimum muon energy cut at at least
$4\,\mathrm{GeV}$~\cite{Blondel:2000gj}.\footnote{For different strategies to
the minimum muon momentum cut and the wrong sign muon background, as well as the
model we use in this paper, see \App~\ref{sec:det}.} This leads to a
significant loss of neutrino events in the range of about $4\,\mathrm{GeV}$ to
$20\,\mathrm{GeV}$, which means that a high muon energy of $E_\mu =
50\,\mathrm{GeV}$ is desirable. The first oscillation maximum then lies at
approximately $3\,000\,\mathrm{km}$. Matter effects are sizable at this baseline
and energy and our limited knowledge of the Earth's matter density profile is an
additional source of errors.

%%%%%%%%%%%%%%%%%%%%%%%%%%%%%%%%%%%%%%%%%%%%%%%%%%%%%%%%%%%%%
\subsection{Benchmark setups}
%%%%%%%%%%%%%%%%%%%%%%%%%%%%%%%%%%%%%%%%%%%%%%%%%%%%%%%%%%%%%
\label{sec:setup}

For the comparison of superbeams and neutrino factories, we
define benchmark setups. For each case we use an initial,
low luminosity experiment and an advanced, high luminosity
setup. For both superbeams and neutrino factories, we use in addition equal
assumptions about target power and running time, since these factors are limited
by similar technological and financial constraints.

As the prototype of a superbeam experiment, we choose the
JHF--Kamioka neutrino project as described in \Ref~\cite{Itow:2001ee},
for which a detailed proposal is available. This experiment is the most advanced
in planning and might be operational by 2007. Using the SuperKamiokande
detector is also very attractive, since this detector has been operating for a
long time and its performance is therefore well known. Furthermore, the proposed
NuMI off--axis beam seems to be quite similar in its characteristics and physics
reach~\cite{Para:2001cu}. Another argument to choose the JHF beam
is that there exists an upgrade strategy both for beam and detector, which
provides a straightforward choice for the advanced, high luminosity setup. Our
benchmark setups are:

\begin{description}
\item[\JHFSK] is the initial, low luminosity version of the
superbeam experiment. The target power is $0.75\,\mathrm{MW}$ and we
assume a running period of 5 years with a neutrino beam. As detector,
we use the SuperKamiokande detector with a fiducial mass of
$22.5\,\mathrm{kt}$ and a baseline of $295\,\mathrm{km}$.
\item[\JHFHK] is the advanced, high luminosity superbeam setup. The
assumed target power is $4\,\mathrm{MW}$ and the assumed running period
is 6 years with an anti neutrino beam and 2 years with a neutrino beam.
As detector, we use the proposed HyperKamiokande detector with a
fiducial mass of $1\,000\,\mathrm{kt}$ and a baseline length of
$295\,\mathrm{km}$. \end{description}

For the neutrino factories, we consider a parent muon
energy of $E_\mu=50\,\mathrm{GeV}$ and a baseline of $3\,000\,\mathrm{km}$.
Many studies indicate that this setup provides, in some sense, an optimal
sensitivity to CP effects (for a summary, see~\cite{Yasuda:2002jk}).
Specifically, we define:

\begin{description}
\item[\NuFactI] is the initial, low luminosity version of a neutrino
factory experiment with an assumed target power of $0.75\,\mathrm{MW}$,
corresponding to $10^{20}$ useful muon decays per year~\cite{FNAL}. The
total running period is $5$ years, $2.5$ years of these with a neutrino beam
and $2.5$ years of these with an anti neutrino beam. As detector, we use a
magnetized iron calorimeter with a fiducial mass of $10\,\mathrm{kt}$.
\item[\NuFactII] is the advanced, high luminosity neutrino factory setup
with an assumed target power of $4\,\mathrm{MW}$, corresponding to
$5.3 \cdot 10^{20}$ useful muon decays per year~\cite{FNAL}.
Corresponding to the advanced superbeam, we assume a total running
period of 8 years, 4 years of these with a neutrino beam and 4 years of these
with an anti neutrino beam. As detector, we use a magnetized iron calorimeter
with a fiducial mass of $50\,\mathrm{kt}$.
\end{description}

The luminosity increase from the initial to the advanced setup is for
the JHF experiments approximately a factor of 95 and for the
neutrino factory experiments roughly a factor of 42. The
statistics of each setup can be inferred from \Tab~\ref{tab:events},
which shows the number of signal and background events in the appearance
channel at $\sin^22\theta_{13}=0.1$, which is the biggest value allowed by
CHOOZ.
\begin{center}
\begin{table}[ht!]
\begin{center}
\begin{tabular}[h]{|l|rrrr|}
\hline
& \JHFSK\ & \JHFHK\ & \NuFactI\ & \NuFactII\ \\
\hline
Signal&139.0&13\,180.0&1\,522.8&64\,932.6\\
Background&23.3&2\,204.6&4.2&180.3\\
\hline
\hline
Total&162.3&15\,384.6&1\,527.0&65\,113.0\\
\hline
\end{tabular}
\mycaption{\label{tab:events} The number of appearance events for the
setups defined in the text. The used oscillation parameters are
$\dm{21}=3.7 \cdot10^{-5}\,\mathrm{eV}^2$,
$\sin^22\theta_{12}=0.8$, $\dm{31}=3 \cdot10^{-3}\,\mathrm{eV}^2$,
$\sin^22\theta_{23}=1$, $\sin^22\theta_{13}=0.1$, and $\deltacp=0$. }
\end{center}
\end{table}
\end{center}
%

%%%%%%%%%%%%%%%%%%%%%%%%%%%%%%%%%%%%%%%%%%%%%%%%%%%%%%%%%%%%%
\subsection{Simulation of experiments}
\label{sec:numerics}
%%%%%%%%%%%%%%%%%%%%%%%%%%%%%%%%%%%%%%%%%%%%%%%%%%%%%%%%%%%%%

Precision measurements at long baseline neutrino experiments
require a detailed understanding of the neutrino beam, the
propagation of the neutrinos in matter, and the detection of
the neutrinos. An evaluation of the physics potential must
therefore model all involved aspects very accurately. Our analysis is based on
event rates and adequate statistical methods, which are described in
\Apps~\ref{sec:appendix}, \ref{sec:det}, and \ref{sec:statistics}. The
implementation, as it is used in this work, is described below:

\subsubsection*{Neutrino production}

Each neutrino beam is characterized by the flux of each neutrino
flavor as a function of the energy (see \fig~\ref{fig:flux}).
A superbeam is mostly composed of $\nu_\mu$, but also contains small
contaminations (typically $<1\%$) of $\nu_e$, $\bar\nu_\mu$, and
$\bar{\nu}_e$.\footnote{For a positive horn current; for a negative horn
current, particles must here be replaced by anti particles.} The spectrum is
dominated by the decay kinematics of pions and their initial momentum
distribution, but K decays contribute as well, especially at higher energies.
The pion and K meson productions are hadronic processes, for which the
momentum distribution has to be determined by Monte--Carlo techniques.
Superbeams may therefore be subject to considerable
uncertainties~\cite{Itow:2001ee,NAKAYA}. We include in our simulation the
$\nu_e$ and $\bar{\nu}_e$ background of the beam and assign an
uncertainty to both its magnitude (referred to as {\em background
normalization}) and spectral shape (referred to as {\em background
tilt}).\footnote{For a more detailed description of the modeling, see
\App~\ref{sec:statistics}.} Anti--muon neutrino beams can be produced by
reversing the horn current, but the operation is not entirely symmetric. The two
polarities of a superbeam are therefore considered to be essentially independent
beams.

At a neutrino factory, the beam is produced by muon decay, \ie,
for muons in the storage ring,
by
\begin{equation*}
\mu^- \rightarrow e^- + \nu_\mu + \bar{\nu}_e.
\end{equation*}
It thus consists of equal numbers of $\nu_\mu$ and
$\bar{\nu}_e$~\cite{Geer:1998iz}. The kinematics of this decay
process rests only on energy and momentum conservation and the
beam of a neutrino factory can therefore be assumed to be exactly
known in terms of the initial muon energy and flux. Compared to the
superbeam, the machine can be operated symmetrically in both polarities.
Therefore, the CP--conjugated beam, emerging from stored $\mu^+$ decays,
can be assumed to be identical for the same energy and muon $\mu^+$ flux.

In general, we include reasonable uncertainties in the total flux
normalizations and shapes for both types of experiments.

\subsubsection*{Neutrino propagation}

The neutrino propagation is calculated by numerically solving the
Schr\"odinger equation for three neutrino flavors in a constant
(average) matter potential~\cite{FHL}. We treat all but the solar
oscillation parameters as quantities which have to be determined
by our experiments. This is a reasonable assumption, since the studied
experiments lead to a significantly better precision of the leading
oscillation parameter measurements than previous ones. The solar
parameters, however, can not be measured very well and are therefore assumed to
be constrained by the KamLAND experiment
within the LMA region~\cite{BARGER,Gonzalez-Garcia:2001zy}. Furthermore, since
the appearance probability has been shown to depend only on the product
$\dm{21}\,\sin^22\theta_{12}$~\cite{FREUND}, we use this product as a fit
parameter. For the matter density, we allow an uncertainty of about
5\%~\cite{Geller:2001ix} by treating it as an additional parameter.

\subsubsection*{Neutrino detection}

The detection of neutrinos involves two main aspects: the neutrino
interaction and the event detection. For the superbeams, we include
the following interaction modes: $\nu_\mu$ charged
current quasi--elastic scattering, $\nu_\mu$ and $\nu_e$
charged current inelastic scattering, and neutral current
scattering for all flavors. For neutrino factories we only distinguish
$\nu_\mu$ charged current events (quasi--elastic and inelastic
scattering) and neutral current events for all flavors. The cross
sections for all of these interactions are taken from
\Refs~\cite{SK1,MESSIER} and displayed in \fig~\ref{fig:xsec}.
For the energies considered for neutrino factories, the cross sections
can be assumed to be exactly known. However, at lower energies
the cross sections may be subject to considerable hadronic
uncertainties~\cite{Kitagaki:1982dx,Paschos:2002mb}. We tested
the influence of these uncertainties and found that, for the
studied experiments, they are rather small.

The event detection introduces additional uncertainties, which are
characterized by additional parameters. We include energy dependent detection
efficiencies in order to take threshold effects and their errors into
account and simulate the energy resolution by Gaussian
averaging over adjacent bins. We also include flavor, charge, and
event misidentification, leading, for example, to a certain fraction
of neutral current events counted as charged current events and to
another fraction of events detected with the wrong flavor.
Thus, we include in addition to beam contaminations, backgrounds from
event misidentification with uncertainties in magnitudes and spectral shapes.
Further on, we will use the term ``background'' to refer to the sum of
all influences leading to uncertainties in the overall normalizations and
spectral shapes.

Finally, we consider the effects of the energy calibration error
and the overall normalization error. For realistic values, we
found, however, that both these errors have only a very small influence on
the performance of the considered experiments.

%%%%%%%%%%%%%%%%%%%%%%%%%%%%%%%%%%%%%%%%%%%%%%%%%%%%%%%%%%%%%
\section{Classification of measurement errors}
\label{sec:classerrors}

The overall precision depends on several potential sources of errors. We include
in our analysis essentially five types of errors of different origin and
meaning, where some of them are well--known and have been included in many
analysis, but some of them are treated for the first time here. For the
sake of completeness, we introduce all relevant error sources in this section.

\subsection{Statistical errors}

\begin{figure}[ht!]
\begin{center}
\includegraphics[height=6cm]{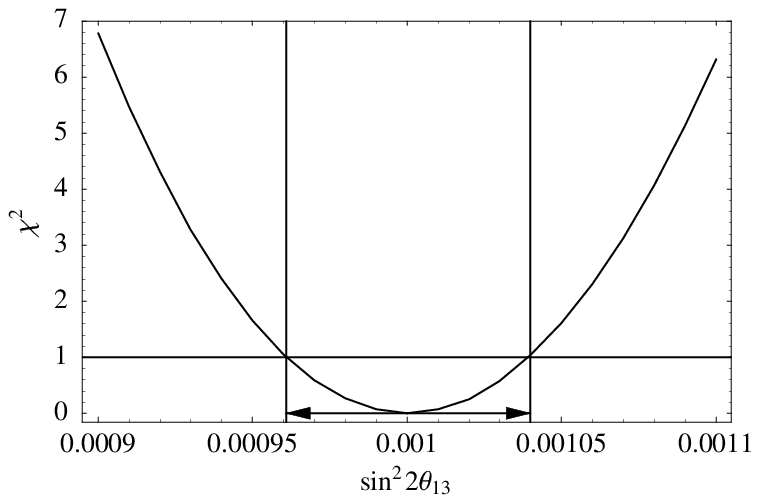}
\end{center}
\mycaption{\label{fig:staterror} The $\chi^2$--function of the statistical
error plotted as function of $\sin^2 2 \theta_{13}$ for the true value
$\sin^2 2 \theta_{13}=0.001$. The arrows indicate the error in $\sin^2 2
\theta_{13}$ read off on the $1 \sigma$ confidence level. In this example, we
choose the \NuFactII\ scenario with the LMA solution in \eq~(\ref{eq:params})
but $\Delta m_{21}^2 = 4.5 \cdot 10^{-5} \, \mathrm{eV}$ and $\sin^2 2
\theta_{12}=1.0$. } \end{figure}

Quite frequently an analysis is based on small or moderate event
rates in the binned energy spectrum with fluctuations in
each bin following a statistical distribution. This requires a
careful statistical treatment, which is described in \App~\ref{sec:statistics}.
In the end, a certain measurement can always be fit to the theoretical
prediction with a suitably defined confidence level. For a given confidence
level and one parameter considered, an interval of allowed values can be
determined. The absolute width of this interval is the statistical error,
which can be quite large even if we do not take other error sources into
account. Our method can thus deal appropriately with small event
rates and also converges to Gaussian statistics for large event rates. As an
example, the $\chi^2$--function is plotted in
\fig~\ref{fig:staterror} as function of $\sin^2 2 \theta_{13}$. The figure
illustrates that all values of $\sin^2 2 \theta_{13}$ for which the
$\chi^2$--function is below the value corresponding to the chosen confidence
level fit the true value $\sin^2 2 \theta_{13}=0.001$.

\subsection{Systematical errors}

\begin{figure}[ht!]
\begin{center}
\includegraphics[height=6cm]{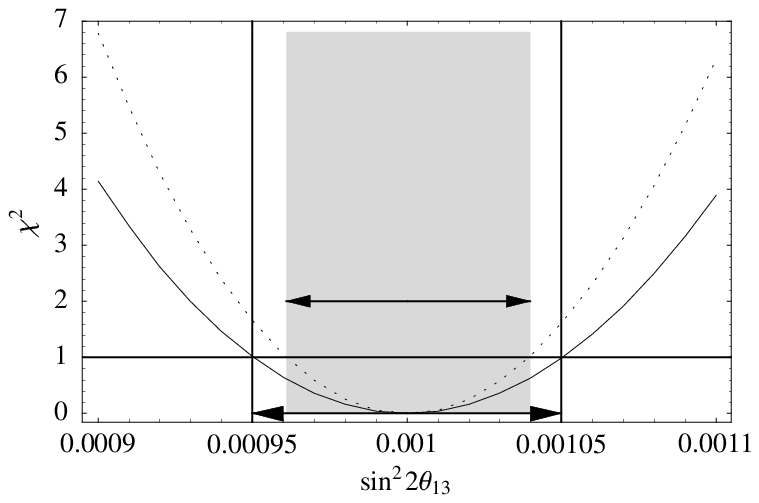}
\end{center}
\mycaption{\label{fig:systerror} The $\chi^2$--function of the statistical
(dotted curve) and statistical plus systematical (solid curve) error as function
of $\sin^2 2 \theta_{13}$ for the true value $\sin^2 2 \theta_{13}=0.001$. The
vertical lines indicate the new overall error in $\sin^2 2 \theta_{13}$ read off
on the $1 \sigma$ confidence level, being somewhat larger than the statistical
error illustrated by the shaded region. In this example, we choose the
\NuFactII\ scenario with the LMA solution in \eq~(\ref{eq:params})
but $\Delta m_{21}^2 = 4.5 \cdot 10^{-5} \, \mathrm{eV}$ and $\sin^2 2
\theta_{12}=1.0$.} \end{figure}

We consider a number of systematical limitations, such as errors
in overall signal and background normalizations, energy dependent
normalizations (tilts), efficiencies, cross sections, the energy
threshold functions. Whenever relevant, we will analyze these
errors separately and talk about ``systematics'' in the context of
errors in signal and background normalizations and tilts, which also
includes the uncertainties in the efficiencies. As usual, the considered
systematical errors make the confidence intervals obtained from
the statistical errors somewhat larger, but not by an order of
magnitude. The typical effect of switching on systematical errors
is illustrated in \fig~\ref{fig:systerror}.

\subsection{Correlation errors}

\begin{figure}[ht!]
\begin{center}
\includegraphics[height=7cm]{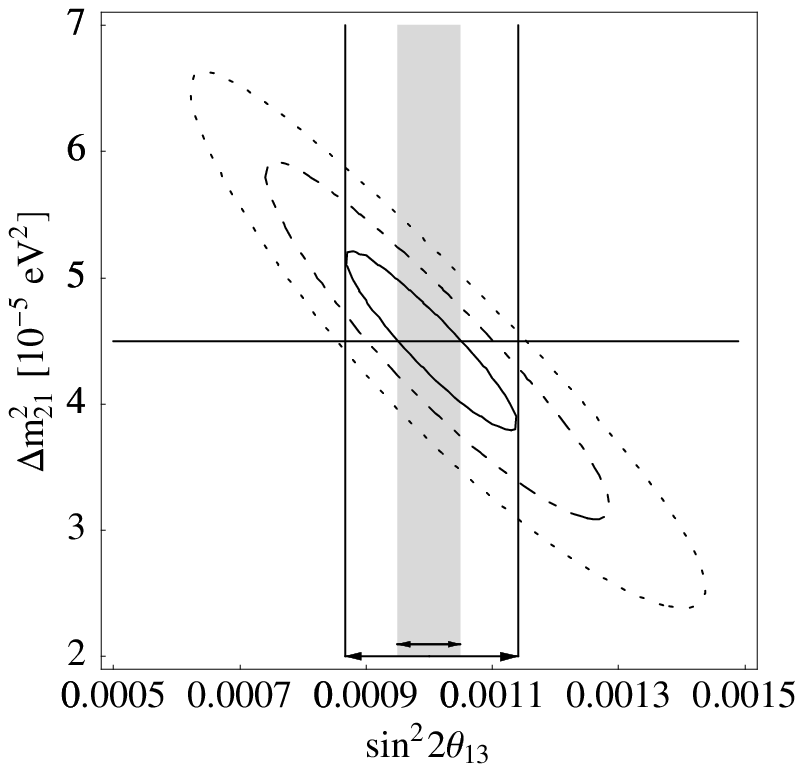}
\end{center}
\mycaption{\label{fig:correrror} The $1 \sigma$ (solid curve), $2 \sigma$ (dashed
curve), and $3 \sigma$ (dotted curve) contours of the $\chi^2$--function
including statistical and systematical errors, which is plotted as function of
$\sin^2 2 \theta_{13}$ and $\Delta m_{21}^2$ for the true values $\sin^2 2
\theta_{13}=0.001$ and $\Delta m_{21}^2 = 4.5 \cdot 10^{-5} \, \mathrm{eV}^2$.
The vertical lines indicate the new overall error in $\sin^2 2 \theta_{13}$,
coming from the correlation with $\Delta m_{21}^2$ and read off on the $1
\sigma$ confidence level. It is considerably larger than the
systematical plus statistical error marked by the shaded region obtained for
the fixed value $\Delta m_{21}^2 = 4.5 \cdot 10^{-5} \, \mathrm{eV}^2$,
indicated by the horizontal line. In this example, we choose the \NuFactII\
scenario with the standard LMA solution in \eq~(\ref{eq:params}) but $\sin^2 2
\theta_{12}=1.0$.} \end{figure}

The underlying oscillation framework is based on trigonometric expressions
where two or more parameters can be correlated in the sense that an
experiment is dominantly sensitive to a certain parameter combination.
Weaker information on other parameter combinations allows typically to
disentangle the parameters, but some correlations survive.
This is nothing but the simple fact that measuring a sum $a+b$ does
not determine the individual values of $a$ and $b$. Weaker information
on other combinations of $a$ and $b$ produces then egg--shaped regions
aligned along $a+b=const.$. Insisting on a measurement of $a$ leads
then, without better knowledge on $b$, to a projection onto the $a$-axis,
which gives a sizable error in $a$. If the value of $b$ were,
however, fixed artificially to some number, the error in $a$ would
artificially appear to be much smaller and would not represent the proper
measurement error of the experiment. In addition, its central value would depend
on the chosen value of $b$. This simple two--parameter correlation example
should illustrate, that two-- and even multi--parameter correlations make long
baseline studies more complicated. \fig~\ref{fig:correrror} illustrates this
problem for a realistic two--parameter correlation between $\sin^2 2
\theta_{13}$ and $\Delta m_{21}^2$.

\subsection{External input}

\begin{figure}[ht!]
\begin{center}
\includegraphics[height=7cm]{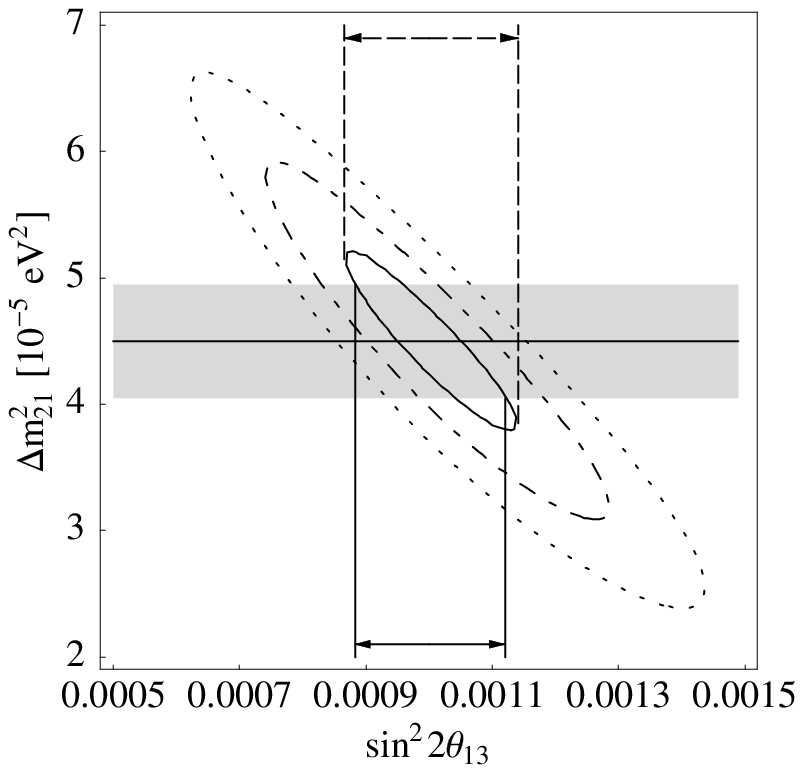}
\end{center}
\mycaption{\label{fig:exterror} The $1 \sigma$ (solid curve), $2 \sigma$ (dashed
curve), and $3 \sigma$ (dotted curve) contours of the $\chi^2$--function
including statistical, systematical, and correlational errors, which is plotted
as function of $\sin^2 2 \theta_{13}$ and $\Delta m_{21}^2$ for the true values
$\sin^2 2 \theta_{13}=0.001$ and $\Delta m_{21}^2 = 4.5 \cdot 10^{-5} \,
\mathrm{eV}^2$. The arrows on the bottom indicate the new overall error in
$\sin^2 2 \theta_{13}$ including the external input, which is in this case a
$10 \%$ precision measurement of $\Delta m_{21}^2$ marked by the shaded region.
This external measurement favors possible values of $\sin^2 2 \theta_{13}$ in
the shaded region, which makes the error in $\sin^2 2 \theta_{13}$ without
external information, indicated by the dashed arrow on the top, shrink. In this
example, we again choose the \NuFactII\ scenario with the LMA solution in
\eq~(\ref{eq:params}) but $\sin^2 2 \theta_{12}=1.0$. } \end{figure}

External information on the measured observables can and should be added to
the analysis to constrain the allowed regions in the parameter space as
good as possible. The atmospheric oscillation parameter measurements are with
the discussed setups much preciser than the pre--existing information, even
compared to the MINOS or ICARUS experiments~\cite{Barger:2001yx} -- assuming
that the old and new measurements are consistent, of course. For those
parameters, external input can therefore be ignored. This is different for the
measurements of the solar mass squared difference and mixing angle, which are
done by the KamLAND experiment. Similarly, the Earth's matter density profile,
as given in the PREM model (Preliminary Reference Earth Model), should be
included as an essential external input. One might take the attitude that
external input could be added later by some global fit. However, it can help to
resolve or reduce correlations in a way which is not possible afterwards. In
some cases, external input even can change the measurement errors by many orders
of magnitude, which again can not be included properly afterwards.
\fig~\ref{fig:exterror} demonstrates how the error in $\sin^2 2 \theta_{13}$
coming from statistics, systematics, and correlations, is reduced by using
external information on $\Delta m_{21}^2$.

\subsection{Degeneracy errors}

Degeneracies occur if two or more separated parameter sets fit the same data.
This sort of error is important because it can quite substantially
affect the error of a measurement, but it is completely different from the ones
discussed above. Dealing with degeneracies, one might, for example, either quote
separate errors for completely separated parameter sets, or take the whole range
covered by the degeneracies as the measurement error. Degeneracies will be
discussed in somewhat greater detail in \Sec~\ref{sec:degerrors}.

%%%%%%%%%%%%%%%%%%%%%%%%%%%%%%%%%%%%%%%%%%%%%%%%%%%%%%%%%%%%%
\section{The oscillation framework}
%%%%%%%%%%%%%%%%%%%%%%%%%%%%%%%%%%%%%%%%%%%%%%%%%%%%%%%%%%%%%
\label{sec:framework}

We assume standard three neutrino mixing. The
leptonic mixing matrix $U$ is parameterized in the same way as
the quark mixing matrix \cite{PDG}. The probabilities for the neutrino
transitions  $\nu_{\alpha}\rightarrow\nu_{\beta}$
in vacuum are for $n$ neutrino flavors given by
\begin{equation}  P_{\alpha \beta}  = \delta_{\alpha \beta} - 4 \;
\underset{i>j}{\sum\limits_{i=1}^{n} \sum\limits_{j=1}^{n}} \re J_{ij}^{\alpha
\beta} \sin^2 \Delta_{ij} - 2 \;
\underset{i>j}{\sum\limits_{i=1}^{n} \sum\limits_{j=1}^{n}}
\im J_{ij}^{\alpha \beta} \sin 2 \Delta_{ij}~,
\label{Plm}
\end{equation}
where
$J_{ij}^{\alpha \beta} \equiv U_{\alpha i} U^{*}_{\alpha j} U^{*}_{\beta i}
U_{\beta j}$ and $\Delta_{ij} \equiv \Delta m^2_{ij} L/(4 E_\nu) \equiv
(m_{i}^2- m_{j}^2) L/(4 E_\nu)$. Moreover, we define
$\Delta \equiv \Delta_{31}$.
We will present results based on a full numerical simulation
including matter effects. It is, however, very useful to have at least a
qualitative, analytical understanding of the effects. The oscillation
probabilities are, in principle, exactly known, but the full expressions
are quite lengthy and do not allow much insight. We therefore use
simplified expressions by expanding the probabilities in \eq~(\ref{Plm})
simultaneously in the small mass hierarchy parameter $\alpha \equiv
\dm{21}/\dm{31}$ and the small mixing angle $\sin 2\theta_{13}$. The terms
can be ordered by their combined power of small quantities as well as matter
effects can be included \cite{Freund:2001ui,FREUND,CERVERA}. The expansion in
$\sin 2\theta_{13}$ is always good, since this mixing angle is known to
be small. The expansion in $\alpha$ is a good approximation as long as
the oscillation governed by the solar mass squared difference is small
compared to the leading atmospheric oscillation, \ie,
$\alpha \, \Delta \lesssim 1$. The expansion in $\alpha$ can therefore be used
for baselines below
\begin{equation}
\label{LLIMIT}
L \lesssim 8000\,\mathrm{km} \left(\frac{E_\nu}{\GeV}\right)
\left(\frac{10^{-4}\eV^2}{\dm{21}}\right),
\end{equation}
which is fulfilled for the studied experiments. The leading terms in small
quantities are for the vacuum appearance probabilities and disappearance
probabilities\footnote{Terms up to the second order, \ie, proportional to
$\sin^2 2\theta_{13}$, $\sin 2\theta_{13}\alpha$, and $\alpha^2$, are taken into
account for $P_{e \mu}$, and terms in the first order are taken into account for
$P_{\mu\mu}$.}
\begin{eqnarray}
\label{eq:PROBVACUUM}
P_{e \mu} & \simeq & \sin^2 2\theta_{13} \, \sin^2 \theta_{23}
\sin^2 {\Delta} \nonumber \\
&\mp&  \alpha\; \sin 2\theta_{13} \, \sin\deltacp  \, \cos\theta_{13} \sin
2\theta_{12} \sin 2\theta_{23}
\sin^3{\Delta} \nonumber \\
&-&  \alpha\; \sin 2\theta_{13}  \, \cos\deltacp \, \cos\theta_{13} \sin
2\theta_{12} \sin 2\theta_{23}
 \cos {\Delta} \sin^2 {\Delta} \nonumber  \\
&+& \alpha^2 \, \cos^2 \theta_{23} \sin^2 2\theta_{12} \sin^2 {\Delta}, \\
\label{eq:DISPROBVACUUM}
P_{\mu \mu} & \simeq & 1 - \cos^2 \theta_{13}
\sin^2 2\theta_{23} \sin^2 {\Delta} \nonumber \\
&+&  2 \alpha  \cos^2 \theta_{13} \cos^2 \theta_{12}
\sin^2 2\theta_{23} {\Delta} \cos{\Delta}.
\end{eqnarray}
The next terms in this expansion can be ignored for a qualitative
discussion, since they are suppressed by an extra power of one of the small
parameters $\alpha$ and $\sin 2\theta_{13}$. The actual numerical values of
$\alpha$ and $\sin^2 2 \theta_{13}$ give each term in
\eqs~(\ref{eq:DISPROBVACUUM}) and (\ref{eq:PROBVACUUM}) a relative weight. In
the LMA case, we have $\alpha\simeq 10^{-2 \pm 1}$.
This means that the first term in \eq~(\ref{eq:PROBVACUUM})
is dominating for $\sin^2  2 \theta_{13} \simeq 0.1$ close to the
CHOOZ bound, which, for example, can make it harder to extract the
CP violating information contained in the $\alpha\cdot\sin 2\theta_{13}$
terms. For smaller values of $\sin 2 \theta_{13} \simeq \alpha$
all terms contribute equally. The $\alpha^2$ term finally dominates for tiny
$\sin 2 \theta_{13} \ll \alpha$, where it again becomes difficult to extract
the CP-parameters from the $\alpha\cdot\sin 2\theta_{13}$ terms in
\eq~(\ref{eq:PROBVACUUM}). One can in general observe in
\eqs~(\ref{eq:PROBVACUUM}) and (\ref{eq:DISPROBVACUUM}) that the CP phase can be
measured in the appearance channel if $\alpha$ is not too small, \ie, in the LMA
case. Note, however, that $\alpha$ should also not be too large compared to
$\sin 2 \theta_{13}$, since in this case the $\alpha^2$ terms start to
dominate, which reduces the relative size of the CP violating effects and
leads to a reduced CP sensitivity even though the event rates are growing. For
neutrino factories, this has been demonstrated in detail in
\Ref~\cite{Freund:2001ui,FREUND}. Note that our analysis takes into account the
complete dependence on the CP phase, involving the CP odd $\sin \deltacp$--term
and the CP even $\cos \deltacp$--term.

The smallness of $\alpha$ and $\theta_{13}$ justifies the categorization of
the parameters into leading, subleading, and subsubleading parameters
\cite{Freund:2001ui}. For $\alpha \rightarrow 0$ (which is equivalent to
$\Delta m_{21}^2 \rightarrow 0$)  and $\theta_{13} \rightarrow 0$,
the appearance channel is suppressed, leading to the two--flavor
neutrino oscillations in the disappearance channel, governed by $\theta_{23}$
and $\Delta m_{31}^2$. We call these parameters {\em leading parameters},
which are already measured by atmospheric neutrino experiments
\cite{Scholberg:1999ar,Fukuda:2000np,Ambrosio:1998wu}. Future long
baseline experiments will allow high precision measurements of these
parameters. For $\Delta m_{21}^2 \rightarrow 0$ and $\theta_{13} \neq 0$
the first term in the appearance probability in \eq~(\ref{eq:PROBVACUUM})
becomes non-zero. Since the sign of $\Delta m_{31}$ also enters this
term after the inclusion of matter effects \cite{FLPR,FHL,Barger:1999fs},
we call $\theta_{13}$ and the sign of $\Delta m_{31}$ {\em subleading
parameters}. They could even be measured very precisely without the
LMA solution. Finally, for $\Delta m_{21}^2 \neq 0$ and $\theta_{13} \neq 0$,
the remaining terms in \eqs~(\ref{eq:PROBVACUUM}) and (\ref{eq:DISPROBVACUUM})
are switched on. These contain the additional parameters $\Delta m_{21}^2$, 
$\theta_{12}$, and $\deltacp$, which we call the {\em subsubleading 
parameters}. Note that every term in \eq~(\ref{eq:PROBVACUUM}) only depends
on the product $\Delta m_{21}^2 \sin 2 \theta_{12}$. Other experiments
(or oscillation channels) are thus required to measure the individual
parameters and we will see that external input, such as on $\Delta m_{21}^2$
from the KamLAND experiment, is essential to improve the precision on $\sin^2 2
\theta_{13}$ and $\deltacp$.

The above formalism will be used in this paper to understand the effects
analytically. The shown results are, however, based on
event rates calculated from the full oscillation formulas, including
sizable corrections due to matter effects
\cite{Wolfenstein:1978ue,Wolfenstein:1979ni,Mikheev:1985gs, Mikheev:1986wj}.
Analytical expressions for the transition probabilities in matter can 
be found in \Refs~\cite{FLPR,FREUND,CERVERA}.

%%%%%%%%%%%%%%%%%%%%%%%%%%%%%%%%%%%%%%%%%%%%%%%%%%%%%%%%%%%%%
\section{Degeneracies}
%%%%%%%%%%%%%%%%%%%%%%%%%%%%%%%%%%%%%%%%%%%%%%%%%%%%%%%%%%%%%
\label{sec:degerrors}

The trigonometric neutrino oscillation probabilities have,
as already mentioned in \Sec~\ref{sec:classerrors}, degenerate
solutions, which require an approach different from the one to the statistical,
systematical, correlational, and external errors discussed in
\Sec~\ref{sec:classerrors}. Degeneracies allow additional
solutions, which may lie close to the best--fit solution or at
completely different positions in the parameter space.
Well--separated degenerate solutions do not affect the shape of
the original fit manifold around a best--fit point and do not change the size of
the measurement error as we have defined it so far. Degeneracies can in
principle be resolved by improving the experimental performance such by
using more than one baseline or better energy resolution. Degeneracies
will be important for this work, since we compare setups
which can not resolve the degeneracies. We will comment on the combination of
experiments in the discussion in \Sec~\ref{sec:conclusion}.

There are three independent two--fold degeneracies, \ie,
an overall ``eight--fold'' degeneracy, which has been discussed
analytically in terms of transition probabilities in
\Ref~\cite{Barger:2001yr}. We study the consequences of these
degeneracies in our event rate based analysis, where we focus
mostly on the measurements of $\sin^2 2 \theta_{13}$ and $\deltacp$.
The three potential degeneracies can already be seen in
\eqs~(\ref{eq:PROBVACUUM}) and (\ref{eq:DISPROBVACUUM}) and are
summarized as follows:

\begin{figure}[ht!]
\begin{center}
\includegraphics[width=16cm]{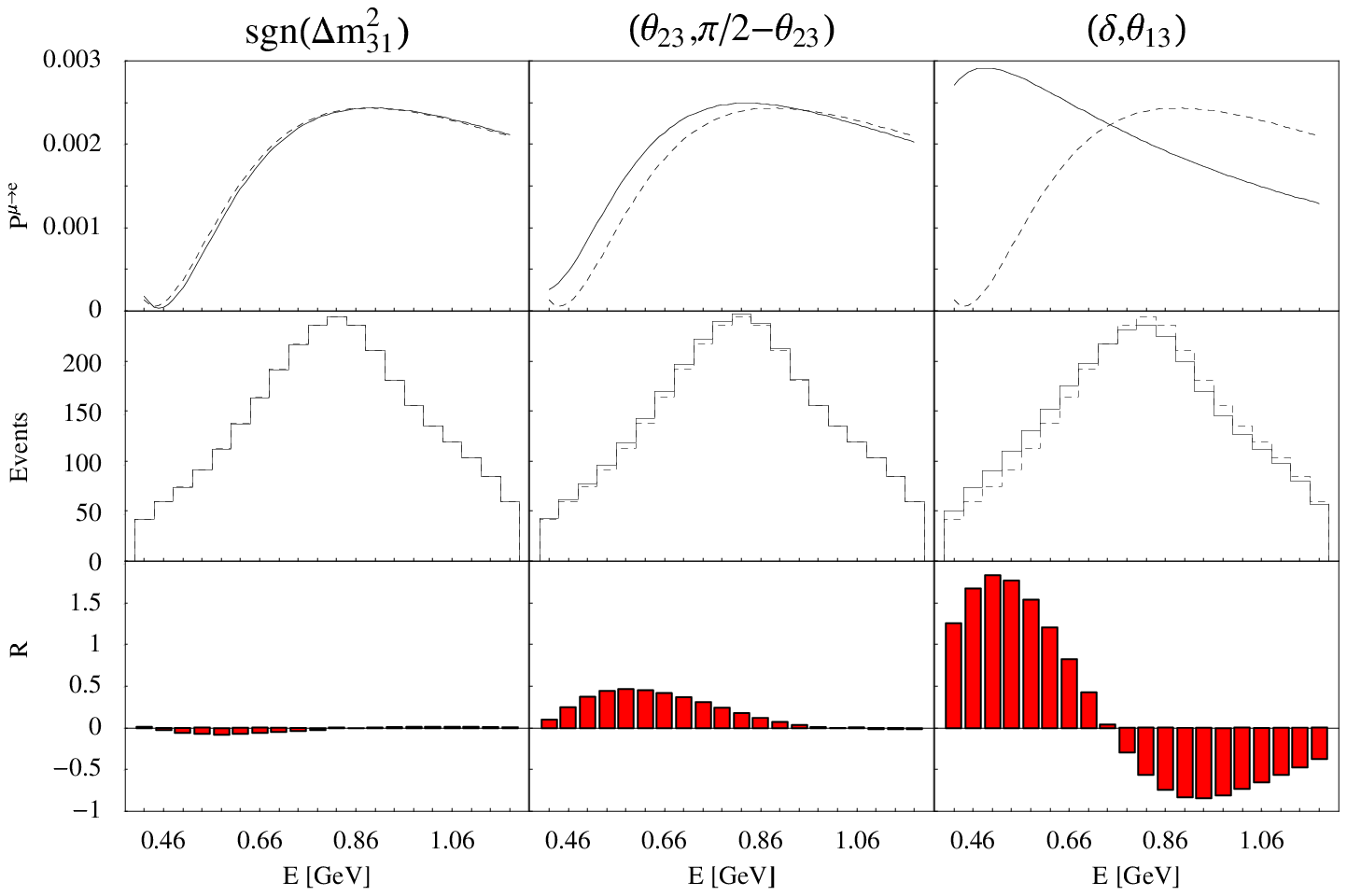}
\end{center}
\mycaption{\label{fig:degsjhf} For JHF-HK: The transition probabilities (first
row), the binned event rates (second row), and the ratios $R \equiv
(N_i-N_i^0)/ \sqrt{(N_i+N_i^0)/2}$ (third row) for the charged current neutrino
events in the appearance channel. The columns
represent the possible degenerate solutions referred to by the labels on the
top. The solid curves show the degenerate solutions and the dashed
curves the the best--fit solutions.
The quantity $R$ gives an idea of the statistical weighting of the individual
event rates $N_i$ in the bins $i$ compared to the best--fit rates $N_i^0$
(figure of merit). For the parameters, we choose the LMA solution but $\Delta
m_{21}^2=4.5 \cdot 10^{-5} \, \mathrm{eV}$, $\sin^2 2 \theta_{12}=1.0$, $\sin^2
2 \theta_{13}=0.01$, $\deltacp=\pi/4$, and $\theta_{23}=\pi/4-0.2$ in order to
be able to observe all degeneracies. } \end{figure}

\begin{figure}[ht!]
\begin{center}
\includegraphics[width=16cm]{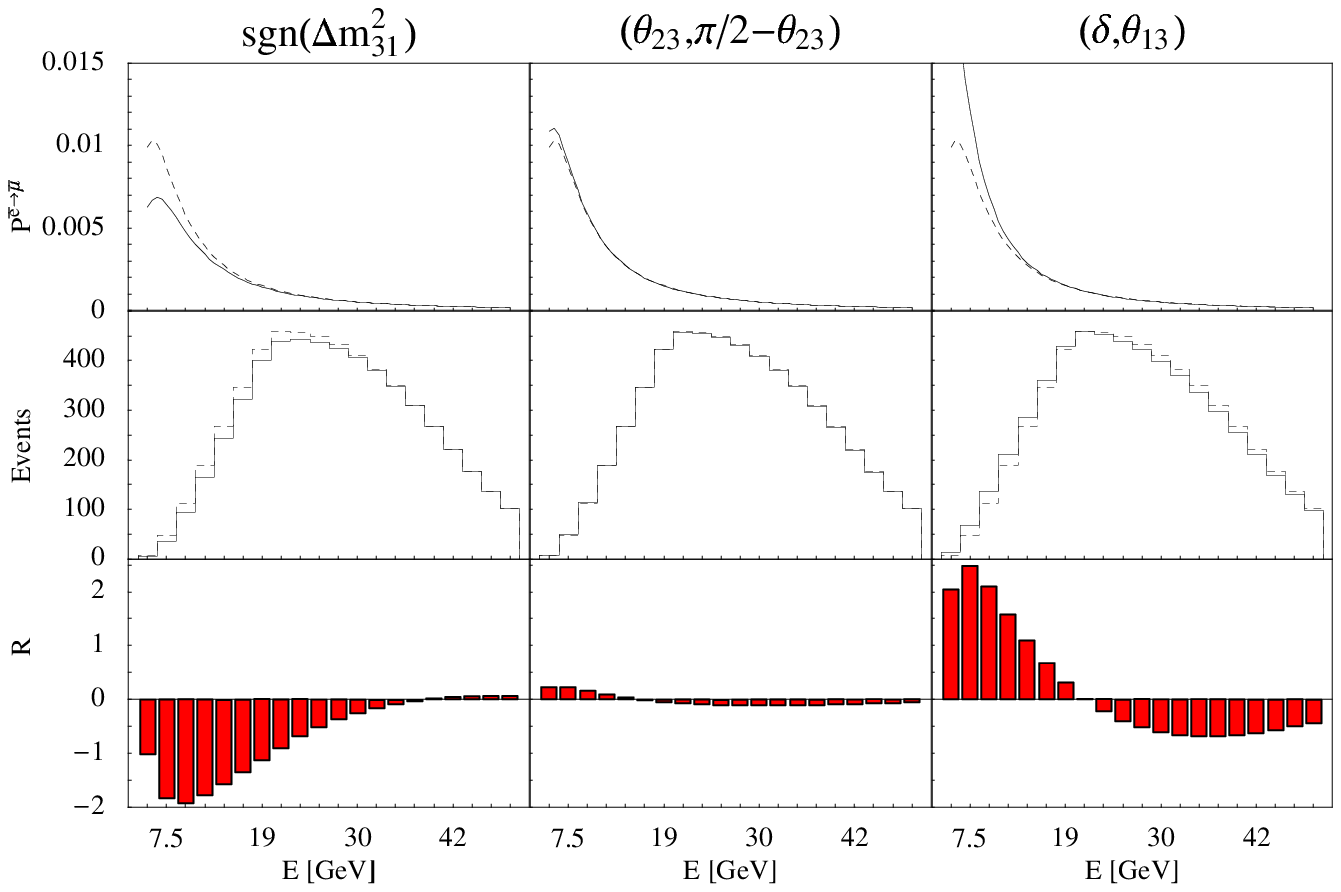}
\end{center}
\mycaption{\label{fig:degsnufact} For NuFact-II: The transition probabilities
(first row), the binned event rates (second row), and the ratios $R \equiv
(N_i-N_i^0)/ \sqrt{(N_i+N_i^0)/2}$ (third row) for the charged current neutrino
events in the appearance channel. The columns
represent the possible degenerate solutions referred to by the labels on the
top. The solid curves show the degenerate solutions and the dashed
curves the the best--fit solutions.
The quantity $R$ gives an idea of the statistical weighting of the individual
event rates $N_i$ in the bins $i$ compared to the best--fit rates $N_i^0$
(figure of merit). For the parameters, we choose the LMA solution but $\Delta
m_{21}^2=4.5 \cdot 10^{-5} \, \mathrm{eV}$, $\sin^2 2 \theta_{12}=1.0$, $\sin^2
2 \theta_{13}=0.01$, $\deltacp=\pi/4$, and $\theta_{23}=\pi/4-0.2$ in order to
be able to observe all degeneracies.} \end{figure}

\begin{figure}[ht!]
\begin{center}
\includegraphics[width=16cm]{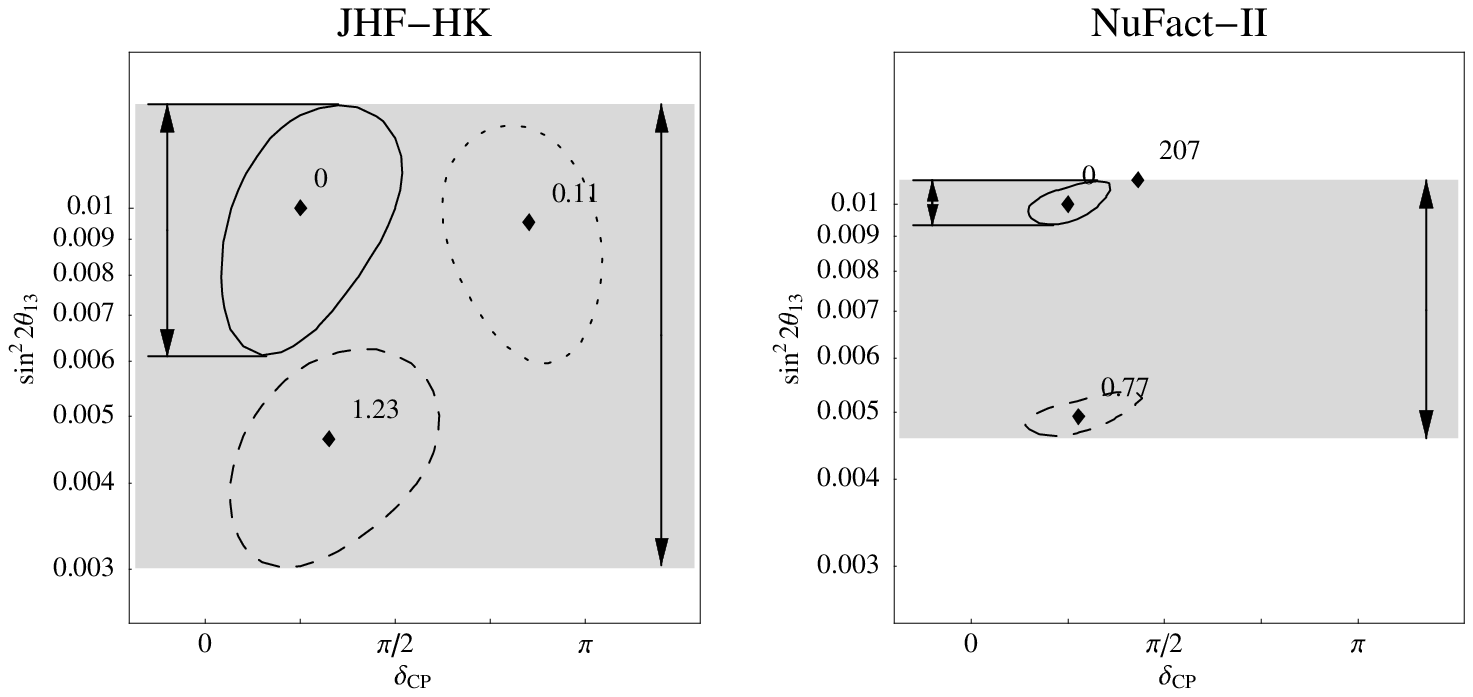}
\end{center}
\mycaption{\label{fig:degerror} The $3 \sigma$ contours of the
$\chi^2$--function, which is plotted as function of $\deltacp$ and $\sin^2 2
\theta_{13}$ for the \JHFHK\ (left--hand plot) and \NuFactII\ (right--hand plot)
experiments. The solid curves refer to the original solution at the best--fit
point, the dashed curves to the degeneracy in $(\theta_{23},\pi/2 -
\theta_{23})$, and the the dotted curve to the degeneracy in $\mathrm{sgn} (
\Delta m_{31}^2 )$. The diamonds mark the local minima with the respective
$\chi^2$--values. The arrows on the left--hand sides of the plots illustrate the
measurement error in $\sin^2 2 \theta_{13}$ from statistical, systematical,
external, and correlational sources, as we had it at the end of
\Sec~\ref{sec:classerrors}. The arrows on the right--hand sides of the plots
mark the overall error, as we would have it for taking the whole range covered
by degeneracies, given by the gray--shaded region. For the oscillation
parameters, we choose the LMA solution with $\Delta m_{21}^2=4.5 \cdot 10^{-5}
\, \mathrm{eV}$, $\sin^2 2 \theta_{12}=1.0$, $\sin^2 2 \theta_{13}=0.01$,
$\deltacp=\pi/4$, and $\sin^2 2 \theta_{23}=\pi/4-0.2$.} \end{figure}

\begin{figure}[ht!]
\begin{center}
\includegraphics[angle=-90,width=16cm]{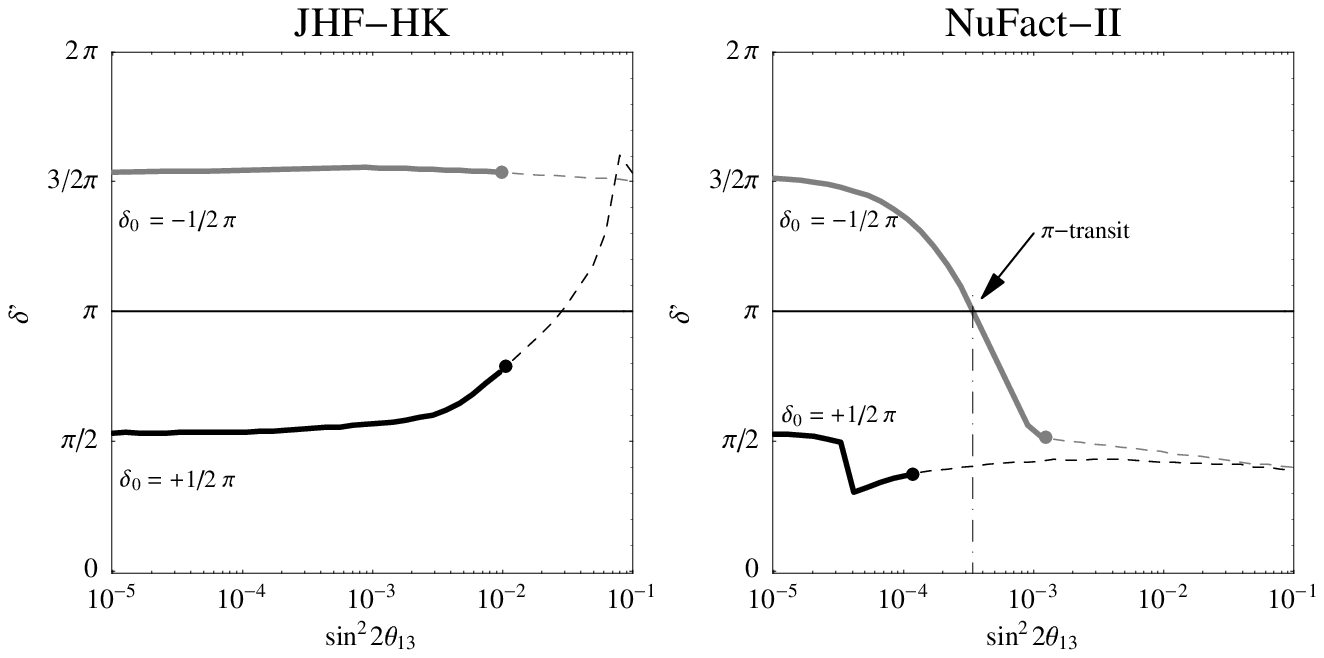}
\end{center}
\mycaption{\label{fig:trees} The position $\delta'$ of the degenerate solution
(in $\mathrm{sgn} ( \Delta m_{31}^2 )$) plotted as function of the true
value of $\sin^2 2 \theta_{13}$, lying approximately at $(\sin^2 2 \theta_{13})'
\simeq \sin^2 2 \theta_{13}$ of the best--fit solution. The plots show the
\JHFHK\ (left--hand plot) and \NuFactII\ (right--hand plot) experiments for the
true values $\delta_0=+1/2 \pi$ (black curve) and $\delta_0=-1/2 \pi$ (gray
curve). The dashed sections of the curves indicate the values of $\sin^2 2
\theta_{13}$ where the degenerate solution can not be observed on the $2 \sigma$
confidence level, whereas it can be observed in the solid sections. For
the oscillation parameters, we choose the LMA solution. } \end{figure}

\begin{description}
\item[The sign of $\Delta m_{31}^2$:] Vacuum oscillations do not depend
on the sign of $\Delta \equiv \Delta m^2_{31}L/4E_\nu$. This degeneracy can, in
principle, be lifted by MSW effects in matter. A degenerate solution for the
opposite sign of $\Delta m_{31}^2$ may survive, however, for a different value
of $\deltacp$ and almost the same value of $\sin^2 2
\theta_{13}$~\cite{Minakata:2001qm}.

\item[Non--maximal mixing  $(\theta_{23},\pi/2-\theta_{23})$:]
For $\theta_{23} \neq \pi/4$, a degenerate solution exists at approximately
$\theta^\prime_{23} = \pi/2-\theta_{23}$. For our setups, it turns out to
be at a similar value of $\deltacp$. This degeneracy can also be seen
in \eqs~(\ref{eq:PROBVACUUM}) and (\ref{eq:DISPROBVACUUM}), where the
factor of $\sin^2 \theta_{23}$ in the first term of \eq~(\ref{eq:PROBVACUUM})
and the factor of $\cos^2 \theta_{23}$ in the fourth term of \eq~(\ref{eq:PROBVACUUM})
are exchanged under the transformation $\theta_{23} \rightarrow \pi/2 -
\theta_{23}$, leading to a possible compensation by changing $\sin^2 2
\theta_{13}$.

\item[The $(\delta,\theta_{13})$ ambiguity:]
\eq~(\ref{eq:PROBVACUUM}) allows that in many cases a combination of
$(\delta,\theta_{13})$ gives the same probabilities as another pair
$(\delta^\prime,\theta_{13}^\prime)$ \cite{Burguet-Castell:2001ez}.
Taking into account the appearance channel only and ignoring matter
effects, the asymmetry of the $\sin \deltacp$ term under a
neutrino--antineutrino transformation opens the possibility of a
CP violation measurement. From this point of view, this degeneracy can be
seen in the total event rates as neutrino and anti neutrino rates equal to the
best--fit point. However, the energy resolution of the experiment allows, in
principle, to resolve the degeneracy (for the dependence on experimental
details, see \Refs~\cite{Burguet-Castell:2001ez,Freund:2001ui} and
\App~\ref{sec:det}).
\end{description}

We illustrate the above ambiguities for our experimental setups for a parameter
set, for which all effects can be seen, \ie, if not otherwise stated, we
choose $\deltacp=\pi/4$, $\sin^2 2 \theta_{13}=0.01$, and $\sin^2 2
\theta_{23}=\pi/4-0.2$. \Figs~\ref{fig:degsjhf} (for \JHFHK) and
\ref{fig:degsnufact} (for \NuFactII) show for the $\nu_e\rightarrow \nu_\mu$
appearance channel the transition probabilities (first row), the
binned event rates (second row), and the the ratios
$R \equiv (N_i-N_i^0)/ \sqrt{(N_i+N_i^0)/2}$ (third row). Every
column corresponds to one type of degeneracy. The dashed and
solid curves correspond for the upper two rows to the best--fit and the
degenerate solution, respectively. The ratio $R$ gives an idea
of the statistical weight of the individual event rates $N_i$ in the bins $i$
compared to the rates $N_i^0$ of the best--fit solution (figure of merit). For
the \JHFHK\ experiment in \fig~\ref{fig:degsjhf}, the degeneracies in
$\mathrm{sgn} (\Delta m_{31}^2)$ and $(\theta_{23},\pi/2-\theta_{23})$
lead only to small statistical deviations from the best--fit spectra
and can, in principle, be observed for our setups. It is interesting
how the very different transition probabilities in the
$(\delta,\theta_{13})$ case lead to similar event rate spectra and
identical total rates. From the $R$ distribution it becomes clear that
this degeneracy can be resolved if the experiment has sufficient energy
resolution. For our experimental setups, it turns out that this degeneracy
is hardly be observed for large $\sin^2 2 \theta_{13}$, while it is quite
important at the sensitivity limit. The neutrino factory is for comparison shown
in \fig~\ref{fig:degsnufact}. The degenerate event rate spectra seem to be very
similar, but the overall event rates are considerably larger and
statistics is thus very good. The $R$ distributions show why only the
$(\theta_{23},\pi/2-\theta_{23})$ degeneracy survives in a fit of the event rate
spectrum. Compared to the \JHFHK\ experiment, the degeneracy in
$\mathrm{sgn}(\Delta m_{31}^2)$ can be resolved at the used value of $\sin^2 2
\theta_{13}$ due to MSW matter effects at this long baseline. The point is that
the matter density resonance energy in Earth (of about $15 \, \mathrm{GeV}$)
lies within the energy window of the detector and its effects can clearly be
seen in the $R$ distribution. We will see in \fig~\ref{fig:trees}, that this
degeneracy can not be resolved anymore for very small values of $\sin^2 2
\theta_{13}$, since the reduced event rates (overall and per bin) do not allow
to use the spectral information to resolve this degeneracy, even though $\sin^2
2 \theta_{13}$ can still be extracted from the total rates. Matter effects play
a minor role for the $(\delta,\theta_{13})$ degeneracy and the
similar qualitative behavior of both experiments is determined by the the fact
that both are operated at a similar value of $L/E$. The
$(\delta,\theta_{13})$ ambiguity therefore looks very similar for the \NuFactII\
and the \JHFHK\ setup, as can be seen from the $R$ distributions.

\Fig~\ref{fig:degerror} shows fits of $\sin^2 2 \theta_{13}$ versus
$\deltacp$  for the \JHFHK\ and \NuFactII\ experiments at the
$3\sigma$ confidence level, where the respective $\chi^2$ values at the local
minima are specified. Besides the best--fit there exist also
degenerate solutions in $\mathrm{sgn} (\Delta m_{31}^2)$ and
$(\theta_{23},\pi/2-\theta_{23})$. The figure illustrates that, for
the chosen parameter values, all of the shown degeneracies
can be observed at the $3 \sigma$ confidence level except from the degeneracy
in $\mathrm{sgn} (\Delta m_{31}^2)$ for the neutrino factory.
The figures also show the pattern which was mentioned before:
The $\mathrm{sgn} ( \Delta m_{31}^2 )$ degeneracy lies more or less
at the same value of $\sin^2 2 \theta_{13}$, but at a different
value of $\deltacp$. The $(\theta_{23},\pi/2-\theta_{23})$
degeneracy lies at a similar value of $\deltacp$ and a different
value of $\sin^2 2 \theta_{13}$. The $\mathrm{sgn} (\Delta m_{31}^2)$
degeneracy will thus especially affect the CP measurement, while
the $(\theta_{23},\pi/2-\theta_{23})$ degeneracy will especially
affect the $\sin^2 2 \theta_{13}$ measurement. We will return to
this issue at the end of this section, in order to explain
how the degeneracies are included in our evaluations.

We mentioned before and illustrated in \fig~\ref{fig:trees}
that the presence of the $\mathrm{sgn} ( \Delta m_{31}^2 )$ degeneracy
depends on the value of $\sin^2 2 \theta_{13}$. \Fig~\ref{fig:trees}
shows, as a function of $\sin^2 2 \theta_{13} \simeq \sin^2 2
\theta_{13}^\prime$, the position of the degenerate solution $\delta^\prime$
resulting from the $\mathrm{sgn} ( \Delta m_{31}^2 )$ degeneracy. The black
curves correspond to true values $\delta_0=\pi/2$ and the gray curves to
$\delta_0=3\pi/2$, which means that the best--fit solutions would be represented
by the horizontal lines $\deltacp\equiv \pi/2$ and $\deltacp\equiv 
3\pi/2$, respectively, independently of $\sin^2 2 \theta_{13}$. The dashed
extrapolated curves indicate the regions where the degeneracy is resolved, \ie
where the degenerate solution is rejected at the $2\sigma$ confidence level. One
can read off from \fig~\ref{fig:trees} that for $\sin^2 2 \theta_{13}=0.01$ this
degeneracy does not show up at this confidence level for the \NuFactII\
experiment, but for the \JHFHK\ experiment. This corresponds to
\fig~\ref{fig:degerror}, where $\sin^2 2 \theta_{13}=0.01$ was used. The
degeneracies are in general resolved at large $\sin^2 2 \theta_{13}$ and they
are cut off at small values of $\sin^2 2 \theta_{13}$ by the sensitivity limit,
which is not shown in \fig~\ref{fig:trees}. It is interesting to note that
the degenerate solution tends to move away from the best--fit solution for
large values of $\sin^2 2 \theta_{13}$, but approaches it for small values of
$\sin^2 2 \theta_{13}$. Especially, it turns out that all starting
values $\delta_0$ lead to solutions attracted by $\pi$/2 (\NuFactII)
or $3\pi/2 $ (\JHFHK) for large values of $\sin^2 2 \theta_{13}$,
which implies that the degenerate solution moves most for the starting
values $3\pi/2$ (\NuFactII) or $\pi/2$ (\JHFHK).\footnote{This can, in
principle, be understood in terms of a CP trajectory diagram in the bi--rate
(neutrino and anti neutrino) plane, such as it is introduced in
\Ref~\cite{Minakata:2001qm}. In such a diagram, the CP trajectories of the
degenerate solutions overlap each other for small values of $\sin^2 2
\theta_{13}$, which means that the degenerate solutions are very close to each
other and not moving much. For large values of $\sin^2  2 \theta_{13}$, the CP
trajectories of the degenerate solutions are well separated, which means that
essentially one point of the best--fit trajectory lies closest to all points of
the degenerate solution trajectory. This point corresponds, depending on the
experiment, to the attractor $\delta=\pi/2$ or $\delta=3 \pi/2$.}
Altogether, there are three important consequences for the
$\sin^2 2 \theta_{13}$ dependence of CP studies, where the degeneracy in
$\mathrm{sgn} ( \Delta m_{31}^2 )$ is very important:
\begin{itemize}
\item
The results will not be affected for large values of $\sin^2 2 \theta_{13}$
since the degeneracy can be resolved.
\item
If there is sensitivity for $\sin^2 2 \theta_{13}$ smaller than about
$10^{-4}$ (\NuFactII) or $10^{-3}$ (\JHFHK), then the degenerate solution
joins with the best--fit point and thus becomes irrelevant.
\item
Especially for a neutrino factory with $\sin^2 2 \theta_{13}$
between about $10^{-3}$ and $10^{-4}$, the degeneracy can move
$\delta$ very far away from its true value especially for $\delta_0=3\pi/2$,
which implies a complex structure in corresponding plots. The degenerate
solution crosses, for example, the line $\delta^\prime=\pi$ at about $\sin^2 2
\theta_{13}=3 \cdot 10^{-4}$ (labeled ``$\pi$--transit'' in
\fig~\ref{fig:trees}), which means that CP violation could not be distinguished
from CP conservation at this transit point (\cf, \figs~\ref{fig:cpviolationsdm}
and \ref{fig:cpviolation} in the next section).
\end{itemize}

Coming back to \fig~\ref{fig:degerror}, we still need to discuss
how to treat degeneracy errors, \ie, how we obtain a total error
including the degeneracies. One could, for example, simply take
the measurement error of $\sin^2 2 \theta_{13}$ at the
best--fit point purely from statistical, systematical, external,
and correlational errors. This optimistic choice is illustrated
by the arrows on the left--hand sides of the plots and simply
ignores the existence of degenerate solutions. However, it is
obvious that without additional external input the degeneracies
can not be resolved. An alternative would be to take the worst case
errors marked by the arrows on the right--hand sides of the plots,
which is identical to the gray--shaded region covering the range
of all possible degeneracies (pessimistic choice). A look at the
two plots shown in \fig~\ref{fig:degerror} tells us that one
must take the complexity of the problem into account. The shown
\JHFHK\ case is obviously better described by the pessimistic
case, since the degenerate solutions are almost touching each other.
For the \NuFactII\ case it would, however, make more sense to
quote the two separated (degenerate) fit regions with their
individual errors. However, the pessimistic choice would
obviously lead to a very different precision, which can be an order
of magnitude worse. It should be obvious that one must look at
the specific case at hand to decide which point of view is more
appropriate. In most cases we will therefore show figures with
and without degenerate solutions and we will discuss the
degeneracy problem wherever appropriate.

%%%%%%%%%%%%%%%%%%%%%%%%%%%%%%%%%%%%%%%%%%%%%%%%%%%%%%%%%%%%%%%%%%
\section{Results I: The measurement of $\sin^2 2 \theta_{13}$}
\label{sec:results1}
%%%%%%%%%%%%%%%%%%%%%%%%%%%%%%%%%%%%%%%%%%%%%%%%%%%%%%%%%%%%%%%%%%

We now present our results concerning the experimental capabilities
to measure $\sin^2 2 \theta_{13}$. This includes an analysis of
the impact factors on the measurement, a discussion of the
$\sin^2 2 \theta_{13}$ sensitivity, and an evaluation of the
precision of the $\sin^2 2\theta_{13}$ measurement. We also investigate
the sensitivity to the sign of $\Delta m_{31}^2$, which is
a closely related issue.

All results of of this and the following section are, unless
otherwise stated, calculated for the current best--fit values
of the atmospheric~\cite{Gonzalez-Garcia:2002mu,Maltoni:2000ib}
and solar neutrino experiments~\cite{Bahcall:2001cb,Penya}.
The ranges are given at the $3 \sigma$ confidence level by
\begin{eqnarray}
\dm{31} & =& 3_{-2}^{+5}\cdot10^{-3}\,\mathrm{eV}^2, \nonumber \\
\sin^2 2\theta_{23} & = & 1_{-0.2}^{+0}, \nonumber \\
\dm{21} & = & 3.7_{-2}^{+42}\cdot10^{-5}\,\mathrm{eV}^2, \nonumber \\
\sin^2 2\theta_{12} & = & 0.8_{-0.2}^{+0.2}, \nonumber \\
\dm{21}\sin^2 2\theta_{12} & =& 3_{-1.8}^{+38}\cdot10^{-5}\,\mathrm{eV}^2,
\label{eq:params}
\end{eqnarray}
where the mean values are throughout the text referred to as the ``LMA
solution''. The indicated ranges represent the current knowledge to be kept in
mind for our analysis. For $\sin^2 2\theta_{13}$ we will only allow values below
the CHOOZ bound \cite{Apollonio:1999ae} and the used value will
be specified in the respective examples. For the CP phase
$\deltacp$, we choose $\deltacp=0$ as standard value
if not otherwise stated. In some cases we use, however,
parameters different from \eq~(\ref{eq:params}),
to demonstrate specific effects. An example are
degenerate solutions in $(\theta_{23},\pi/2-\theta_{23})$,
which are only observable for $\sin^2 2 \theta_{23} < 1$.

%%%%%%%%%%%%%%%%%%%%%%%%%%%%%%%%%%%%%%%%%%%%%%%%%%%%%%%%%%%%%%%%%%%%%%%5
\subsection{Impact factors on the measurement of $\sin^2 2 \theta_{13}$}
%%%%%%%%%%%%%%%%%%%%%%%%%%%%%%%%%%%%%%%%%%%%%%%%%%%%%%%%%%%%%%%%%%%%%%%5

\label{sec:impactfactors}

\begin{figure}[ht!]
\begin{center}
\includegraphics[width=16cm]{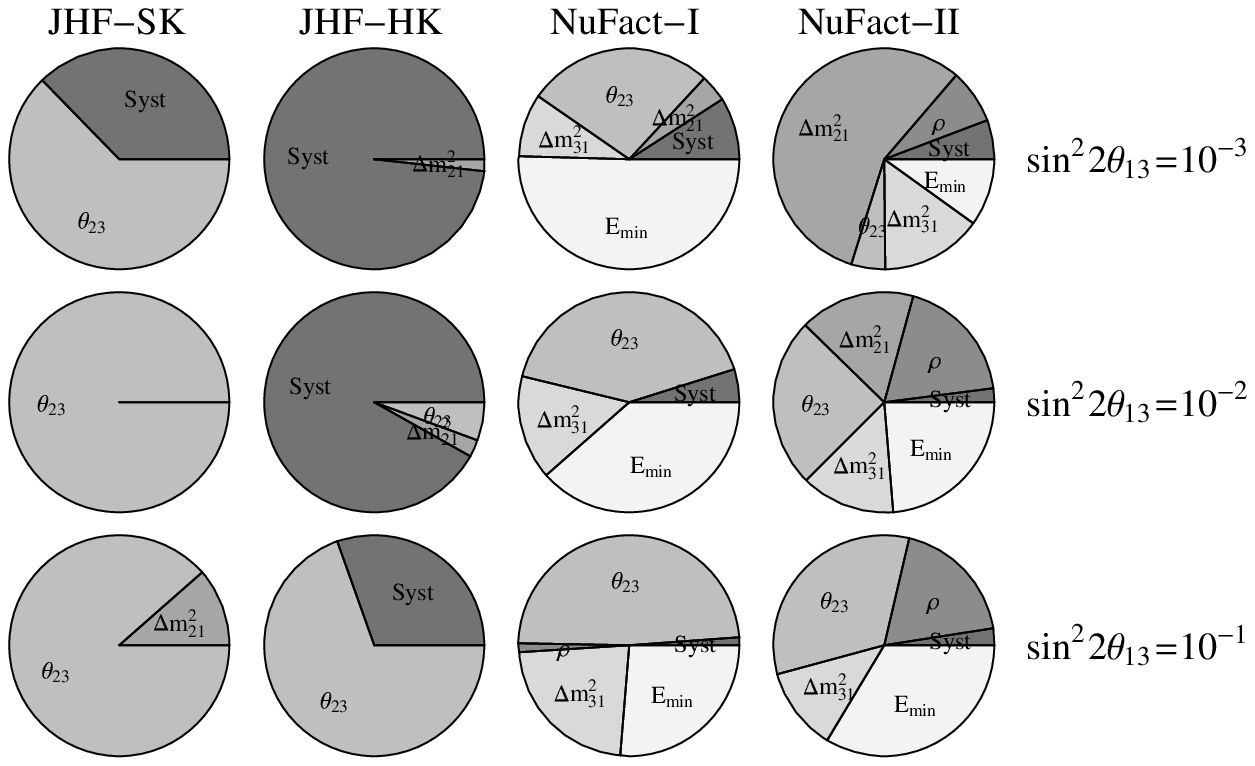}
\end{center}
\mycaption{\label{fig:errorsall} The relative contributions of the impact factors
on the measurement of $\sin^2 2\theta_{13}$ for different experiments and
different values of $\sin^2 2 \theta_{13}$. The error sources $\xi$ are the
systematical errors (``Syst''), the correlations with
the large (``$\Delta m_{31}^2$'') and small (``$\Delta m_{21}^2$'') mass squared
differences, the correlation with the mixing angle $\theta_{23}$
(``$\theta_{23}$''), the energy threshold function for neutrino factories
(``$E_{\min}$''), and the uncertainty in the matter density
(``$\rho$''). For the oscillation parameters, we choose the LMA solution.}
\end{figure}

In order to identify the relevant and controllable main contributions to the
errors of the measurement of $\sin^2 2 \theta_{13}$, we define the impact factor
$\mathrm{IF}_{\xi}$ of an error source $\xi$ as\footnote{Since the CP phase can
not be controlled or optimized in this sense, we will not include the
respective correlation in the impact factor analysis.}
\begin{equation}
  \mathrm{IF}_{\xi} \equiv 1-\frac{(\delta \sin^2 2
\theta_{13})_{\mathrm{Error \, source \, \xi \, off}}}{(\delta \sin^2 2
\theta_{13})_{\mathrm{All \, error \, sources \, on}}},
\label{eq:impactfactor}
\end{equation}
where $\delta \sin^2 2 \theta_{13}$ is the total error of the
measurement of $\sin^2 2 \theta_{13}$ chosen at the $2 \sigma$
confidence level, including the statistical, systematical, correlational,
and external errors (\cf, \fig~\ref{fig:exterror}).
The impact factor specifies the relative impact of a certain source of
errors $\xi$ on the total error of the measurement, \ie, the error
source $\xi$ is switched off compared to taking into account all error sources.
In this section, we especially focus on the relative contributions of
systematical and correlational errors in order to determine the main
sources limiting the precision of the studied experiments. For the
moment, we do not take into account degeneracies and focus on the
best--fit solution, since we are interested in how the size of the
individual solutions is reduced (\eg, as shown in \fig~\ref{fig:degerror}).
\Fig~\ref{fig:errorsall} shows for the four different setups the
relative contributions of the impact factors for three different
values of $\sin^2 2 \theta_{13}$. As error sources $\xi$ we include
systematical errors\footnote{In this analysis, ``systematical errors''
refer to the overall uncertainties in signal and background normalizations
and spectral tilts, which also includes uncertainties in the efficiencies.},
the correlations with the mass squared differences and $\theta_{23}$,
the energy threshold function (for details, see \App~\ref{sec:det}),
and the uncertainty in the matter density (assumed to be $5 \%$).
Note that we include the external input from the KamLAND experiment for the
solar parameters, which means that without this information the pie slices of
the small mass squared difference could be larger.

Before we analyze the behavior of the different setups in
\fig~\ref{fig:errorsall} in detail, let us point towards the
effects of the different values of $\sin^2 2\theta_{13}$.
In general, the smaller $\sin^2 2\theta_{13}$ is, the smaller
is the number of events we can expect to see in the
appearance channels, which are dominating the precision of the $\sin^2
2\theta_{13}$ measurement. This loss in statistics leads to a larger relative
impact of the systematical errors.

\begin{description}
\item[\JHFSK]
This setup leads only to a very low number of events and is thus
statistics dominated. No single impact value contributes more than
about 7~\% to the absolute value. This comes essentially from the
large statistical error in $\sin^2 2 \theta_{13}$
without any systematics and correlations, which means that the
latter only modify the absolute error in a minor way. Note that for
the JHF to Kamiokande experiments the baseline is too short for
matter effects to play an important role.

\item[\JHFHK]
This setup leads to a very large number of events, \ie, very good
statistics. Thus, especially for small values of $\sin^2 2 \theta_{13}$,
the main impact comes from systematics. A closer analysis of the
systematical impact factors shows that the systematical error is
dominated by the background uncertainties of the charged current
events. The remaining contributions are minor effects of the signal and
background normalizations, as well as of the tilts. For this experiment,
it is thus important to improve the understanding of background
uncertainties, especially for the charged current events.

\item[\NuFactI]
The initial stage neutrino factory leads to a moderate number of
events with a somewhat limited statistics dominating most effects.
The main impact factors are the correlations with the leading oscillation
parameters $\theta_{23}$ and $\Delta m_{31}^2$ as well as the energy
threshold function. Especially for small values of $\sin^2 2 \theta_{13}$,
the loss of low energy events due to the threshold function dominates.
An extra detector component with a good coverage in the threshold energy
regime or another improvement in the threshold properties would considerably
help to improve the precision. Matter effect uncertainties are negligible
at the achievable level of precision, even though the baseline is
quite long in this setup.

\item[\NuFactII]
This setup has much better statistics than the initial stage neutrino factory.
It is a high precision instrument for which the relative impact of all sources
becomes quite important. Three points are especially interesting to observe:
First, for small values of $\sin^2 2 \theta_{13}$, the correlation with
the small mass squared difference dominates. The precision of this experiment
depends therefore in this case on the precision of an external measurement,
such as the KamLAND experiment. Better knowledge of $\Delta m_{21}^2$ than
assumed would considerably help to improve the performance. Second, the energy
threshold function dominates the errors for large $\sin^2 2 \theta_{13}$,
probably because the low energy events have more relative weight. Again, a
second detector with good coverage around the threshold energy or another
improvement of the threshold properties would improve the precision. Third,
because of the long baseline and the high precision, the matter density
uncertainty is important, especially for large $\sin^2 2
\theta_{13}$ where the statistics is best. \end{description}

%%%%xxxxxxxx

\begin{figure}[ht!]
\begin{center}
\includegraphics[width=15cm]{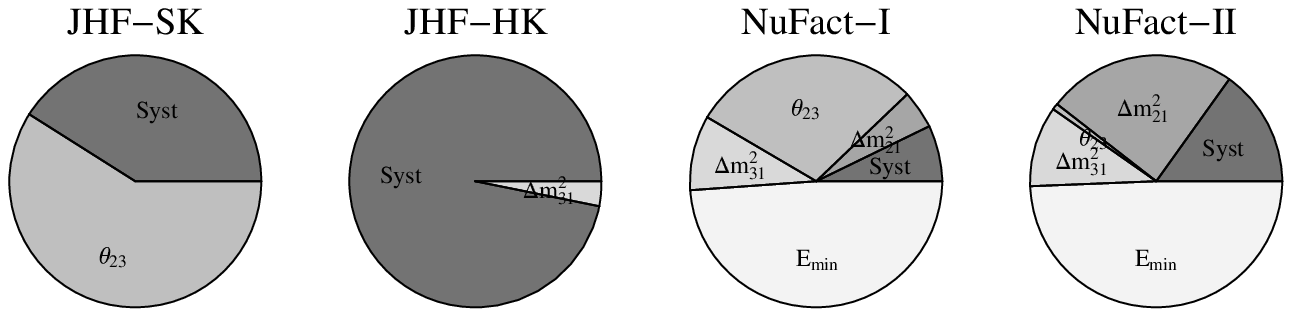}
\end{center}
\mycaption{\label{fig:errors0} The relative contributions of the impact factors on
the measurement of $\sin^2 2 \theta_{13}$ for different experiments at the
sensitivity limit ($\sin^2 2 \theta_{13}=0$). The error sources $\xi$ are the
systematical errors (``Syst''), the correlations with
the large (``$\Delta m_{31}^2$'') and small (``$\Delta m_{21}^2$'') mass squared
differences, the correlation with the mixing angle $\theta_{23}$
(``$\theta_{23}$''), the energy threshold function for neutrino factories
(``$E_{\min}$''), and the uncertainty in the matter density (``$\rho$''). For
the other oscillation parameters, we choose the LMA solution.}
\end{figure}

In \fig~\ref{fig:errors0}, the impact factors are in addition shown for the
sensitivity limit in $\sin^2 2 \theta_{13}$, since the behavior in $\sin^2 2
\theta_{13}$ can not necessarily extrapolated to this limit.
However, we do not
observe a change in qualitative behavior from the plots in the first row of
\fig~\ref{fig:errorsall} for all experiments except from the \NuFactII\
experiment. This part of the result is not very surprising, since for all
experiments except from \NuFactII\ the sensitivity limit lies
close to (or upper) $\sin^2 2 \theta_{13}=0.001$, which corresponds to the first
row of \fig~\ref{fig:errorsall}. The \NuFactII\ experiment, however, performs
better and therefore the sensitivity limit is much lower. One can see from
\fig~\ref{fig:errors0} that in this case again the effect of the threshold
function is the biggest impact factor.

\begin{figure}[ht!]
\begin{center}
\includegraphics[width=10cm]{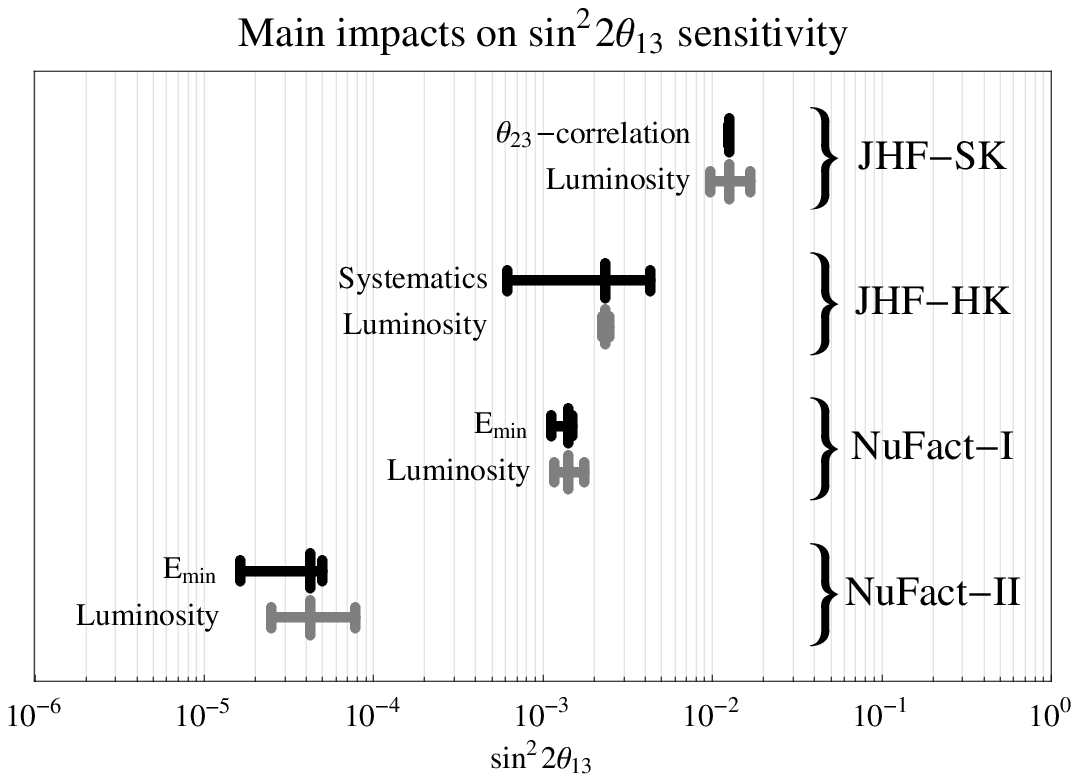}
\end{center}
\mycaption{\label{fig:mainimpacts} The $\sin^2 2 \theta_{13}$ sensitivity on
the $2 \sigma$ confidence level for the different experiments on the right--hand
side of the plot. Each bar represents the range of the sensitivity
under a variation of the main impact factor (black bar) or
a variation of the luminosity (gray bar), as labeled on the left--hand
side of the respective bar. The left vertical dash on each bar corresponds to
the optimistic choice, the middle vertical dash to our standard choice, and the
right vertical dash to the pessimistic choice, where the meaning of each choice
depends on the impact factor and is explained in the main text. For the
oscillation parameters, we choose the LMA solution.}
\end{figure}

In order to summarize our results, we show in \fig~\ref{fig:mainimpacts} the
$\sin^2 2 \theta_{13}$ sensitivity on the $2 \sigma$ confidence level for all
experiments. Each bar represents the range of the sensitivity under a variation
of the main error impact or under a variation of the luminosity (number of
stored muons times detector mass). The left dash on each bar
corresponds to the optimistic choice, the middle dash to our standard choice,
and the right dash to the pessimistic choice. For the luminosity, we choose
twice the luminosity of our standard assumption as the optimistic choice and
half the luminosity of our standard assumption as the pessimistic choice. For
the main impact factors, we proceed similarly with the systematical
uncertainties and double it or divide it by two. For the threshold function,
represented by ``$E_{\min}$'', we approximate it by a step function at $4 \,
\mathrm{GeV}$ (neutrino factory only) for the optimistic choice or replace it by
a linear function climbing from $4$ to $25 \, \mathrm{GeV}$ for the pessimistic
choice instead of $4$ to $20 \, \mathrm{GeV}$ for the standard choice. For the
$\theta_{23}$--correlation, we eliminate it for the optimistic choice by
assuming $\theta_{23}$ to be known from external input or keep it for the
pessimistic choice, since $\theta_{23}$ is measured by the experiment itself.
One can easily see that the \JHFSK\ experiment is statistics limited, \ie,
changing the luminosity has the biggest impact. The \JHFHK\ experiment, however,
is systematics limited. By improving the knowledge on the background
uncertainties by a factor of two, the experiment would perform much better than
by doubling the luminosity. The \NuFactI\ experiment would equally profit from
higher statistics and a special low energy detector in order to lift the
$(\delta,\theta_{13})$ ambiguity, which is for our setups especially important
at the sensitivity limit. Similarly, a special low energy detector would help
to improve the \NuFactII\ experiment, where the effect would be
even better than doubling the luminosity.

\subsection{The $\sin^2 2 \theta_{13}$ sensitivity}
\label{sec:theta13sens}

\begin{figure}[ht!]
\begin{center}
\includegraphics[width=16cm]{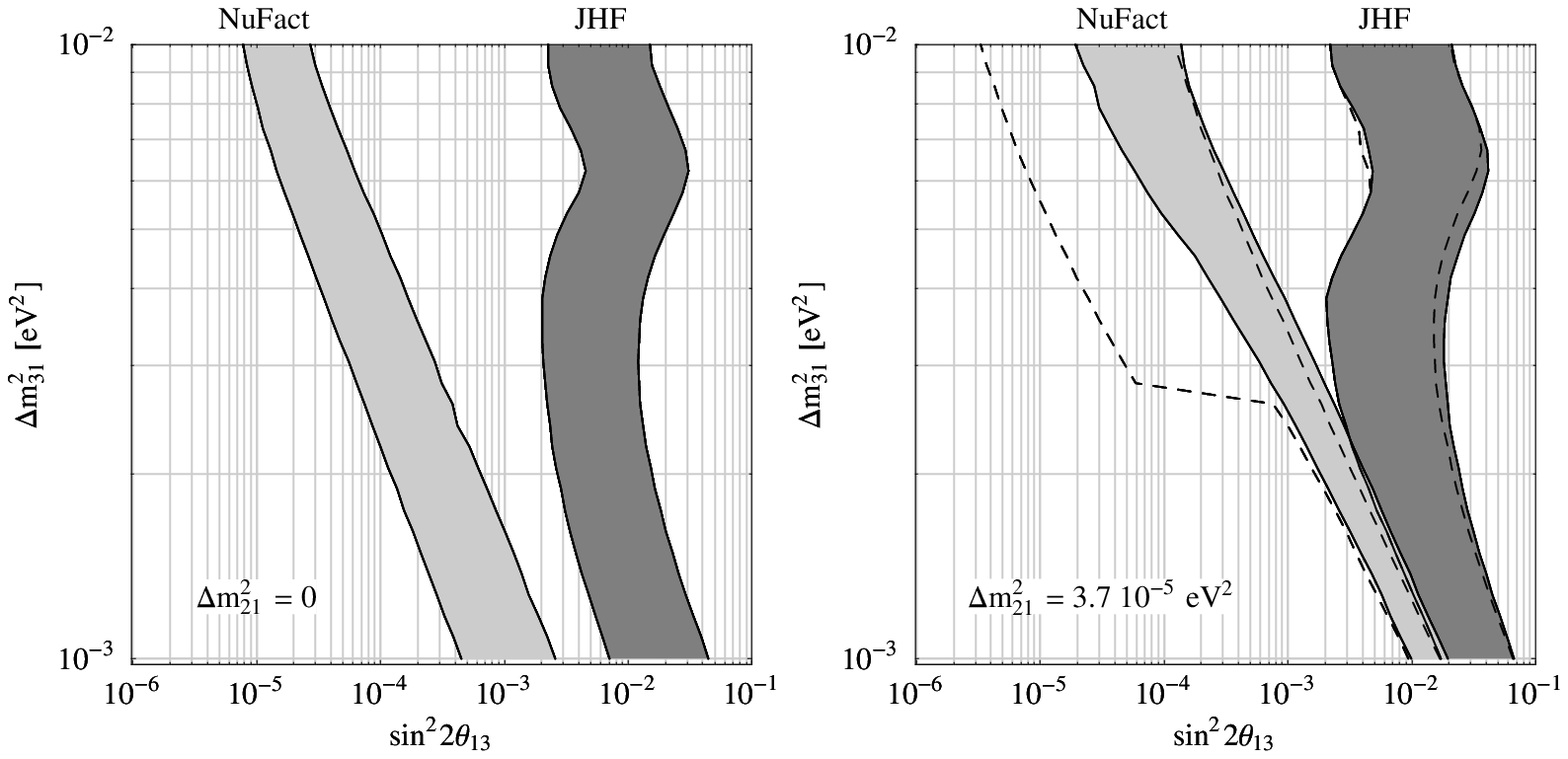}
\end{center}
\mycaption{\label{fig:th13limit} The $\sin^2 2 \theta_{13}$ sensitivity limit at
the 90\%~confidence level, plotted as function of the true values of $\sin^2
2 \theta_{13}$ and $\Delta m_{31}^3$ for the four different experiments. The
shaded regions mark the regions of all superbeam--like (gray) or neutrino
factory--like (black) experiments, where the left curve of each region is drawn
for the initial stage experiment and the right curve for the advanced stage
experiment. The left--hand plot is calculated for $\dm{21}=0$ and $\sin^2  2
\theta_{23}=1.0$ and therefore degenerate solutions do not appear. The
right--hand plot is drawn for $\dm{21}=3\cdot10^{-5}\,\mathrm{eV}^2$ and $\sin^2
2\theta_{23}=0.8$ with all degeneracies taken into account (solid curves). The
dashed curves in the right--hand plot correspond to not taking into account the
$(\theta_{23},\pi/2-\theta_{23})$ degeneracy. For the other oscillation
parameters, we choose the LMA solution.} \end{figure}

Here we analyze the $\sin^2 2 \theta_{13}$ sensitivity limit of our experimental
setups, \ie, the minimum value of $\sin^2 2 \theta_{13}$ the experiments could
distinguish from $\sin^2 2 \theta_{13}=0$ at the $90 \%$ confidence level. Since
we want to demonstrate the effects of the degenerate solutions, we show in
\fig~\ref{fig:th13limit} two plots for different sets of parameters, where the
$\sin^2 2 \theta_{13}$ sensitivity limit is plotted at 90\%~confidence level
over the true values of $\sin^2 2 \theta_{13}$ and $\Delta m_{31}^3$ for the
four different experiments (curves enveloping the shaded regions marking the
regions of all superbeam-- or neutrino factory--like experiments). The
left--hand plot
is produced for $\dm{21}=0$ and $\sin^2  2 \theta_{23}=1.0$, implying that CP
effects are switched off, and thus the $(\delta,\theta_{13})$ degeneracy, as
well as the $(\theta_{23},\pi/2-\theta_{23})$ degeneracy because
of $\sin^2  2 \theta_{23}=1.0$. The right--hand plot is
calculated for $\dm{21}=3\cdot10^{-5}\,\mathrm{eV}^2$ and $\sin^2
2\theta_{23}=0.8$ including these degeneracies by taking the worst case result
over all degenerate solutions, since the experiments themselves can not resolve
them. The dashed curves in this plot show the effect of the
$(\theta_{23},\pi/2-\theta_{23})$ degeneracy being switched off, which means
that the difference between the left--hand plot and the dashed curves in the
right--hand plot essentially comes from $\Delta m_{21}^2$ effects, which are
also controlling the $(\delta,\theta_{13})$ degeneracy. In
\Sec~\ref{sec:degerrors} we have shown that this degeneracy is, because of the
energy resolution, for large $\sin^2 2 \theta_{13}$ not present in the neutrino
factory experiments. However, at the
sensitivity limit there is quite poor statistics and almost no energy
information, which means that the degenerate solution can not be resolved. As it
is illustrated in \fig~\ref{fig:degerror}, the $\mathrm{sgn} ( \Delta m_{31}^2
)$ degeneracy, lying at a similar value of $\sin^2 2 \theta_{13}$, does not
influence the results very much. In summary, the JHF experiments are almost
exclusively influenced by the $\Delta m_{21}^2$ effects and the
$(\delta,\theta_{13})$ degeneracy, reducing the performance by an order of
magnitude for small values of $\Delta m_{31}^2$. Though being systematics
limited, they have less problems with degeneracies in general. The neutrino
factory experiments, however, especially suffer for a large number of events
from the degenerate solutions. For large $\Delta m_{31}^2$, the
$(\theta_{23},\pi/2-\theta_{23})$ degeneracy dominates and for small $\Delta
m_{31}^2$ the $(\delta,\theta_{13})$ degeneracy. It is important to note that
the transition between both cases is close to the best--fit value of $\Delta
m_{31}^2$, meaning that a somewhat smaller value of $\Delta m_{31}^2$ could also
change our impact factor analysis in the last section slightly, since the
$(\delta,\theta_{13})$ ambiguity is especially influenced by the threshold
effects.

\subsection{The $\sin^2 2 \theta_{13}$ precision}
\label{sec:theta13prec}

\begin{figure}[ht!]
\begin{center}
\includegraphics[angle=-90,width=11cm]{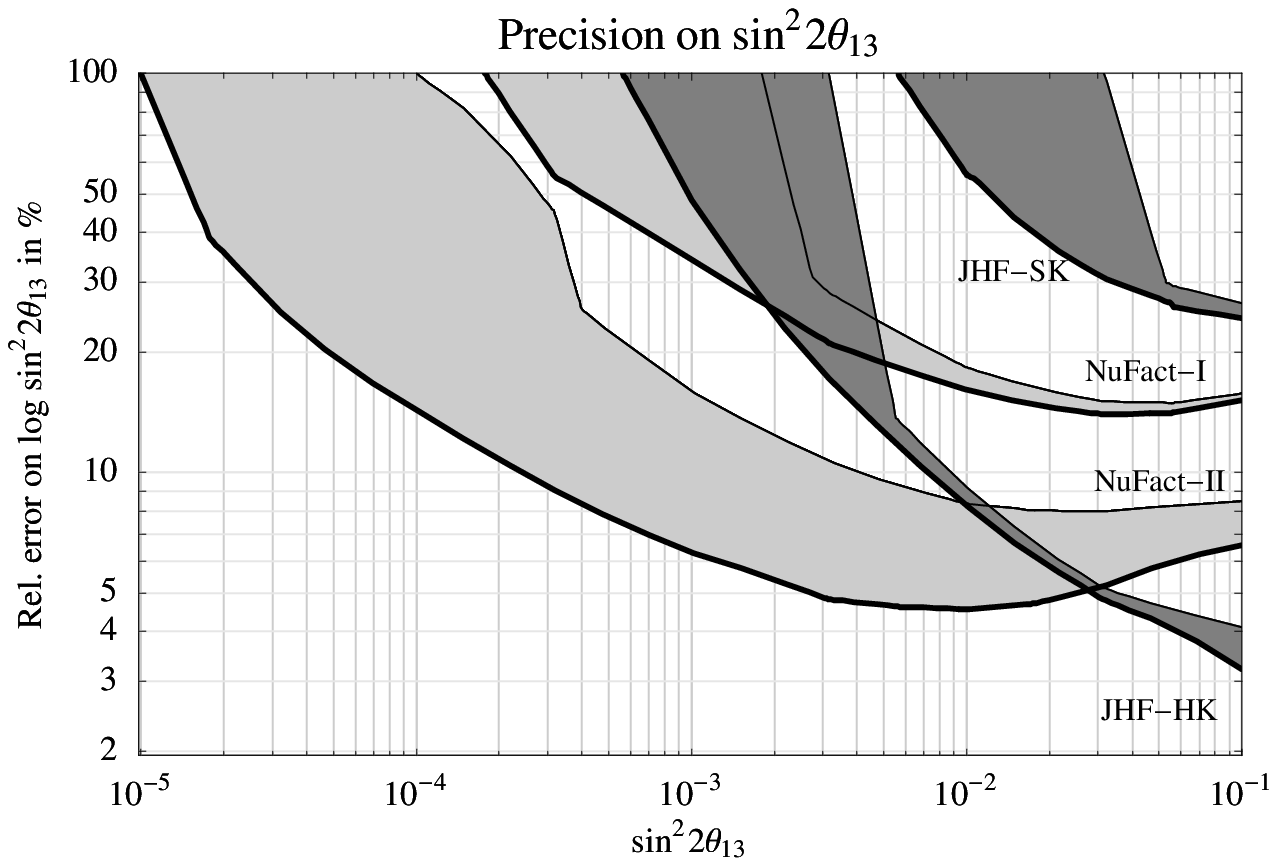}
\end{center}
\mycaption{\label{fig:th13accuracy} The relative error on $\log \sin^2 2
\theta_{13}$, plotted as function of the true value of $\sin^2 2
\theta_{13}$ for the four different experiments. The shaded regions
indicate the error for all possible values of $\deltacp$, which means that the
lower edge (thick curve) of each region corresponds to the best case and the
upper edge (thin curve) to the worst case. For the oscillation parameters, we
choose the LMA solution.} \end{figure}

Compared to the $\sin^2 2 \theta_{13}$ sensitivity, the $\sin^2 2 \theta_{13}$
precision describes how precisely one can measure $\sin^2 2 \theta_{13}$, which
depends on the true value of $\sin^2 2 \theta_{13}$, of course.
\fig~\ref{fig:th13accuracy} shows the relative error on $\log \sin^2 2
\theta_{13}$ over the true value of $\sin^2 2 \theta_{13}$ for the
four different experiments.
The shaded regions indicate the error for all
possible values of $\deltacp$, which means that the lower edge of each
region corresponds to the best case and the upper edge to the worst case. The
lower (thick) curves would also correspond to the case of $\Delta m_{21}^2
\rightarrow 0$ or $\alpha \rightarrow 0$ in \eq~(\ref{eq:PROBVACUUM}), which makes
CP effects disappear. Especially, they would apply for the SMA, LOW,
and VAC solutions. Degeneracies are not taken into account for
this plot, \ie, we focus on the best--fit solution. This is quite a reasonable
assumption, since $\sin^2 2 \theta_{23}=1$ (no $(\theta_{23},\pi/2-\theta_{23})$
degeneracy), $\sin^2 2 \theta_{13}$ is almost everywhere larger than the
sensitivity limit (little effect of the $(\delta,\theta_{13})$ degeneracy), and
the $\mathrm{sgn} ( \Delta m_{31}^2 )$ degeneracy does not affect the $\sin^2 2
\theta_{13}$ measurement very much, as it is illustrated in
\fig~\ref{fig:degerror}. All errors become very large at the respective
sensitivity limits, because at the sensitivity limits the values of $\sin^2 2
\theta_{13}$ cannot be distinguished from zero anymore. As an important result,
the \NuFactII\ experiment performs best except from large values of $\sin^2 2
\theta_{13}$, where the \JHFHK\ setup is better. As we have shown in
\Sec~\ref{sec:impactfactors}, the neutrino factory measurement results are most
strongly influenced by the matter density uncertainty for large $\sin^2 2
\theta_{13}$. One can show that the upward bending of the neutrino factory
curves for large $\sin^2 2 \theta_{13}$ especially comes from the increasing
importance of the matter density uncertainty.

\subsection{The sign of $\Delta m_{31}^2$ sensitivity}
\label{sec:signdm23}

In addition to the measurement of $\sin^2 2 \theta_{13}$, one can ask for the
sensitivity to the sign of $\Delta m_{31}^2$, \ie, what the smallest value of
$\sin^2 2 \theta_{13}$ is, for which the sign of $\Delta m_{31}^2$ could be
measured. As we know from \Sec~\ref{sec:degerrors}, there exists a degeneracy in
the sign of $\Delta m_{31}^2$. Since the two possible cases $\dm{31}<0$ and
$\dm{31}>0$ to be distinguished in this measurement correspond to the
ambiguous solutions of this $\mathrm{sgn} ( \Delta m_{31}^2)$ degeneracy, the
determination of the mass hierarchy is a much more difficult subject than
previously thought (\cf, \Refs~\eg~\cite{CERVERA,FHL,Barger:1999fs}). It
requires sufficiently strong matter effects, which implies the need for a very
long baseline ($>1\,000\,\mathrm{km}$) together with an energy close to
the MSW--resonance energy of about $10$ to $15
\,\mathrm{GeV}$~\cite{Dick:2000fn}. Because the degenerate solution also implies
a different value of $\deltacp$, smaller CP--effects in the second oscillation
maximum would eliminate the CP--phase as a parameter of the measurement and
could help to measure the sign of $\Delta m_{31}^2$. For a neutrino factory,
this translates into a baseline of $\simeq 7\,000\,\mathrm{km}$. For superbeam
experiments, it seems to be questionable if it is possible to satisfy all
these requirements, because the possibility of a measurement in the second
oscillation maximum is strongly limited by statistics.
\begin{figure}[ht!]
\begin{center} \includegraphics[width=8cm]{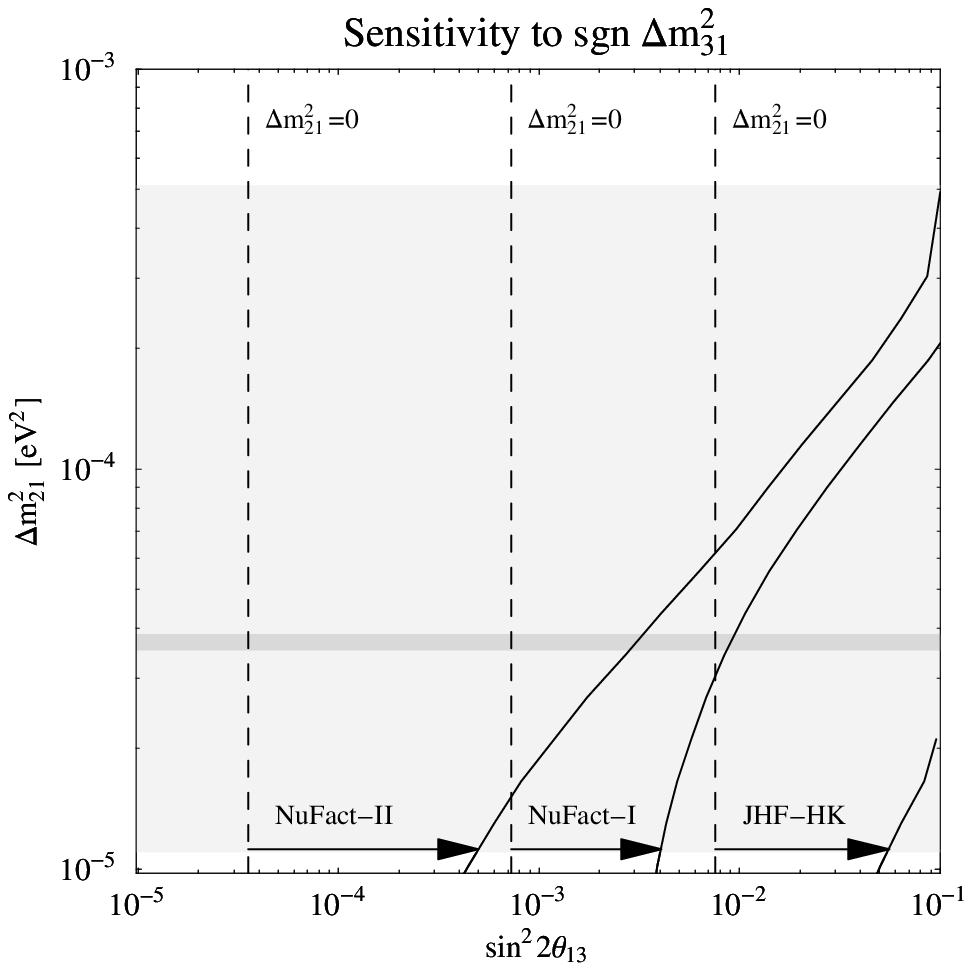} \end{center}
\mycaption{\label{fig:signsens} The sensitivity to the sign of $\Delta m_{31}^2$,
plotted as function of $\sin^2 2 \theta_{13}$ and $\Delta m_{21}^2$ for the
experiments labeled at the respective curves. The arrows indicate the difference
to $\Delta m_{21}^2=0$. The gray shaded regions refer to the allowed LMA region
for $\Delta m_{21}^2$ (light gray) and the best--fit value for $\Delta m_{21}^2$
(dark gray). The vertical dashed lines show the asymptotic sensitivity for
$\dm{21}\rightarrow 0$. }
\end{figure}
For our setups, we show the sensitivity to the sign of
$\Delta m_{31}^2$ in \fig~\ref{fig:signsens} as functions of $\sin^2 2
\theta_{13}$ and $\Delta m_{21}^2$. Since CP effects are
proportional to $\Delta m_{21}^2$, or $\alpha$ in \eq~(\ref{eq:PROBVACUUM}),
the $\mathrm{sgn} ( \Delta m_{31}^2 )$ degeneracy disappears for small values of
$\Delta m_{21}^2$. For large values of $\Delta m_{21}^2$, however, it can spoil
the sign of $\Delta m_{31}^2$ sensitivity as much as that there can be no
sensitivity at all. The figure demonstrates that, depending on the actual value
of $\Delta m_{21}^2$ within the allowed LMA region, this sensitivity can vary
over many orders of magnitude in $\sin^2 2 \theta_{13}$.
For this figure, any allowed degenerate solution fitting the data is taken
into account and may spoil the result if it reduces the
$\chi^2$--function compared to the best--fit solution. In the next section, we 
will re--discover the $\mathrm{sgn} ( \Delta m_{31}^2 )$ degeneracy in a
different context, \ie, the CP measurement performance.

%%%%%%%%%%%%%%%%%%%%%%%%%%%%%%%%%%%%%%%%%%%%%%%%%%%%%%%%%%%%%%%%%%
\section{Results II: The measurement of $\deltacp$}
\label{sec:results2}

The measurement of $\deltacp$ can be treated from two different points of
view. First, one can ask for which values of $\deltacp$ and $\sin^2 2
\theta_{13}$ an experiment could distinguish CP violation from CP conservation,
\ie, $\deltacp = 0$ or $\deltacp = \pi$. This we call CP violation
sensitivity. Second, one can discuss the precision of the measurement of
$\deltacp$. In this case, we are talking about
the $\deltacp$ precision. Since both problems are interesting and quite
different from each other, we will investigate both of them.

We learned in \Sec~\ref{sec:degerrors} that the $\mathrm{sgn} ( \Delta m_{31}^2
)$ degeneracy may allow different solutions at different values of
$\deltacp$. In addition, all solutions in $\deltacp$ for which the energy
spectrum fits the one of the true value of $\deltacp$ are lying on a circle
from $0$ to $2 \pi$. Moreover, there is little theoretical evidence for which
value of $\deltacp$ should be preferred, \ie, any value may be equally likely.
Thus, we need to define the precision of the CP measurement in a different way
than the one of the $\sin^2 2 \theta_{13}$ measurement. We choose
\begin{equation}  \mathrm{Coverage \, in \,} \deltacp
\equiv \frac{\sum\limits_I \Delta^{(I)}(\deltacp)}{2 \pi},
\label{eq:covdcp}
\end{equation}
where $I$ represents all possible intervals where solutions are allowed
and $\Delta^{(I)}(\deltacp)$ the corresponding interval lengths on the
$\deltacp$ circle at the chosen confidence level. It represents the coverage
of the circle of all allowed CP phases with possible solutions, corresponding to
adding the intervals of the errors in direction of the $\delta$--axis in
\fig~\ref{fig:degerror}. If this quantity is unity, all possible values of
$\deltacp$ are allowed, meaning that the relative measurement error is $100
\%$. The smaller it is, the better is the performance of the experiment.

\subsection{Impact factors on the measurement of $\deltacp$}
\label{sec:impactfactorscp}

\begin{figure}[ht!] \begin{center}
\includegraphics[width=7.5cm]{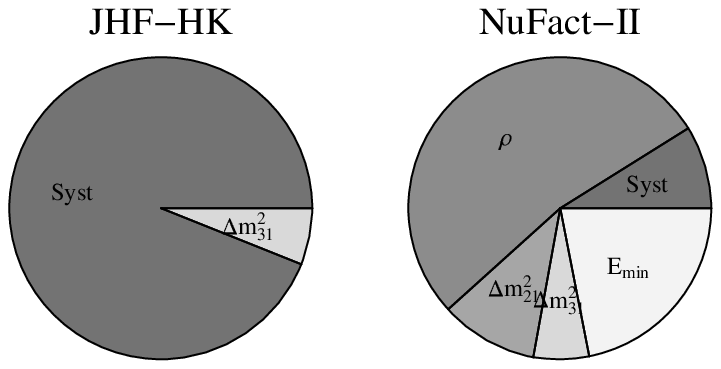}
\end{center}
\mycaption{\label{fig:errorsCP} The relative contributions of the impact factors
on the measurement of $\deltacp$ for $\sin^2 2 \theta_{13}=0.01$ and
$\deltacp=\pi/4$. The error sources $\xi$ are the systematical errors
(``Syst''), the correlations with the large (``$\Delta
m_{31}^2$'') and small (``$\Delta m_{21}^2$'') mass squared differences, the
energy threshold function for neutrino factories (``$E_{\min}$''), and the
uncertainty in the matter density (``$\rho$''). For the other oscillation
parameters, we choose the LMA solution.}
\end{figure}

Before we come to the discussion of the CP performance of the experiments, we
show the main impact factors for the CP measurement.
We define the impact factors for the measurement of $\deltacp$ similar to
\eq~(\ref{eq:impactfactor}) in \Sec~\ref{sec:impactfactors} for $\sin^2 2
\theta_{13}$. Note, however, that since in this case the error of the
measurement is defined as fraction of the $2 \pi$ interval in
\eq~(\ref{eq:covdcp}), the impact factor is also defined in terms of the
relative change of this fraction. Because we will see below that
a sensible CP measurement is, at least for the LMA best--fit case, only
possible for the \JHFHK\ and \NuFactII\ experiments, we focus on these two
experiments. In addition, the impact factors depend, of course, on the values of
$\sin^2 2 \theta_{13}$ and $\deltacp$. In order to identify and compare the
impact factors on the errors of these two experiments, we choose a parameter set
which allows a quite good and comparable CP measurement in both cases, \ie,
$\sin^2 2 \theta_{13}=0.01$ and $\deltacp=\pi/4$. In addition, we again focus
on the best--fit solution ignoring degeneracies. The result of this analysis is
shown in \fig~\ref{fig:errorsCP}. 

For \JHFHK\ we do not observe any surprising results, the
measurement of $\deltacp$ is, as the one of $\sin^2 2 \theta_{13}$, dominated
by systematics. For \NuFactII, however, the uncertainty in the matter effects
takes over the main contribution to the error (impact factor about $36 \%$).
The importance of the matter density uncertainty on the
CP measurement has, for example, been pointed out in \Ref~\cite{Shan:2001br}. In
addition, the impact of the energy threshold function is quite large
(impact factor about $15 \%$).

\begin{figure}[ht!]
\begin{center}
\includegraphics[width=10cm]{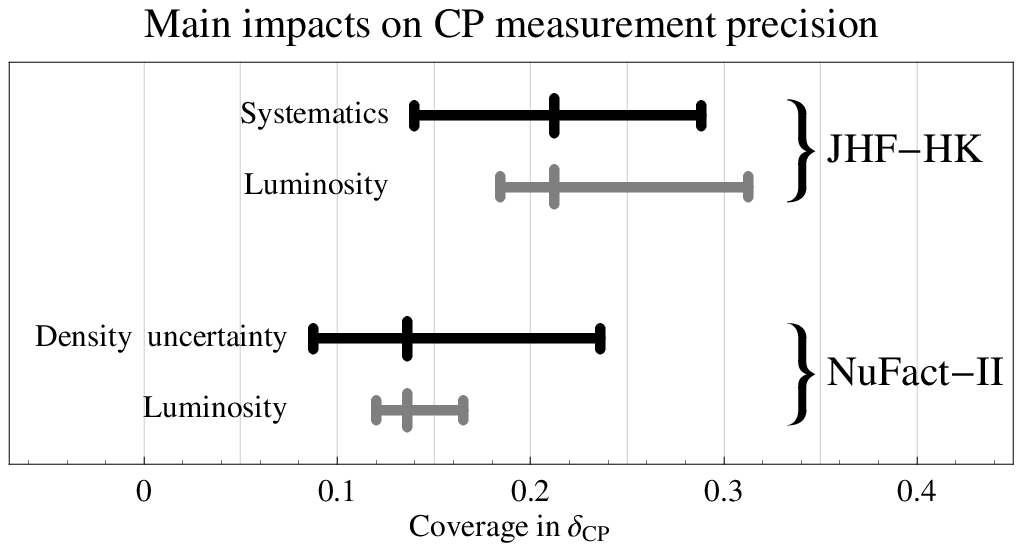}
\end{center}
\mycaption{\label{fig:mainimpactsCP} The coverage in $\deltacp$, as
defined in \eq~(\ref{eq:covdcp}), at the $2 \sigma$ confidence level for the
\JHFHK\ and \NuFactII\ experiments. Each bar represents
the range of the sensitivity under a variation of the main impact factor (black
bar) or under a variation of the luminosity (gray bar), as labeled on the
left--hand side of the bar. The left vertical dash on each bar corresponds to
the optimistic choice, the middle vertical dash to our standard choice, and the
right vertical dash to the pessimistic choice, where the meaning of each choice
is described in the main text. For the oscillation parameters, we choose the LMA
solution with $\deltacp=\pi/4$ and $\sin^2 2 \theta_{13}= 0.01$, since the two
experiments perform similarly well at these parameter vales.} \end{figure}

To summarize our results, \fig~\ref{fig:mainimpactsCP} shows the coverage in
$\deltacp$, as defined in \eq~(\ref{eq:covdcp}), on the $2 \sigma$ confidence
level for the \JHFHK\ and \NuFactII\ experiments. Each
bar in this figure represents the range of the coverage in $\deltacp$ under a
variation of the main error impact or under a variation of the luminosity. The
vertical dashes on each bar correspond to the optimistic choice (left), the
standard choice (middle), and the pessimistic choice (right). For the
luminosity, we choose twice the luminosity of our standard assumption for the
optimistic choice and half the luminosity of our standard assumption for the
pessimistic choice. For the main impact factors, we proceed similarly with the
systematical and matter density uncertainties and double it or divide it by two.
The figure shows that the \JHFHK\ experiment could be equally improved by
increasing the luminosity and reducing the systematical errors, \ie, mainly the
background uncertainties. For the \NuFactII\ experiment, though it is statistics
dominated, the matter density uncertainty is the main impact factor and even
improving the luminosity cannot compensate the need for additional knowledge
about the Earth's matter density profile.

\subsection{The sensitivity to CP violation}
\label{sec:cpviolation}

\begin{figure}[ht!]
\begin{center}
\includegraphics[angle=-90,width=16cm]{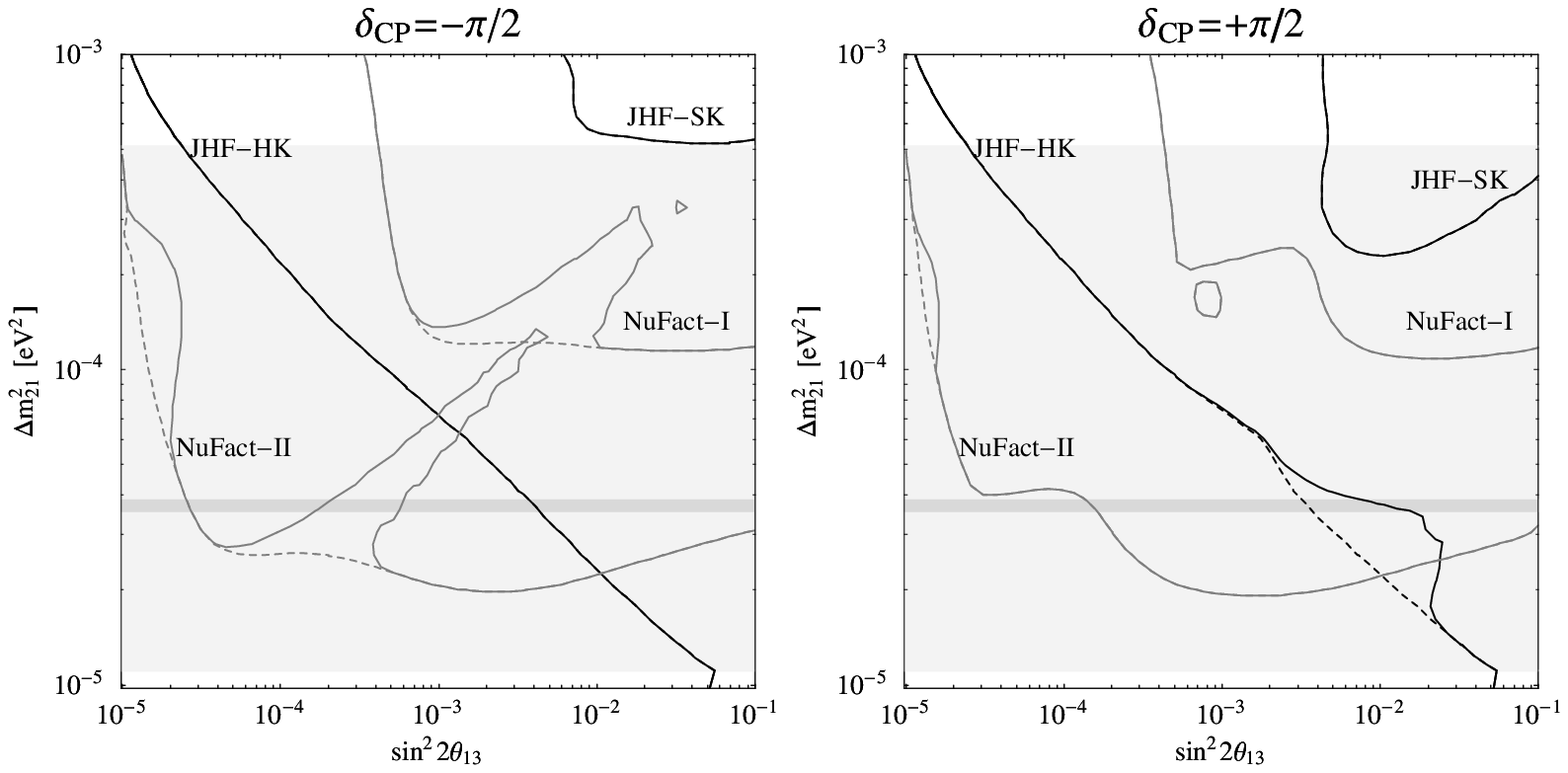}
\end{center}
\mycaption{\label{fig:cpviolationsdm}The sensitivity to CP violation, as it is defined
in the text, for all experiments (as labeled in the plots) at the $2 \sigma$
confidence level, plotted as function of $\sin^2 2 \theta_{13}$ and $\Delta
m_{21}^2$. The left--hand plot shows the case of $\deltacp = - \pi/2$ and the
right--hand plot the case of $\deltacp = + \pi/2$. Solid curves refer to taking
into account all degenerate solutions and dashed curves to taking into
account the best--fit manifold only. The gray shaded regions refer to the
allowed LMA region for $\Delta m_{21}^2$ (light gray) and the best--fit value
for $\Delta m_{21}^2$ (dark gray). For the other oscillation parameters, we
choose the LMA values.  } \end{figure}

\begin{figure}[ht!] \begin{center}
\includegraphics[width=16cm]{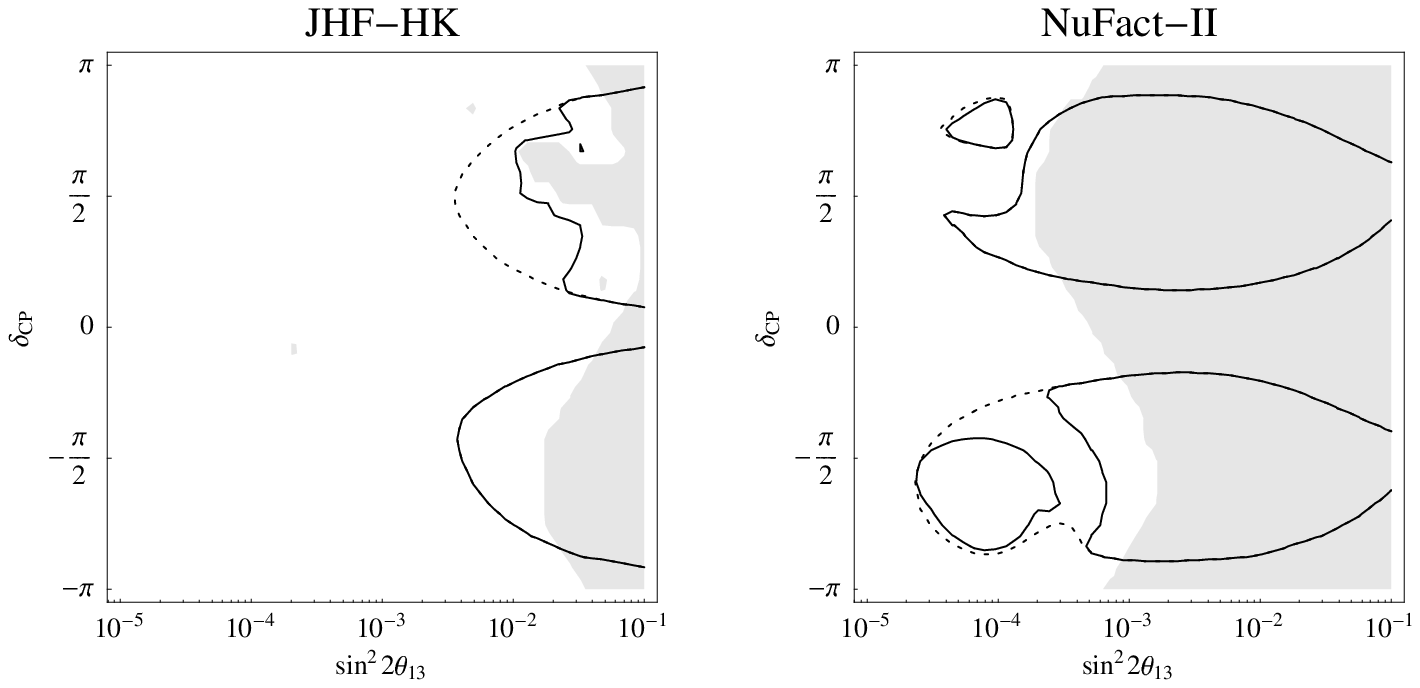} \end{center}
\mycaption{\label{fig:cpviolation} The sensitivity to CP violation, as it is
defined in the text, for the \JHFHK\ (left--hand plot) and \NuFactII\
(right--hand plot) experiments at the $2 \sigma$ confidence level, plotted as
functions of $\sin^2 2 \theta_{13}$ and $\deltacp$. Solid curves refer to the
case of taking into account all degenerate solutions and dashed curves to
taking into account the best--fit manifold only. The gray shading marks the
regions where no degenerate solutions are present at the considered confidence
level. For the oscillation parameters, we choose the LMA values.} \end{figure}

In this section, we focus on the identification of CP violation. We define the
CP violation sensitivity as the ability of the experiment to distinguish CP
violation from CP conservation on the $2 \sigma$ confidence level, where CP
conservation refers to $\deltacp=0$ or $\deltacp=\pi$, \ie, we define the
overall $\chi^2$--function $\chi^2_{\mathrm{CP \, violation}}(\delta_0)$ at the
best--fit point $\delta_0$ as
\begin{equation*}
\chi^2_{\mathrm{CP \, violation}}(\delta_0) \equiv \min \left[
\chi^2(\delta_0,\delta=0) ,\chi^2(\delta_0,\delta=\pi) \right].
\end{equation*}
Similarly to the $\sin^2 2 \theta_{13}$ sensitivity, we take into
account all degenerate solutions by the worst case strategy. Since the CP
violation sensitivity depends very much on $\sin^2 2 \theta_{13}$ and $\Delta
m_{21}$, as well as the value of $\deltacp$ itself, we show the results in
two figures. \fig~\ref{fig:cpviolationsdm} shows the regions of CP violation
sensitivity plotted as functions of $\sin^2 2 \theta_{13}$ and $\Delta m_{21}^2$
for $\deltacp=-\pi/2$ (left plot) and $\deltacp=+\pi/2$ (right plot)
for the different experiments (regions closed by contours to the left), where
the dashed curves show what happens without taking into account degenerate
solutions. First, we observe that there is no CP violation sensitivity for the
\JHFSK\ and \NuFactI\ setups at the LMA best--fit value of $\Delta m_{21}^2$.
Therefore, we will further on not discuss the measurement of $\deltacp$ in
the context of these two experiments anymore. Second, the \NuFactII\ experiment
performs better than the \JHFHK\ experiment for most values of $\Delta m_{21}^2$
close to the LMA solution best--fit value. However, the difference between these
two experiments can be quite small even within the allowed LMA region indicated
by the light gray shading. Third, the \NuFactII\ curve is again bend upwards for
large values of $\sin^2 2 \theta_{13}$, such as it was in
\fig~\ref{fig:th13accuracy}, an effect coming from the matter density
uncertainty. For an explanation of the degeneracy effects leading to the gap in
the left--hand plot, we refer to \fig~\ref{fig:trees} and the respective
discussion in \Sec~\ref{sec:degerrors} (\cf, position of the ``$\pi$--transit
point''). The counterpart to \fig~\ref{fig:cpviolationsdm} is
\fig~\ref{fig:cpviolation}, showing the CP violation sensitivity over $\sin^2 2
\theta_{13}$ and $\deltacp$ for the LMA small mass squared difference.
Again, the dashed lines show the case of not taking into account the
degeneracies and how they make the result worse. It is obvious in
these plot that there is no CP violation sensitivity around the CP conserving
values $0$ and $\pi$ and the CP violation sensitivity is limited by the $\sin^2
2 \theta_{13}$ sensitivity to the left. The \NuFactII\ experiment performs well
in a large range of $\sin^2 2 \theta_{13}$, while the \JHFHK\ setup is better at
around the CHOOZ bound. The reason is again the matter density uncertainty,
making the contours bend inwards for the \NuFactII\ experiment. The upper
irregularity in the \NuFactII\ plot seems to come from a numerical coincidence,
the lower irregularity can be explained by degeneracy effects similar to the
last example. We also refer to \fig~\ref{fig:trees} and the respective
discussion in \Sec~\ref{sec:degerrors} for an explanation.

\subsection{The $\deltacp$ precision}
\label{sec:precisiondeltacp}

\begin{figure}[ht!]
\begin{center}
\includegraphics[width=11.5cm]{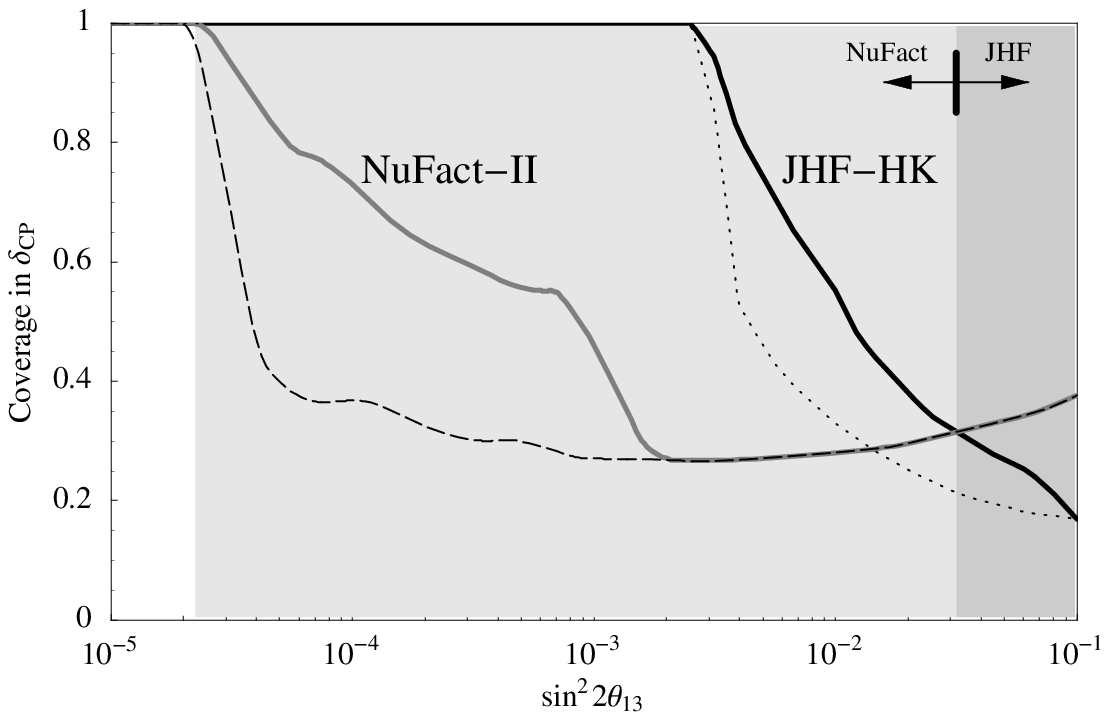}
\end{center}
\mycaption{\label{fig:cpviolationprec} The coverage in $\deltacp$, as defined
in \eq~(\ref{eq:covdcp}), plotted as function of $\sin^2 2 \theta_{13}$ for
the \JHFHK\ and \NuFactII\ experiments. The curves are computed for the worst
case in $\deltacp$. Solid curves refer to the case of
taking into account all degenerate solutions, dashed curves refer to taking into
account the best--fit manifold only. The light gray shading marks the
region, where the \NuFactII\ experiment performs better and the dark gray
shading the region, where the \JHFHK\ experiment performs better.
For the oscillation parameters, we choose the LMA values.}
\end{figure}

In this section, we especially discuss the precision of the CP measurement,
for which we introduced the quantity ``Coverage in $\deltacp$'' in
\eq~(\ref{eq:covdcp}) to characterize the error on $\deltacp$. It is plotted
in \fig~\ref{fig:cpviolationprec} over $\sin^2 2 \theta_{13}$ for the \NuFactII\
and \JHFHK\ experiments, where the worst case over all values of $\deltacp$
is taken.

The dashed and dotted curve correspond to the case of not taking into account
degenerate solutions. The \NuFactII\ experiment performs better in the light
gray shaded region, whereas the \JHFHK\ experiment does better in the dark gray
shaded region. However, the upward bending of the \NuFactII\ curve close to
the CHOOZ bound again comes from the matter density uncertainty.
The bump in the \NuFactII\ curve below about $\sin^2 2
\theta_{13}=10^{-3}$ essentially comes from the
$\mathrm{sgn} ( \Delta m_{31}^2 )$ degeneracy, such as the gaps in the last two
plots. Here we also refer to \fig~\ref{fig:trees} and the respective discussion
in \Sec~\ref{sec:degerrors} for an explanation.

%%%%%%%%%%%%%%%%%%%%%%%%%%%%%%%%%%%%%%%%%%%%%%%%%%%%%%%%%%%%%%%%%%
\section{Summary and conclusions}
\label{sec:conclusion}

\begin{figure}[ht!]
\begin{center}
\includegraphics[angle=-90,width=11cm]{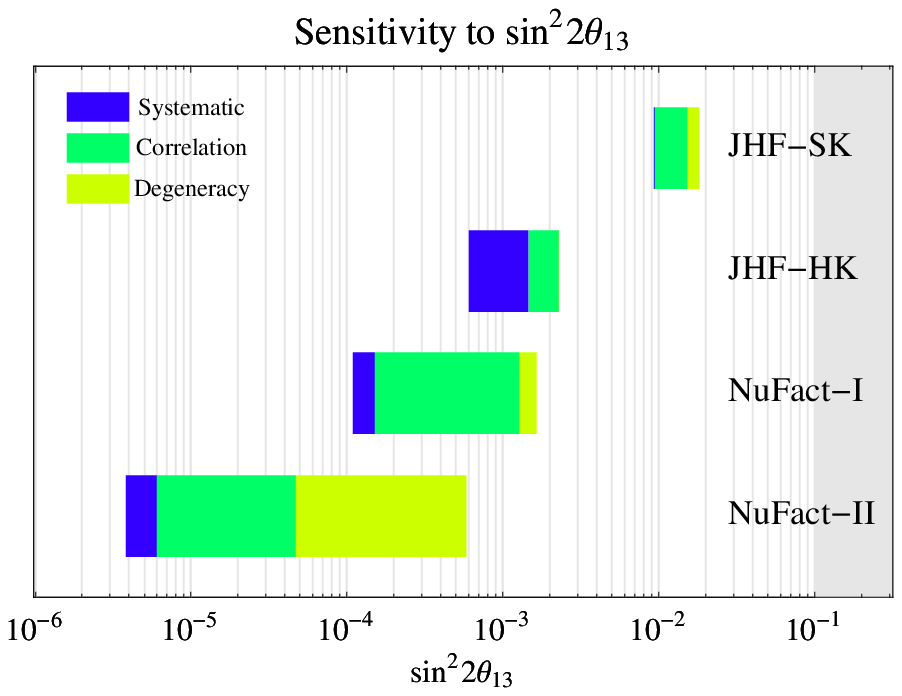}
\end{center}
\mycaption{\label{fig:th13summary} The sensitivity to $\sin^2 2
\theta_{13}$ at the 90\%  confidence level for all experiments and for $\Delta
m_{31}^2 = 3 \cdot 10^{-3} \, \mathrm{eV}^2$. The plot shows the effect of
successively switching on the different error sources.  For the oscillation
parameters, the choose the LMA values but $\sin^2 2 \theta_{23}=0.8$. }
\end{figure}

\begin{figure}[ht!]
\begin{center}
\includegraphics[angle=-90,width=11cm]{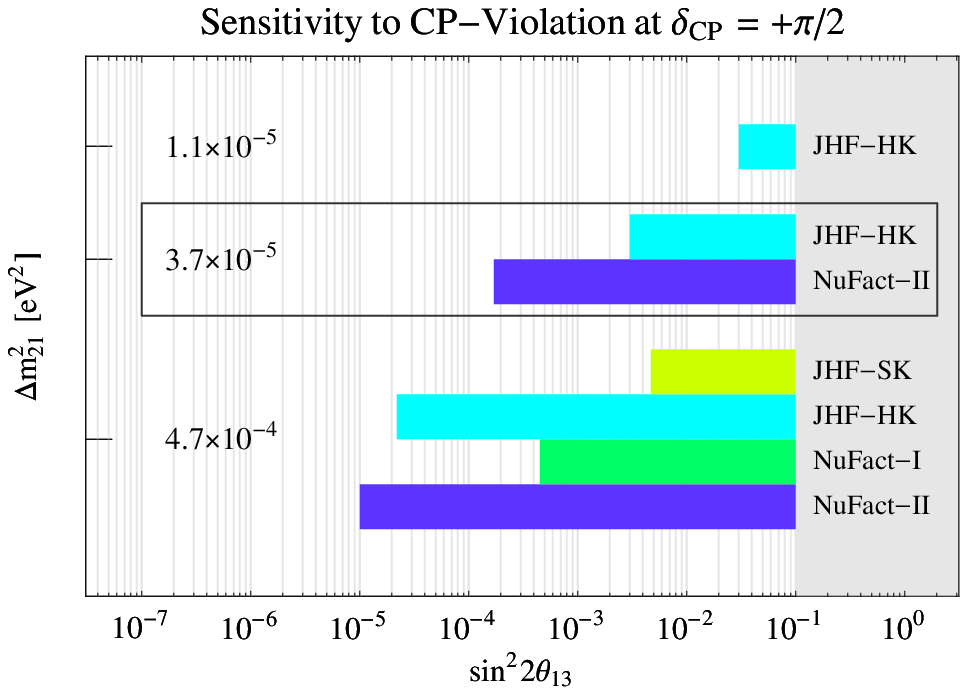}
\end{center}
\mycaption{\label{fig:cpsummary} The sensitivity to CP violation
for $\deltacp=+\pi/2$ at the 90\%  confidence level, plotted as function of
$\sin^2 2 \theta_{13}$. The plot shows the ranges of the sensitivity to CP
violation for several values of $\Delta m_{21}^2$, where the top row corresponds
to the lower bound, the bottom row to the upper bound, and the middle row to the
best--fit value of the LMA region. If no bar is drawn for an experiment, it does
not have any CP violation sensitivity. For the parameters, the choose the LMA
values.  }
\end{figure}

In summary, we have studied different superbeam and neutrino
factory setups in order to assess and compare their physics potential for
the first time on an equal footing. We have tried to model the experimental
details and systematical uncertainties of each class of experiments as realistic
as possible, because these are crucial to the performance of each experiment.
We have performed a sophisticated statistical analysis
based on event rates, including the \emph{full} multi--parameter correlations
among all oscillation parameters as well as the appearance of degenerate
solutions. As additional input, we have assumed that the solar parameters are
measured by the KamLAND experiment with a good accuracy. This comprehensive
analysis is necessary to obtain reliable estimates of the
physics reach of each long baseline experiment. In some cases, we have even
observed a worsening of the results by orders of magnitude by switching on the 
sources of errors. For example, the
impact of the degenerate solutions can completely destroy the sensitivity to the
sign of $\Delta m_{31}^2$ even for a high luminosity neutrino factory, such as
our \NuFactII\ setup.

In order to perform an appropriate comparison among the experiments, aspects
equivalent for both machines have been described as by identical parameters as
much as possible. We have, for example, assumed that the \JHFSK\  and \NuFactI,
as well as the \JHFHK\ and \NuFactII\ setups have pairwise equal running
periods. For example, we have for the \JHFSK\ and \NuFactI\ experiments taken
parameters which seem to be realistic for a first experiment of the considered
category. For the \JHFHK\ and \NuFactII\ experiments, we have in addition chosen
parameters which we consider realistic for a fully developed setup. However,
such a study always depends on the current understanding or expectations. Our
comparison allows moderate extrapolation if some of the parameters are
different.

The most important aspects of our comparative results are summarized in
\figs~\ref{fig:th13summary} and~\ref{fig:cpsummary}, where the potentials to
measure $\sin^2 2\theta_{13}$ and the CP phase $\deltacp$ are illustrated for
the four scenarios. \Fig~\ref{fig:th13summary} demonstrates that the
$\sin^2 2\theta_{13}$ sensitivity is reduced from an initially only
statistically limited value (left edge) to a final, realistic value (right edge)
including systematical errors, correlations, and degeneracies, which are
successively switched on\footnote{Note, however, that these
errors add up in a non--commutative way, which means that the shown
contributions depend on the order in which they are displayed. They are
switched on from the left to the right.}. The sensitivities of all experiments
are excellent compared to the existing bounds. The \JHFSK\ setup can explore
$\sin^22\theta_{13}$ down to $\simeq 2\cdot10^{-2}$, whereas the \JHFHK\ and
\NuFactI\ setups are sensitive down to $\simeq 10^{-3}$. The \NuFactII\
experiment can finally constrain  $\sin^22\theta_{13}$ to values as
small as $\simeq 5\cdot10^{-4}$, which could be
considerably improved by about one order of magnitude by a
second baseline or almost maximal mixing in the atmospheric sector to lift the
$(\theta_{23},\pi/2-\theta_{23})$ degeneracy.

The most important impact factors controlling the sensitivity to
$\sin^2 2 \theta_{13}$ have been presented in \Sec~\ref{sec:impactfactors}.
First, we have clearly seen that the \JHFSK\ setup is limited by 
statistics (\ie, the luminosity) and that systematical errors are of minor
importance. Second, the \JHFHK\ experiment performs about one order of magnitude
better and systematical uncertainties and correlations become the limiting
factors. Third, the \NuFactI\ setup is on a purely statistical basis about one
order of magnitude better than the \JHFHK\ experiment, but both options become
rather similar after the inclusion of all errors. The largest reduction in the
sensitivity comes from the correlations, \ie, especially the correlation with
$\deltacp$. Finally, the \NuFactII\ setup is on a purely statistical level much
better than all the other options. However, taking into account realistic
errors, the final $\sin^2 2\theta_{13}$ sensitivity
limit is then only less than one order of magnitude better than that of the
\NuFactI\ setup, especially due to the correlation and degeneracy errors. Note,
however, that the sensitivity losses by the correlations and
degeneracies can, in principle, be partially compensated by combining
different experiments. The route to a setup such as \NuFactII\ will
certainly involve at least one superbeam and an initial stage neutrino
factory. Having different baselines and energies, these experiments in the
staging scenario would help to improve the
sensitivity limits by adding partially complementary information. The final
sensitivity limit would then be somewhere between the systematic sensitivity
limit (right edge of the black bars in \fig~\ref{fig:th13summary}) and the full
sensitivity limit (right edge of the total bars in \fig~\ref{fig:th13summary}).
From this point of view, the initial stage neutrino factory could help to improve
the performance of a superbeam much better than the \JHFHK\ setup, since it is
less limited by systematics. In this staging scenario, the \NuFactII\
experiment could provide improved results on the $\sin^2 2\theta_{13}$
sensitivity by about one order of magnitude. This argumentation demonstrates
that the final assessment of the physics potential of a single setup also
depends on its predecessors as subsequent milestones in the staging
scenario. It is therefore important to keep in mind that the physics
scopes of planned experiments should be coordinated in a way that they 
are providing at least some complementary information.

An issue closely connected to the measurement of $\sin^2 2 \theta_{13}$, is the
determination of the mass hierarchy, \ie, the measurement of $\mathrm{sgn} (
\Delta m_{31}^2 )$. Surprisingly, we have found that the
degeneracy between $\deltacp$ and the sign of $\dm{31}$ is for all experiments
the critical factor which is limiting the sensitivity. Its effects are strongest
for the two superbeam experiments because of the short baselines and low
energies, which means that matter effects are small. It therefore seems
questionable if one can perform this measurement with superbeams at all. Not
only the superbeams, but also the neutrino factory setups are limited by this
degeneracy. For the largest allowed values of $\dm{21}$ within the LMA
region, it can lead to a complete loss of sensitivity. At the LMA best--fit
value, the smallest value of $\sin^22\theta_{13}$ for which the \NuFactI\
setup can determine the sign of the mass hierarchy, is about $10^{-2}$, whereas
it is $5\cdot10^{-3}$ for the \NuFactII\ scenario. The influence of is
degeneracy could be reduced by a measurement at a second very long baseline of
$\simeq7\,000\,\mathrm{km}$~\cite{FLPR,FHL,Freund:2001ui}.

\Fig~\ref{fig:cpsummary} shows the sensitivity to CP violation  as a function of
$\sin^2 2\theta_{13}$, which we have defined as the ability to distinguish
maximal CP violation from CP conservation. Therefore, it represents
the limits to CP effects for our setups. It is important to note that the
minimal value of $\sin^2 2\theta_{13}$ with sensitivity to CP violation
considerably depends on the value of $\Delta m^2_{21}$ within the LMA range. For
the LMA central value $\Delta m^2_{21}= 3.7\cdot 10^{-5}~\mathrm{eV^2}$, only
the \JHFHK\ and \NuFactII\ experiments are sensitive at all. In
general, the \NuFactII\ experiment performs best down to
$\sin^22\theta_{13}\simeq10^{-4}$ and is at the solar best--fit point about an
order of magnitude better than the \JHFHK\ setup. For small values of $\Delta
m_{21}^2$, however, only the \JHFHK\ setup is sensitive at all, which is a
consequence of the short baseline and therefore the irrelevance of the matter
density uncertainty. At the other end of the LMA range at the largest
allowed values of $\Delta m_{21}^2$, all setups can measure CP violation and the
\JHFHK\ and \NuFactII\ experiments perform quite similar. Nevertheless, the
accuracy of the \NuFactII\ is slightly better because of the large number of
events. The critical experimental factor limiting the CP sensitivity is,
for superbeams, mainly the systematical uncertainty and, for neutrino
factories, the matter density uncertainty. This also explains the surprisingly
good performance of the \JHFHK\ experiment.

In this work, we have not discussed the optimization problem for our setups in
great detail. However, for superbeams we identified the background
uncertainties as the critical factor to be reduced. A potential measurement at a
second, much longer baseline seems to be difficult in this case because of the
huge loss in event numbers. For the neutrino factories, we have found that
they would benefit from a special low energy detector for events in the range 
of about $4$ to $20\,\mathrm{GeV}$. In addition, a reduction in the
uncertainty of the matter density would help. A second, very long baseline is
probably necessary to have stronger matter effects and determine the
mass hierarchy.

%%%%%%%%%%%%%%%%%%%%%%%%%%%%%%%%%%%%%%%%%%%%%%%%%%%%%%%%%%%%%%%%%%%%
%%%%                      Acknowledgments                      %%%%
%%%%%%%%%%%%%%%%%%%%%%%%%%%%%%%%%%%%%%%%%%%%%%%%%%%%%%%%%%%%%%%%%%%%

\vspace*{7mm}
\subsubsection*{Acknowledgments}
We wish to thank M.~Messier, T.~Nakaya, and
C.~Pe$\tilde{\mathrm{n}}$a--Garay for providing us input data and
useful information. Furthermore, we thank M.~Freund and N.~Kaiser for
useful discussions and comments.

%%%%%%%%%%%%%%%%%%%%%%%%%%%%%%%%%%%%%%%%%%%%%%%%%%%%%%%%%%%%%%%%%%%%%
%%%%                       References                            %%%%
%%%%%%%%%%%%%%%%%%%%%%%%%%%%%%%%%%%%%%%%%%%%%%%%%%%%%%%%%%%%%%%%%%%%%

%%%%%%%%%%%%%%%%%%%%%%%%%%%%%%%%%%%%%%%%%%%%%%%%%%%%%%%%%%%%%%%%%%%%%
\newpage
\begin{appendix}
%%%%%%%%%%%%%%%%%%%%%%%%%%%%%%%%%%%%%%%%%%%%%%%%%%%%%%%%%%%%%
\section{Calculation of event rates}\label{sec:appendix}
%%%%%%%%%%%%%%%%%%%%%%%%%%%%%%%%%%%%%%%%%%%%%%%%%%%%%%%%%%%%%

The possible types of events in any detector can be classified by the flavor of
the detected neutrino and its interaction type. For the experiments used
in this work, these interaction types (IT) are neutral current (NC), total
charged current (CC) and quasi--elastic charged current (QE) interactions. The
relative abundance of these modes strongly depends on the primary
energy of the incident neutrino, which means that, for example, QE
interactions are only important at energies well below $10\,\mathrm{GeV}$. In
this appendix, we will demonstrate how the interaction modes are implemented
in the calculation of event rates.

\subsection*{Calculation of raw event rates}

For the calculation of event rates, the first step is to compute the number of
events for each IT in the fiducial mass of the detector for each neutrino flavor
and energy bin. The second step is to include the detector effects coming from
the insufficient knowledge used in the event reconstruction. We combine these
two steps in order to obtain the differential event rate spectrum for each
flavor and IT as seen by the detector, which we call a ``channel''. The channels
for all ITs then have to be combined in a way which takes into account
that the ITs or flavors can, either because of
physical reasons (\eg, the flavor--blindness of NC interactions) or because of
detector effects (\eg, charge misidentification), not be measured separately.

The master formula for the differential event rate for each channel, \ie,
the final flavor $f$ and the interaction type IT, is given by
%%%%%%%%%%%%%%%%%%%%%%%%%%%%%%%%%%%
\begin{eqnarray}
\label{eq:master}
\frac{dn_{f}^{\mathrm{IT}}}{dE'}=&&N\,\sum_{f_i}\int \int dE\,d\hat{E}\quad
\underbrace{\Phi_{f_i} (E)}_{\mathrm{Production}} \times \nonumber\\
&&\underbrace{\frac{1}{L^2} P_{(f_i\rightarrow f)}(E,L,\rho;\theta_{23},\theta_{12},\theta_{13},
\Delta m^2_{31},\Delta m^2_{21},\deltacp)}_{\mathrm{Propagation}}
\times \nonumber \\ &&\underbrace{\sigma^{\mathrm{IT}}_f(E)
k_f^{\mathrm{IT}}(E-\hat{E})}_{\mathrm{Interaction}} \times \nonumber \\
&&\underbrace{ T_f(\hat{E}) V_f(\hat{E}-E')}_{\mathrm{Detection}}, \end{eqnarray}
%%%%%%%%%%%%%%%%%%%%%%%%%%%%%%%%%%%
where $f_i$ is the initial flavor of the neutrino, $E$ is the incident
neutrino energy, $\Phi_{f_i} (E)$ is the flux of the initial flavor at the
source, $L$ is the baseline length, $N$ is a normalization factor, and $\rho$ is
the matter density. The interaction term is composed of two factors, which
are the total cross section $\sigma^{\mathrm{IT}}_f(E)$ for the flavor $f$ and
the interaction type IT, and the energy distribution of the secondary particle
$k_f^{\mathrm{IT}}(E-\hat{E})$ with $\hat{E}$ the energy of the
secondary particle. The detector properties are modeled by the
threshold function $T_f(\hat{E})$, coming from the the limited resolution or
the cuts in the analysis, and the energy resolution function $V_f(\hat{E}-E')$
of the secondary particle. Thus, $E'$ is the {\em reconstructed} neutrino
energy.

Since it is rather cumbersome to numerically solve this double integral,
we use an approximation. We evaluate the integral over
$\hat{E}$, where the only terms containing $\hat{E}$ are
$k_f^{\mathrm{IT}}(E-\hat{E})$,  $ T_f(\hat{E})$, and $ V_f(\hat{E}-E')$, and
define
\begin{eqnarray} R_f^{\mathrm{IT}}(E,E')\,\epsilon_f^{\mathrm{IT}}(E') \equiv
\int d\hat{E} \quad T_f(\hat{E})\,k_f^{\mathrm{IT}}(E-\hat{E})
\,V_f(\hat{E}-E'). \end{eqnarray}
We now approximate $R_f^{\mathrm{IT}}$ by the analytical expression
\begin{equation}
R_f^{\mathrm{IT}}(E,E')=\frac{1}{\sigma_E
\sqrt{2\pi}}\exp{\frac{(E-E')^2}{2\sigma_E^2}}.
\end{equation}
The values for the effective relative energy resolution $\sigma_E$ and
the effective efficiency $\epsilon_f^{\mathrm{IT}}$ can be found and are
explained in Appendix~\ref{sec:det}.
\begin{figure}[ht!] \begin{center}
\includegraphics[angle=-90]{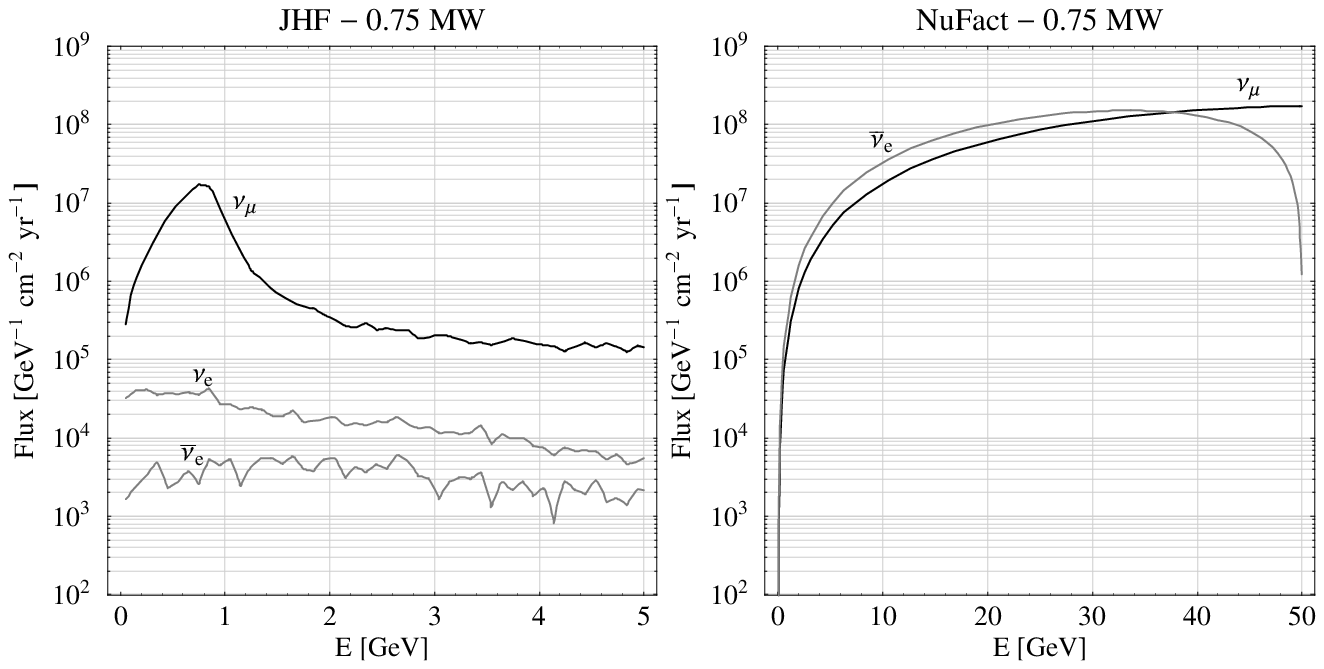} \end{center} \mycaption{\label{fig:flux}
Fluxes of the JHF beam (left--hand side, taken from \Ref~\cite{NAKAYA}) and
neutrino factory (right--hand side) at a distance of $1\,\mathrm{km}$ with a
target power of $0.75\,\mathrm{MW}$, corresponding to our initial, low
luminosity setups \JHFSK\ and \NuFactI. The anti neutrino beam spectrum for
the JHF beam is different and not shown. For a neutrino factory, the anti
neutrino beam has an identical flux spectrum.} \end{figure}
\begin{figure}[ht!]
\begin{center}
\includegraphics[angle=-90]{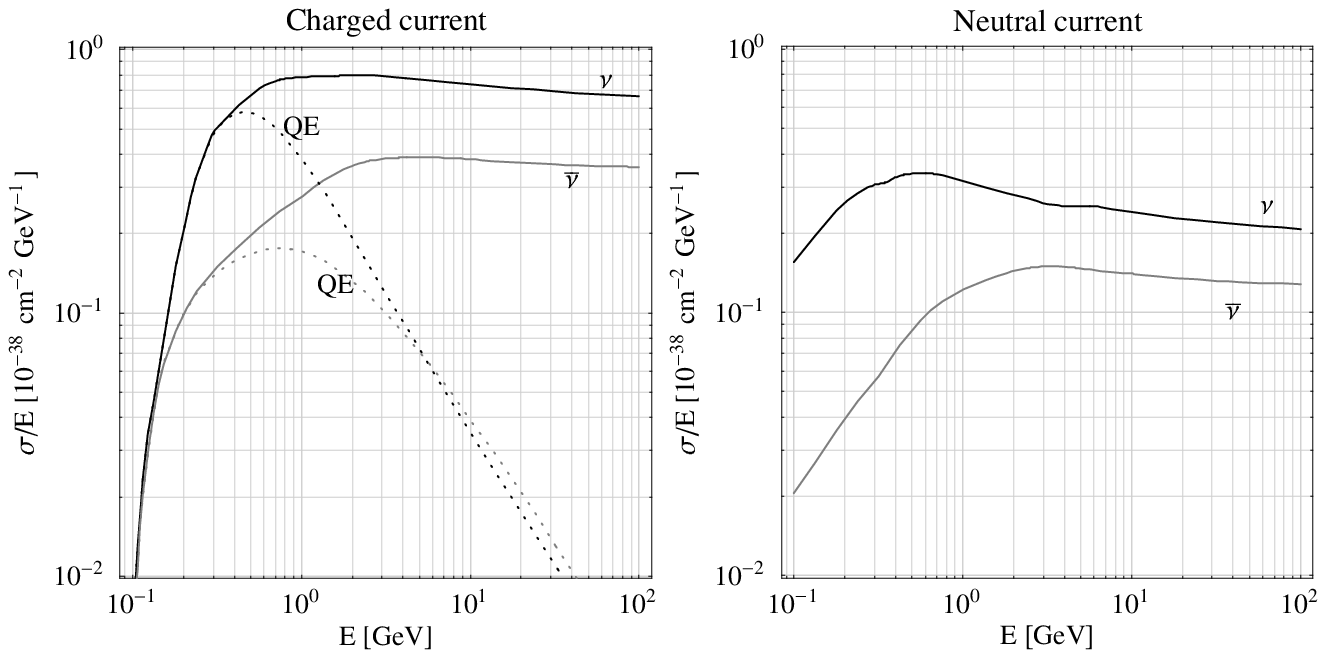}
\end{center}
\mycaption{\label{fig:xsec} The cross sections for the total charged current
(solid curves, left--hand plot), quasi--elastic charged current (dotted
curves, left--hand plot), and neutral current (right--hand plot) neutrino (black
curves) and anti neutrino (gray curves) interactions~\cite{MESSIER}. }
\end{figure}

%%%%%%%%%%%%%%%%%%%%%%%%%%%%%%%%%%%%%%%%%%%%%%%%%%%%%%%%%%%%%%%%%%%%%%%%%%%%%
\section{Detector description}
\label{sec:det}

\subsection{The JHF detector}

The detector for the JHF--beam is the Super--Kamiokande (\JHFSK) detector or
its $1 \,\mathrm{Mt}$ version Hyper--Kamiokande (\JHFHK). We assume that the
$1 \,\mathrm{Mt}$ detector has the same characteristics as the
Super--Kamiokande detector. This type of water Cherenkov detector has an
excellent capability to identify electrons and muons based on the topology of
the Cherenkov ring. Furthermore, the lepton momentum can be measured with a high
precision~\cite{Itow:2001ee}.

Within each considered setup, the signal events are given by $\nu_\mu$ and
$\nu_e$ CC interactions. The backgrounds are the $\nu_\mu$ NC events and the
$\nu_e$ CC events already contained in the beam. Another source 
of backgrounds for the appearance measurement is the misidentification of
muon neutrinos as electron neutrinos. The background
rejection factors and the signal efficiencies are taken from Table~2 in
\Ref~\cite{Itow:2001ee}. They are assumed to be constant, which should be a
reasonable approximation.
\begin{center}
\begin{table}[ht!] \begin{center} \begin{tabular}[h]{|l|lll|} \hline \hline
\multicolumn{4}{|l|}{Disappearance}\\
\hline
{\small Signal}&$0.9\otimes(\ruu)_\mathrm{QE}$&&\\
&&&\\
{\small Background}&$0.0056\otimes(\rux)_\mathrm{NC}$&&\\
\hline
\hline
\multicolumn{4}{|l|}{Appearance}\\
\hline
{\small Signal}&$0.505\otimes(\rue)_\mathrm{CC}$&&\\
&&&\\
{\small
Background}&$0.0056\otimes(\rux)_\mathrm{NC}$&$3.3\cdot10^{-4}\otimes(\ruu)_\mathrm{
CC}$&\\
{\small Beam
background}&$0.505\otimes(\ree)_\mathrm{CC}$&$0.505\otimes(\reeb)_\mathrm{CC}$&\\
\hline \hline \end{tabular}
\mycaption{\label{tab:jhf} The efficiencies for the signals and backgrounds of
the \JHFSK\ and \JHFHK\ experiments.} \end{center} \end{table}
\end{center}

\begin{figure}[ht!]
\begin{center}
\includegraphics[width=10cm]{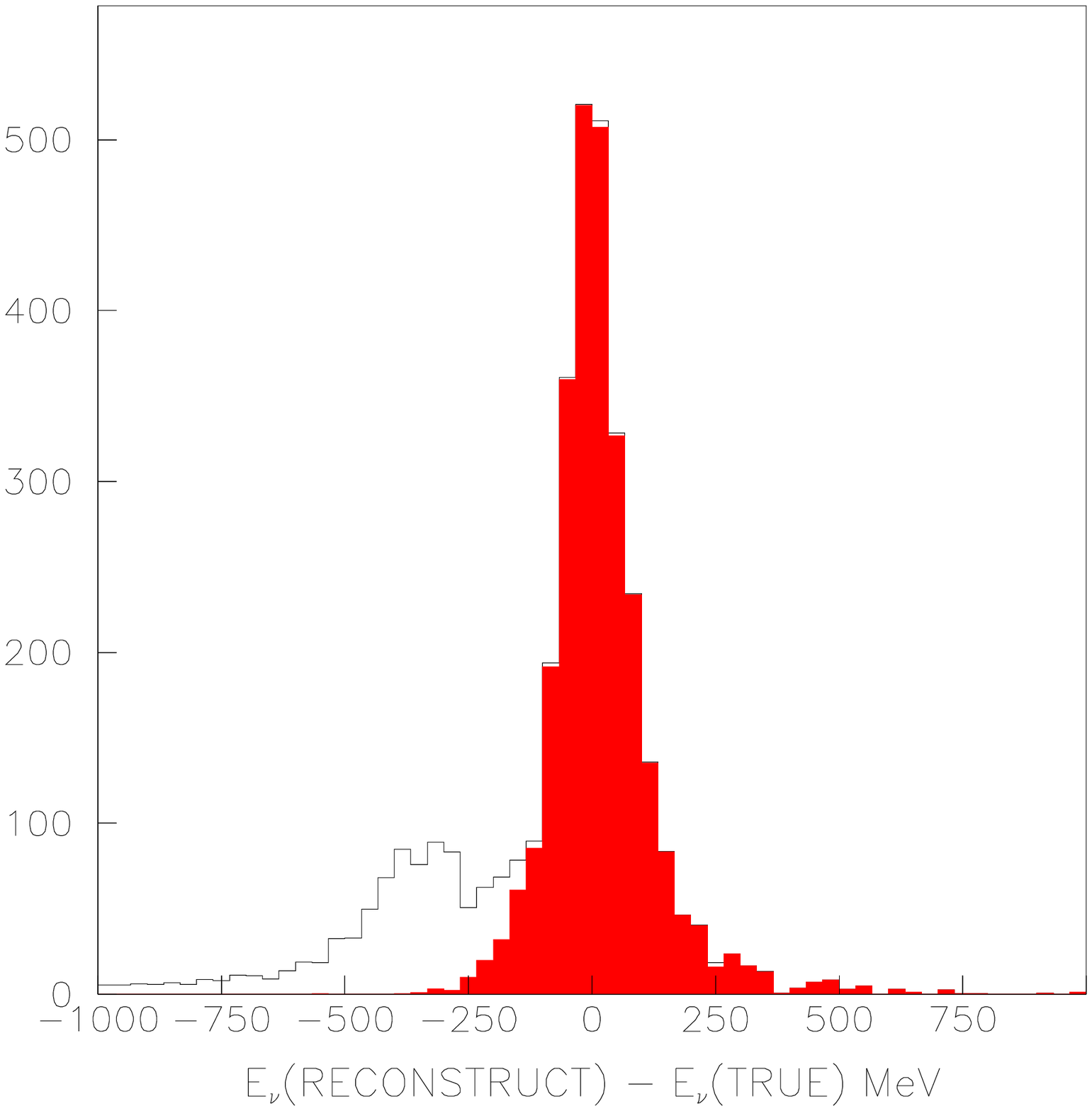}
\end{center}
\mycaption{\label{fig:skenergy} The energy resolution of the $\nu_\mu$ events
for the JHF--beam (figure taken from \Ref~\cite{Itow:2001ee}). The filled
histogram corresponds to the QE events and the open histogram to the
non--QE events.}
\end{figure}

Since a water Cherenkov detector is only sensitive to leptons, there is no
measurement of the hadronic energy deposition of a neutrino interaction.
Therefore, the analysis of the energy spectrum has to be constrained to the
QE event sample. The energy resolution for this sample is shown in
\fig~\ref{fig:skenergy} (filled histogram).
The accuracy of the energy reconstruction is limited by the Fermi--motion of the
nucleons, inducing a width of about $80-100 \, \mathrm{MeV}$~\cite{NAKAYA}, as
it can be also read off \fig~\ref{fig:skenergy}. The distribution is,
up to very good accuracy, Gaussian~\cite{NAKAYA}. The Fermi--motion
is energy independent and causes by far the largest error in the
energy reconstruction.  We use a constant width of $85 \, \mathrm{MeV}$
in our analysis, \ie, $\sigma_E = 0.085 \, \mathrm{GeV}$, and we have checked
that the results of the \JHFHK\ experiment depend only very little on this width
in the range of about $80-100 \, \mathrm{MeV}$. For the \JHFSK\ setup, the
energy resolution is not important since the statistics is too low to extract
spectral information.

In order to use the spectral information, we separate the signal event samples
into QE events and non--QE events. In the disappearance channel, there are
enough events to completely discard the non-QE events, as it is done
in~\cite{Itow:2001ee}. For the appearance channel, we adopt the following
strategy in order to optimally exploit the information in the data: We
use the total number of all CC events in the range $0.4- 1.2\,\mathrm{GeV}$, as
well as the QE spectrum with a free normalization in order to avoid a double
counting of QE events. This separation of rates and spectra is a well-known
technique, such as it is used in the solar analysis of Super--Kamiokande in
\Ref~\cite{Bahcall:2001cb}.

\subsection{The NuFact detector}

\begin{figure}[ht!]
\begin{center}
\includegraphics[width=12cm]{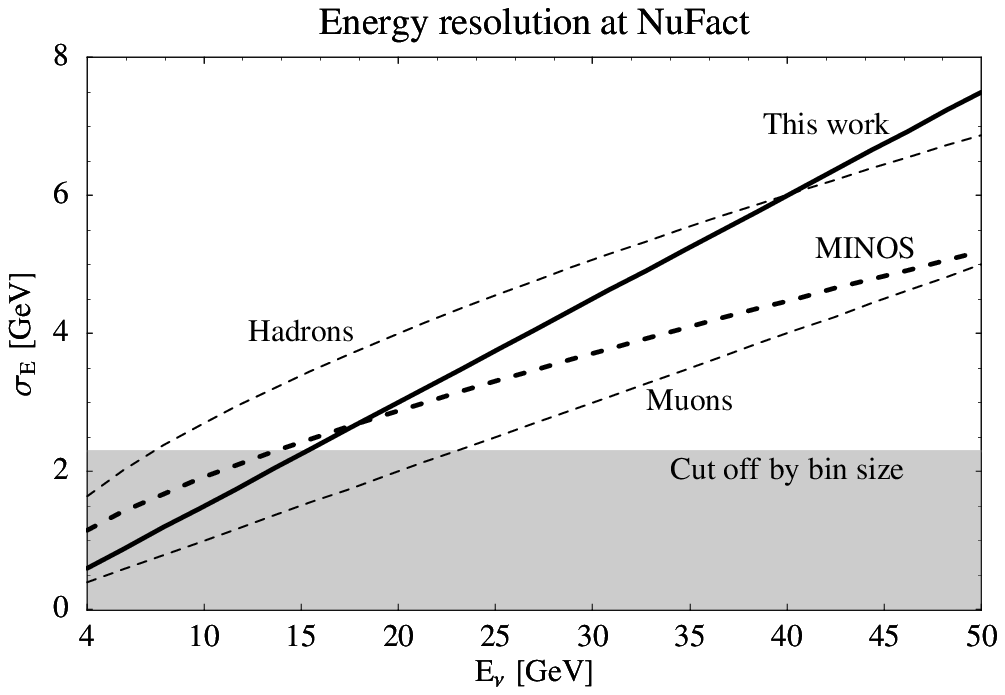}
\end{center}
\mycaption{\label{fig:nfenergy} The energy resolution $\sigma_E$ of $\nu_\mu$
events for a NuFact detector. The solid black curve represents the
energy resolution used in this work. The thick dashed curve shows the
resolution for the MINOS detector~\cite{Ables:1995wq}, assuming that the neutrino energy is
equally distributed between hadrons and muons. The thin dashed curves show the
hypothetical resolution of MINOS, if the neutrino energy went either
only to hadrons (upper curve) or muons (lower curve) only. The gray--shaded
area indicates the region which is cut off by the bin width used in our work.}
\end{figure}

The detector we consider for a neutrino factory is a magnetized iron
calorimeter, similar to the MINOS~\cite{Ables:1995wq} or
MONOLITH~\cite{Agafonova:2000xm} detectors. This detector type allows to measure
both the leptonic and hadronic energy deposition. In addition, the
background rejection for NC events is very good and the accuracy of the muon
charge measurement with the magnetic field of $\sim 1 \, \mathrm{T}$ is very
high. \fig~\ref{fig:nfenergy} shows the energy resolution $\sigma_E$, \ie, the
width of a Gaussian energy distribution curve, plotted as function of the
neutrino energy $E_\nu$. 
The Gaussian modeling of the energy resolution should be fairly
good, especially since the non--Gaussian tails are expected to be
small~\cite{Ables:1995wq}. The figure shows the curves for our analysis (thick
solid curve), the MINOS detector for an equal distribution of the energy between
hadrons and muons (thick dashed curve), and the MINOS detector for the
hypothetical case of all neutrino energy going to hadrons (upper thin dashed
curve) or muons (lower thin dashed curve) only.  As it can be seen from this
figure, we use the linear, quite conservative approximation $\sigma_E = 0.15
\, E_\nu$ of the energy resolution, since below $15 \, \mathrm{GeV}$ the energy
resolution is constrained by the bin size in our model. In order to have a more
accurate modeling of the problem and a better sensitivity to physical
quantities, one could even use smaller energy bins at low energies together with
more accurate parameterization for the energy resolution.

Another important issue for the NuFact detector is the minimum muon
momentum cut together with the background level. There existe essentially
 two detailed studies of detectors~\cite{Albright:2000xi,Cervera:2000vy},
 which find slightly different results.
One strategy is to optimize the signal to noise ratio, which is especially
important for poor statistics, \ie, a rather small number of events. The
authors of \Ref~\cite{Cervera:2000vy}, for example, use a minimum muon momentum
cut at $7.5 \, \mathrm{GeV}$ and a $q_t$ cut at $1\,\mathrm{GeV}$. These tight
cuts result in a low wrong-sign muon background level of $\sim 10^{-6}$ and a
small efficiency below $10\,\mathrm{GeV}$. Another possibility is to use a
lower muon momentum cut at $4 \,\mathrm{GeV}$, such as it is done in
\Ref~\cite{Albright:2000xi}. This results in a much higher background level of
$\sim 10^{-4}$ and better efficiencies at low and high energies. Since a
larger number of events in the low--energy regime is for the planned
precision measurements more important than maximum background suppression,
we follow an intermediate strategy, interpolating between those two cases. We
use an efficiency linearly rising from zero to one between $4 \,
\mathrm{GeV}$ and $20 \, \mathrm{GeV}$ together with an intermediate background
level of $10^{-5}$. This corresponds to a detector in between the setups studied
in \Refs~\cite{Albright:2000xi,Cervera:2000vy}. In
\Refs~\cite{CERVERA,Cervera:2000vy} it is also shown and clearly stated that the
S/N ratio does not strongly change for a cut on the muon momentum at
more than $5\,\mathrm{GeV}$ and a cut on $q_t$ at more than $0.5 \,
\mathrm{GeV}$. \begin{figure}[ht!] \begin{center}
\includegraphics[width=16cm]{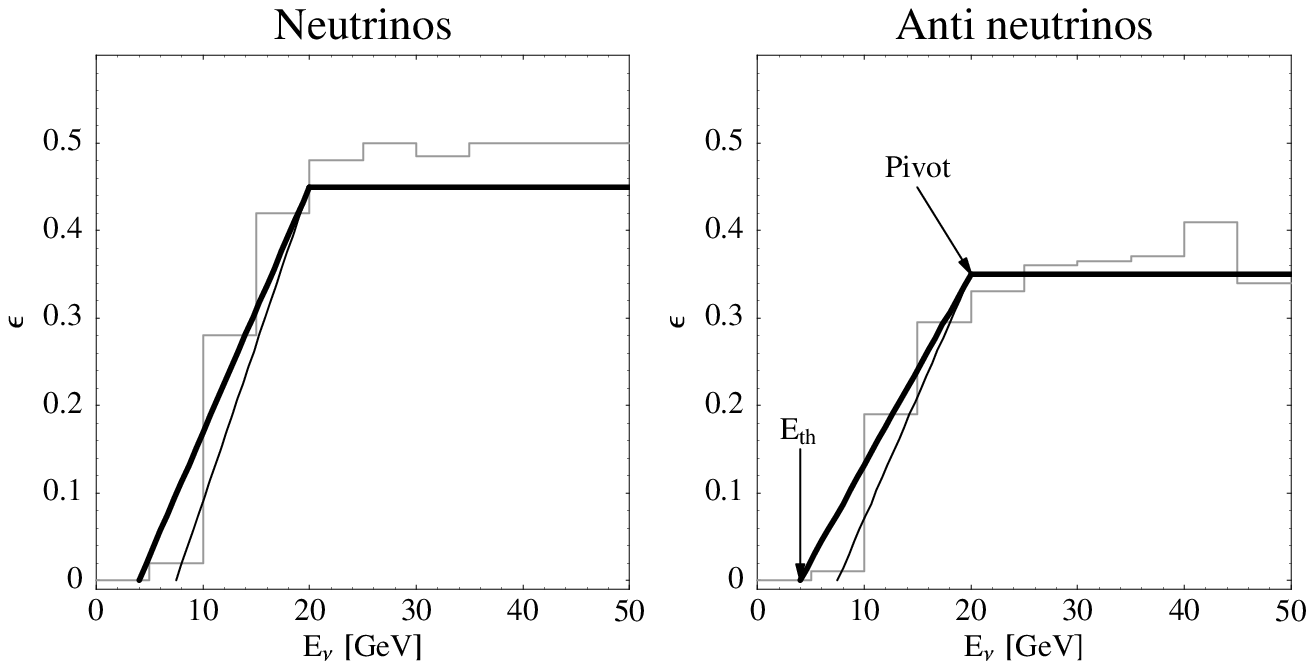}
\end{center}
\mycaption{\label{fig:nfeff} The signal efficiencies as functions of the
neutrino energy used for our NuFact setups (thick black curves). The gray
histograms are taken from \Ref~\cite{CERVERA}. The thin curves represent the
best linear fits between $0$ and $20 \, \mathrm{GeV}$ to the gray histograms.}
\end{figure}
In \fig~\ref{fig:nfeff}, the
efficiency functions used in this work are plotted as functions of the energy
as thick curves, whereas the gray histograms correspond to the efficiency
functions used in \Ref~\cite{CERVERA}. The thin curves represent linear
interpolations to these histograms. In this case, the efficiencies are zero at
$7.5\,\mathrm{GeV}$ and rise to their final values at $20\,\mathrm{GeV}$.

The signal at a neutrino factory consists of the $\nu_\mu$ disappearance and
the wrong-sign muon appearance signal. The backgrounds for these signals are
a certain fraction of the NC events for all flavors and the
misidentified fraction of the $\nu_\mu$--CC events. The values used in
this work are shown in \Tab~\ref{tab:nufact}.
\begin{center}
\begin{table}[ht!]
\begin{center}
\begin{tabular}[h]{|l|lll|}
\hline
\hline
\multicolumn{4}{|l|}{Disappearance -- $\mu^-$ stored}\\
\hline
Signal&$0.45\otimes(\ruu)_\mathrm{CC}$&&\\
&&&\\
Background&$10^{-5}\otimes(\rux)_\mathrm{NC}$&&\\
\hline
\multicolumn{4}{|l|}{Disappearance -- $\mu^+$ stored}\\
\hline
Signal&$0.35\otimes(\ruub)_\mathrm{CC}$&&\\
&&&\\
Background&$10^{-5}\otimes(\ruxb)_\mathrm{NC}$&&\\
\hline
\hline
\multicolumn{4}{|l|}{Appearance -- $\mu^-$ stored}\\
\hline
Signal&$0.45\otimes(\reub)_\mathrm{CC}$&&\\
&&&\\
Background&$5\cdot10^{-6}\otimes(\rux)_\mathrm{NC}$ &
$5\cdot10^{-6}\otimes(\ruu)_\mathrm{CC}$ & \\ \hline
\multicolumn{4}{|l|}{Appearance -- $\mu^+$ stored}\\ \hline
Signal&$0.35\otimes(\reu)_\mathrm{CC}$&&\\
&&&\\
Background&$5\cdot10^{-6}\otimes(\ruxb)_\mathrm{NC}$&
$5\cdot10^{-6}\otimes(\ruub)_\mathrm{CC}$&\\ \hline \hline
\end{tabular}
\mycaption{\label{tab:nufact} The high energy efficiencies for the signals and
backgrounds of the \NuFactI\ and \NuFactII\ experiments.} \end{center}
\end{table}
\end{center}
These high energy efficiencies are taken from \Ref~\cite{CERVERA} and the low
energy efficiencies from \Ref~\cite{Albright:2000xi}. The background
level is an interpolation between the two mentioned references to take into
account the lower muon momentum cut. The high energy efficiencies of
\Ref~\cite{Albright:2000xi} are $\sim 50\%$ for both neutrinos and
anti neutrinos, which are considerably larger than the ones used in this work.

\begin{figure}[ht!]
\begin{center}
\includegraphics[height=15.5cm,angle=-90]{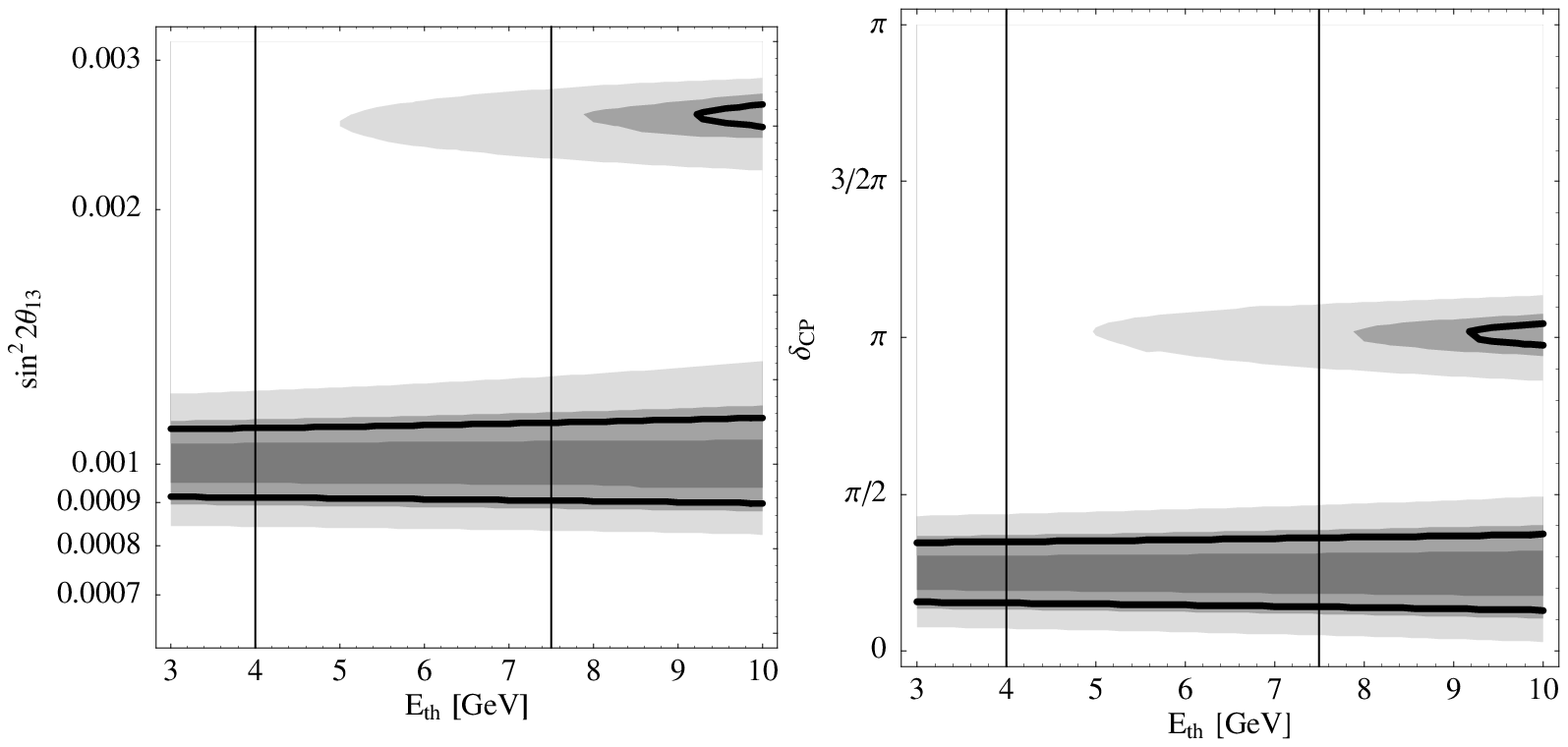}
\end{center}
\mycaption{\label{fig:depT} The allowed ranges in
$\sin^22\theta_{13}$ (left--handed panel) and $\deltacp$ (right--handed panel)
as functions of the threshold energy $E_{th}$ for the starting values
$\sin^22\theta_{13}=10^{-3}$, $\deltacp=\pi/4$, and the LMA solution. The black
solid curves show the results at the $90\%$ CL used in this work. The shading
marks the $1\,\sigma$ (dark gray), $2\,\sigma$ (medium gray), and $3\,\sigma$
(light gray) allowed ranges. The left vertical lines corresponds to the black
solid curves in \fig~\ref{fig:nfeff} and the right vertical lines to the thin
black curves in~\fig~\ref{fig:nfeff}. The shaded ranges are the results of
two--dimensional fits of $\sin^22\theta_{13}$ and $\deltacp$, which means that
multi--parameter correlations are neglected in this plot.} \end{figure}

\begin{figure}[ht!]
\begin{center}
\includegraphics[width=16cm]{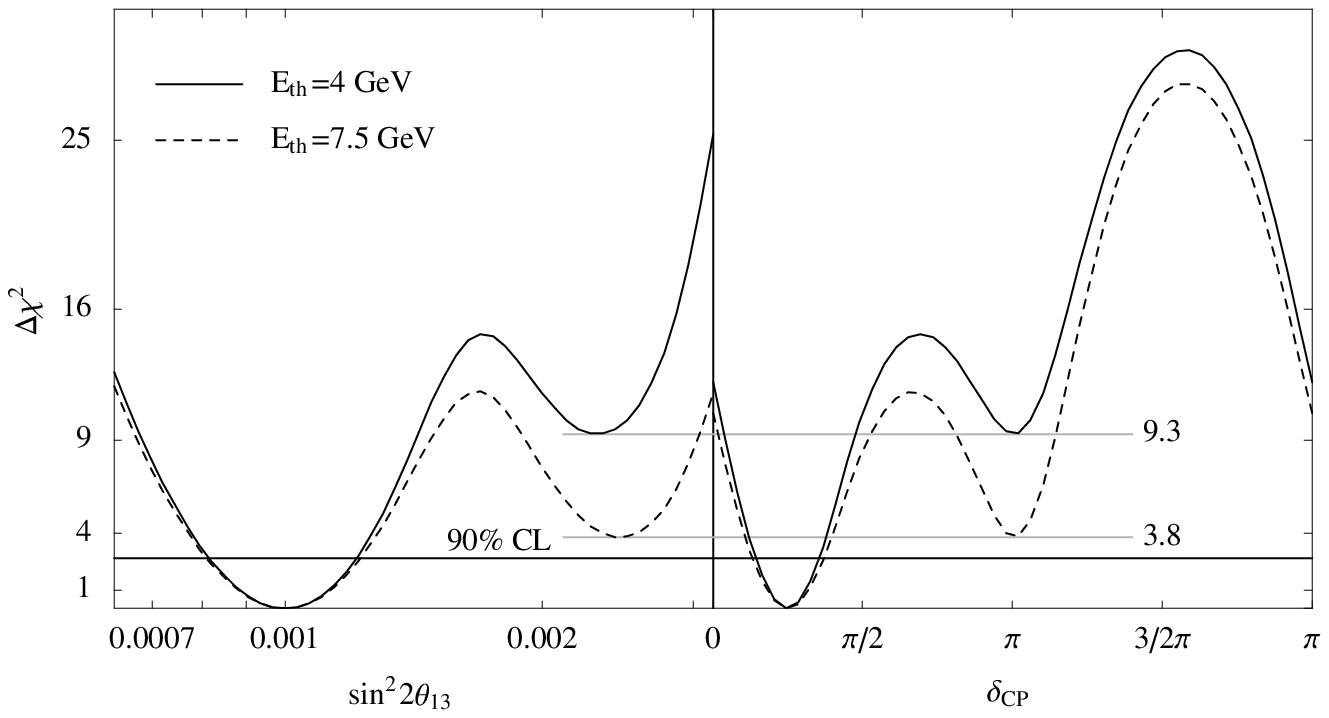}
\end{center}
\mycaption{\label{fig:fdepT} The $\Delta\chi^2$--functions of a full analysis,
as they are used throughout this work, as functions of $\sin^22\theta_{13}$
(left--handed panel) and $\deltacp$ (right--handed panel). The used parameter
values are $\sin^22\theta_{13}=10^{-3}$, $\deltacp=\pi/4$, and the LMA solution.
The solid curves correspond to the thick black curves
in \fig~\ref{fig:nfeff} and the dashed curves correspond to the
thin black curves in \fig~\ref{fig:nfeff}. The horizontal lines show the
confidence level used in this work.} \end{figure}

In order to illustrate the effects of the energy threshold function, which is
determined by the muon momentum cut, we show in \fig~\ref{fig:depT} the
dependence of our results for $\deltacp$ and $\sin^22\theta_{13}$ on the muon
momentum cut $E_{th}$. We parameterize the threshold function by keeping the
point at $20\,\mathrm{GeV}$ in \fig~\ref{fig:nfeff}, labeled as pivot, fixed and
moving the point labeled by $E_{th}$.
In \fig~\ref{fig:depT} the allowed ranges in $\deltacp$ (right--handed panel)
and in $\sin^22\theta_{13}$ (left--handed panel) are shown as functions of
the threshold energy $E_{th}$. The starting values used are the
LMA values, $\deltacp=\pi/4$, and $\sin^22\theta_{13}=10^{-3}$. The allowed
ranges are determined by projection of the \emph{two}--dimensional\footnote{In
this figure, multi--parameter correlations are neglected compared to our full
analysis.} $\deltacp -\sin^22\theta_{13}$ allowed region onto the $\deltacp$ and
$\sin^22\theta_{13}$ axis, respectively. The gray shading corresponds to the
$1\,\sigma$ (dark gray), $2\,\sigma$ (medium gray), and $3\,\sigma$ (light gray)
allowed ranges. The thick black curves represents the $90\%$ CL used in this
work. The left vertical lines mark the value for the threshold function used in
this work and corresponds to the thick black curves in \fig~\ref{fig:nfeff}. The
right vertical lines mark the value for the threshold function given by the thin
black curves in \fig~\ref{fig:nfeff}, which very closely approximates the gray
histogram with the efficiencies used in~\Ref~\cite{CERVERA}.
\fig~\ref{fig:fdepT} clearly demonstrates that changing the threshold value
mainly changes the confidence level at which the second solution of the
$(\delta,\theta_{13})$ ambiguity appears. For the $90\%$ CL used in this work,
this dependency is obviously very small, since neither does the range around the
true solution increase much, nor does the second solution appear at the $90\%$
CL.
This can be also seen in \fig~\ref{fig:fdepT}, where the
results of a full analysis, as used throughout this work, are shown.
It shows $\Delta\chi^2$ as function of $\sin^22\theta_{13}$
(left--handed panel) and $\deltacp$ (right--handed panel). The solid
curves show the results obtained with $E_{th}=4 \, \mathrm{GeV}$, as used in
this work, and indicated by the thick black curves in \fig~\ref{fig:nfeff}. The
dashed curves represent the result obtained with $E_{th}=7.5\,\mathrm{GeV}$,
corresponding to the thin black curves in \fig~\ref{fig:nfeff}. The solid and
dashed curves are practically identical around the true solution. Only at the
degenerate solution from the $(\delta, \theta_{13})$ ambiguity a
noteworthy difference occurs. It results in a lowering of the minimum
$\Delta\chi^2$ of the degenerate solution from $9.3$ to $3.8$, which means
that the degenerate solution now appears at the $2\,\sigma$ level instead of
the $3\,\sigma$ confidence level. Still, it does not appear at the $90\%$ CL
used in this work. In general, the qualitative impact of the threshold value is
small and its quantitative influence strongly depends on the selected confidence
level. In both cases, $E_{th}=4\,\mathrm{GeV}$ and $E_{th}=7.5\,\mathrm{GeV}$,
the degenerate solution is present, but at different confidence levels. 
Furthermore, a change of the oscillation parameters, such as especially of
$\dm{21}$, can change the confidence level at which the degenerate solution 
appears. Finally, it is not yet clear which strategy -- low momentum cut with
a high background level or vice versa -- will be most efficient for the
quantities of interest. This discussion shows that more detailed studies of 
detector properties and detector concepts  will be needed in order to optimally
balance the threshold versus background level for optimum sensitivities.

%%%%%%%%%%%%%%%%%%%%%%%%%%%%%%%%%%%%%%%%%%%%%%%%%%%%%%%%%%%%%%%%%%%%%%%%%%%%%%
\section{Statistics and the treatment of systematical errors}
\label{sec:statistics}

In order to evaluate the physics potential of each setup, we compute spectral
event rate spectra for a certain set of (true) oscillation parameters
$\lambda_0$ and compare it to a parameter set $\lambda$ by a $\chi^2$--analysis.
In this paper, this analysis aims to obtain statistically reliable statements
about possible measurements of $\theta_{13}$, $\mathrm{sgn} (\dm{31}^2)$, and
$\deltacp$. The fit to the data is performed including all available
information, \ie, using all types of signals. These are the appearance
and disappearance channels for the neutrino beam and appearance and
disappearance channels for the anti neutrino beam (which are not available for
the \JHFSK\ setup). In this appendix, we refer to the types of signals by
$\alpha$. It turns out to be extremely important to include the disappearance
signals and the energy information in the analysis, although we are only
interested in the measurements of $\theta_{13}$, $\mathrm{sgn} (\dm{31}^2)$, and
$\deltacp$. Since these measurements require excellent knowledge about the
atmospherical oscillation parameters $\dm{31}$ and $\theta_{23}$, and the
precision from other experiments, such as the MINOS
experiment~\cite{Barger:2001yx}, is not comparable to the one of the experiments
described in this work, the experiments themselves will supply this knowledge.
In addition, the spectral information in the appearance channels helps to
improve the resolution of degeneracies (\eg, see \figs~\ref{fig:degsjhf}
and~\ref{fig:degsnufact}).

We use the full set of the standard three--flavor oscillation parameters, which
are $\dm{21}$, $\theta_{12}$, $\dm{31}$, $\theta_{23}$, $\theta_{13}$, and
$\deltacp$. The appearance probability in \eq~(\ref{LLIMIT}), however,
only depends on the solar parameter product
$\pi_\mathrm{sol} \equiv \sin2\theta_{12}\cdot\dm{21}$. Therefore, we use
this product $\pi_\mathrm{sol}$ as the relevant parameter reducing the number
of oscillation parameters to five. In summary, we call our set of oscillation
parameters $\lambda \equiv \{\pi_\mathrm{sol},\dm{31},
\theta_{23},\theta_{13},\deltacp \}$. This is the only approximation
introduced for the treatment of the transition probabilities.

As basis for a statistical analysis, we use the standard  $\chi^2$-function for
Poissonian distributed quantities, such as given in~\cite{PDG}:
\begin{eqnarray} \chi^2&=&
\sum_{i=1}^b\left( 2 [\langle x_i \rangle - x_i ]+2
  \thinspace x_i \log \frac{x_i}{\langle x_i \rangle
    }\right).
\label{eq:chi2}
\end{eqnarray}
Here, $b$ is the number of bins, $x_i$ is the number of events in the $i$th
bin, and $\langle x_i \rangle $ is the expectation value of the number of
events in the $i$th bin. Though this method is in this form quite simple, it
is not easy to include systematical errors and external information in a
consistent and simple way. To implement systematical errors and external
information, we parameterize the considered problem linearly and restrict it to
a certain interval given by its uncertainty.

For each bin $i$, the total number of events is the sum of
signal and background events, \ie, \begin{equation} x_i=s_i+b_i,
\end{equation}
where $s_i$ is the number of signal events and $b_i$ is the number of background
events, such as it is defined in the previous appendix. We introduce
systematical errors in $x_i$ and $b_i$ by modifying the signal $s_i^0$ and
background $b_i^0$ raw event rates from the previous appendix by
\begin{eqnarray}
s_i=s_i(n_s,t_s)=s_i^0\,(1+n_s+t_s\cdot
E_i/(E_\mathrm{max}-E_\mathrm{min}))\,,\nonumber\\
b_i=b_i(n_b,t_b)=b_i^0\,(1+n_b+t_b\cdot E_i/(E_\mathrm{max}-E_\mathrm{min})),
\end{eqnarray}
where $E_i$ is the average energy in the $i$th bin, $E_\mathrm{max}$ is the
maximum energy, and $E_\mathrm{min}$ is the minimum energy of all bins. The
auxiliary parameter $n$ is called ``normalization'' and describes the effect of
an overall change in the magnitude of signal or background, whereas the
parameter $t$ is called ``tilt'' and describes a linear distortion in the
spectral shape of the signal or background. This distortion can come from
limited knowledge of the background or an energy calibration error. The total
number of events in the $i$--th bin $x_i$ now becomes
\begin{equation}
x_i=x_i(n_s,t_s,n_b,t_b)=s_i(n_s,t_s)+b_i(n_b,t_b). \end{equation} Note that
$n^\alpha_s$, $t^\alpha_s$, $n^\alpha_b$, and $t^\alpha_b$ are independent
parameters and we have a different set of these four parameters for
each channel $\alpha$. Finally, the $\chi^2$--function does not only depend on
the five oscillation parameters and the matter density $\rho$, but also on the
auxiliary parameters $n^\alpha_s,t^\alpha_s,n^\alpha_b$ and $t^\alpha_b$:
\begin{equation}
\chi^2_\alpha=\chi^2_\alpha(\lambda,\rho;n^\alpha_s,t^\alpha_s,n^\alpha_b,t^\alpha_b)\,.
\end{equation}

In order to implement the uncertainties of the auxiliary
parameters, we assume that the values of
$n^\alpha_s$, $t^\alpha_s$, $n^\alpha_b$, $t^\alpha_b$ are known with
some accuracy $\sigma$, \ie, are distributed in a Gaussian way with the mean
value $0$ and the standard deviation $\sigma$. The standard deviation $\sigma$
corresponds to the systematical uncertainty of these quantities, whereas the
statistical uncertainty is already taken into account by the form of the
$\chi^2$-function in \eq~(\ref{eq:chi2}). Because of the likelihood principle,
\ie, the $\chi^2$--function in terms of the likelihood $\mathcal{L}$ given by
$-2\log \mathcal{L}$, the restriction to a Gaussian with mean $\mu$ and standard
deviation $\sigma$ leads to the additional term $(x-\mu)^2/\sigma^2$ in the
$\chi^2$-function. For each of the auxiliary parameters, we include this
additional term and minimize with respect to it:
\begin{equation}
\hat{\chi}^2_\alpha(\lambda,\rho)=\min_{n^\alpha_s,t^\alpha_s,n^\alpha_b,
t^\alpha_b}\left(\chi^2_\alpha(\lambda,\rho;n^\alpha_s,t^\alpha_s,n^\alpha_b,
t^\alpha_b) +\frac{(n^\alpha_s)^2}{\sigma^2_{n^\alpha_s}}
+\frac{(t^\alpha_s)^2}{\sigma^2_{t^\alpha_s}}
+\frac{(n^\alpha_b)^2}{\sigma^2_{n^\alpha_b}}
+\frac{(t^\alpha_b)^2}{\sigma^2_{t^\alpha_b}}\right).
\label{eq:normtilts}
\end{equation}
In this
equation, $\hat{\chi}^2_\alpha(\lambda,\rho)$ is the marginalized (with respect
to $n^\alpha_s,t^\alpha_s, n^\alpha_b,t^\alpha_b$)  distribution of
$\chi^2_\alpha$. With this step, we have eliminated a part of the nuisance
parameters which have been assumed to be fully uncorrelated. In principle, it is
also possible to perform this marginalization for all nuisance parameters and
the projection onto the parameter of interest $\eta$ in one step. However, it
saves computation time to use a two--step method, since the first step does not
require the re--computation of oscillation probabilities. For both steps, we use
a fast ``direction set'' method to find the minimum. The algorithm
and its implementation in C are taken from~\cite[pp. 420--425]{NumRep}. The
advantages of using this minimization algorithm are its high speed and its
excellent accuracy compared to a brute force grid--based methods.
Furthermore, it does not require the computation of derivatives of the function
to be minimized, because derivatives of numerical functions are very
unstable and quite slow to be computed. However, such a heuristics does not
completely run independently without using some expert knowledge from the field.

Similarly to the inclusion of the normalizations and tilts, we can use the same
procedure for the matter density $\rho$, where the mean value and uncertainty
are taken from seismic wave measurements, and the solar parameters
$\pi_\mathrm{sol}$, where the mean value and uncertainty are expected to be
measured by the KamLAND experiment. Finally, we project onto the parameter
of interest $\eta$ in order to obtain
\begin{equation}
\chi^2_\mathrm{F}(\eta)=\min_{\bar{\lambda},\rho}\left(
\sum_{\alpha}\hat{\chi}^2_\alpha(\lambda,\rho) +
\frac{(\rho-\rho^0)^2}{\sigma^2_{\rho}}+
\frac{(\pi_\mathrm{sol}-\pi^0_\mathrm{sol})^2}{\sigma^2_{\pi_\mathrm{sol}}}
\right), \end{equation}
where $\bar{\lambda}$ refers to the set of oscillation parameters without
$\eta$. We use for all
setups~\cite{BARGER,Gonzalez-Garcia:2001zy,Geller:2001ix}
$$
\sigma_{\pi_\mathrm{sol}}=0.15\,\pi_\mathrm{sol}^0\quad
\mathrm{and}\quad\sigma_\rho=0.05\,\rho^0. $$
The $1\sigma$ uncertainties for the auxiliary parameters are for all
setups given in \Tab~\ref{tab:aux}.
In our analysis, it turns out that for all these errors a variation by a
factor of two does not affect the final results very much -- with the exception
of $\sigma_{n_b}$ for the appearance channels of the \JHFHK\ setup. For a more
comprehensive discussion, we refer to \Secs~\ref{sec:impactfactors}
and~\ref{sec:theta13sens}.

\begin{center}
\begin{table}[hbt!]
\begin{center}
\begin{tabular}[h]{|l|llll|}
\hline
&$\sigma_{n_s}$&$\sigma_{t_s}$&$\sigma_{n_b}$&$\sigma_{t_b}$\\
\hline
\hline
Disappearance -- NuFact&0.01&0.05&0.05&0.05\\
Appearance -- NuFact&0.01&0.05&0.2&0.05\\
\hline
Disappearance -- JHF&0.05&0.025&0.2&0.025\\
Appearance -- JHF&0.05&0.025&0.05&0.05\\
\hline
\end{tabular}
\mycaption{\label{tab:aux} Systematical uncertainties for our setups from
\Refs~\cite{Agafonova:2000xm,GEISER,Cervera:2000vy, Blondel:2000gj} for the
neutrino factory experiments and from \Refs~\cite{Itow:2001ee,SK1,SK2,SK3,SK4}
for the JHF experiments.}
\end{center} \end{table} \end{center}

The error or sensitivity for the parameter of interest $\eta$ is defined as
the range of $\eta$ where $\chi^2_\mathrm{F}(\eta)\leq\chi^2_\mathrm{crit}$. Here
$\chi^2_\mathrm{crit}$ is determined by the $\chi^2$--distribution with one degree
of freedom, \ie, $\chi^2_\mathrm{crit}=1$ on $1 \sigma$ confidence level,
$\chi^2_\mathrm{crit}=2.71$ for $90 \%$ confidence level, and
$\chi^2_\mathrm{crit}=4$ for $2 \sigma$ confidence level. Specifically, the
sensitivity limit for $\eta=\sin^2 2\theta_{13}$ is defined by
$\chi^2_\mathrm{F}(\eta,\eta^0=0)$, the CP violation sensitivity is given by
$\min \left\{\chi^2_\mathrm{F}(\deltacp,\deltacp^0=0),
\chi^2_\mathrm{F}(\deltacp,\deltacp^0=\pi)\right\}$, and the sensitivity
to the sign of $\dm{31}$ is obtained by minimizing the
$\chi^2$--function over all possible parameter sets with the sign of $\dm{31}$
opposite to the sign at the starting point $\lambda^0$.

\end{appendix}

\end{document}